\documentclass[twocolumn]{aastex631}

\shorttitle{The Way of Water}
\shortauthors{F. Perrotta et al.}

\begin{document}

\title{The Way of Water: \\ 
ALMA resolves H$_2$O emission lines in a strongly lensed dusty star-forming galaxy at z $\sim$ 3.1}

\author[0000-0001-9808-0843]{Francesca Perrotta}
\affiliation{SISSA, Via Bonomea 265, 34136 Trieste, Italy}

\author[0000-0002-1847-4496]{Marika Giulietti}
\affiliation{SISSA, Via Bonomea 265, 34136 Trieste, Italy}\affiliation{IRA-INAF, Via Gobetti 101, 40129 Bologna, Italy}

\author[0000-0002-0375-8330]{Marcella Massardi}
\affiliation{Italian ARC, IRA-INAF, Via Piero Gobetti 101, 40129 Bologna, Italy}\affiliation{SISSA, Via Bonomea 265, 34136 Trieste, Italy}

\author[0000-0003-3248-5666]{Giovanni Gandolfi}
\affiliation{SISSA, Via Bonomea 265, 34136 Trieste, Italy}\affiliation{IFPU, Via Beirut 2, 34014 Trieste, Italy}\affiliation{INFN-Sezione di Trieste, via Valerio 2, 34127 Trieste, Italy}

\author[0000-0002-3515-6801]{Tommaso Ronconi}
\affiliation{SISSA, Via Bonomea 265, 34136 Trieste, Italy}\affiliation{IFPU, Via Beirut 2, 34014 Trieste, Italy}

\author[0000-0001-7883-496X]{Maria Vittoria Zanchettin}
\affiliation{SISSA, Via Bonomea 265, 34136 Trieste, Italy}\affiliation{OATS-INAF, Via Tiepolo 11, 34143 Trieste, Italy}

\author[0000-0002-9948-0897]{Quirino D' Amato}
\affiliation{SISSA, Via Bonomea 265, 34136 Trieste, Italy}\affiliation{IRA-INAF, Via Gobetti 101, 40129 Bologna, Italy}

\author[0000-0002-6444-8547]{Meriem Behiri}
\affiliation{SISSA, Via Bonomea 265, 34136 Trieste, Italy}

\author[0000-0001-9301-5209]{Martina Torsello}
\affiliation{SISSA, Via Bonomea 265, 34136 Trieste, Italy}

\author[0000-0003-3103-9170]{Francesco Gabrielli}
\affiliation{SISSA, Via Bonomea 265, 34136 Trieste, Italy}

\author[0000-0003-3127-922X]{Lumen Boco}
\affiliation{SISSA, Via Bonomea 265, 34136 Trieste, Italy}\affiliation{IFPU, Via Beirut 2, 34014 Trieste, Italy}

\author[0000-0003-1394-7044]{Vincenzo Galluzzi}
\affiliation{IRA-INAF, Via Gobetti 101, 40129 Bologna, Italy}\affiliation{OATS-INAF, Via Tiepolo 11, 34143 Trieste, Italy}

\author[0000-0002-4882-1735]{Andrea Lapi}
\affiliation{SISSA, Via Bonomea 265, 34136 Trieste, Italy}\affiliation{IRA-INAF, Via Gobetti 101, 40129 Bologna, Italy}\affiliation{IFPU, Via Beirut 2, 34014 Trieste, Italy}\affiliation{INFN-Sezione di Trieste, via Valerio 2, 34127 Trieste,  Italy}

\correspondingauthor{F. Perrotta}
\email{perrotta@sissa.it}

\begin{abstract}
We report ALMA high-resolution ($\lesssim$ 0.3 arcsec) observations of water emission lines $p-{\rm{H_2O}} (2_{02}-1_{11}$), $o-{\rm{H_2O}} (3_{21}-3_{12})$, $p-{\rm{H_2O}} (4_{22}-4_{13})$, in the strongly lensed galaxy HATLASJ113526.2-01460 at redshift z $\sim$ 3.1. From the lensing-reconstructed maps of water emission and line profiles, we infer the general physical properties of the ISM in the molecular clouds where the lines arise.  We find that the water vapor lines $o-{\rm{H_2O}} (3_{21}-3_{12})$, $p-{\rm{H_2O}} (4_{22}-4_{13})$  are mainly excited by FIR pumping from dust radiation in a warm and dense environment, with dust temperatures ranging from 70 K to $\sim 100$ K, as suggested by the line ratios. The $p-{\rm{H_2O}} (2_{02}-1_{11})$ line instead, is excited by a complex interplay between FIR pumping and collisional excitation in the dense core of the star-forming region. This scenario is also supported by the detection of the medium-level excitation of CO resulting in the line emission CO (J=8-7). Thanks to the unprecedented high resolution offered by the combination of ALMA capabilities and gravitational lensing, we discern the different phases of the ISM and locate the hot molecular clouds into a physical scale of $\sim$ 500 pc. We discuss the possibility of J1135 hosting an AGN in its accretion phase.  Finally, we determine the relation between the water emission lines and the total IR luminosity of J1135, as well as the SFR as a function of water emission intensities, comparing the outcomes to local and high-$z$ galactic samples from the literature.   
 \end{abstract}


\section{Introduction}
\label{sec:intro}
A key goal of modern physical cosmology is to understand the process by which an initially uniform dense gas distribution gave birth to gravitationally bound galaxies and their stellar populations. It is now well established that, over a relatively short period around redshift z $\sim$ 2-3 (the 'cosmic noon'), the overall galaxy population formed about half of their current stellar mass.  At these early cosmic times, a  major contribution to the Star Formation Rate Density (SFRD) came from the dustiest and gas-rich objects, nourishing SFRs as high as $10^3\, M_\odot$ yr$ ^{-1}$ (\citealt{Blain1996}, \citealt{Casey2014}). Because of this huge dust content, these objects appear heavily obscured in the optical/UV bands, where the photons from the most massive newborn stars are absorbed and reprocessed by the dust component of molecular clouds, producing a balancing extremely bright continuum flux in the far-infrared (FIR) and sub-millimeter (sub-mm) bands ($S_{870 \mu\,m} \geq 1$ mJy), which justifies their denomination of Sub-Millimeter Galaxies (SMGs) (\citealt{Carilli}, \citealt{Blain2002}).

Dusty Star-Forming Galaxies have been identified as the 
progenitors of massive quiescent
early-type galaxies. In fact, they are
found to be abundant at high redshift ($z > 1$), they are the main contributors to the total star formation
history at  $1 <z < 4$,  and have been detected up to z $\sim 6$ (\citealt{Simpson2014}).  

Observationally, these high-redshift SMGs appear to have a large reservoir of molecular gas, that fuels their extreme star formation activity, which in turn requires an efficient cooling process via continuum and line photon emission. Together with the continuum emission observations, spectroscopic analysis can shed light on the basic cooling mechanisms necessary for their substantial star formation process. Cooling through molecular/atomic emission lines becomes effective as long as collisions with collisional partners result in the excitation of atomic or molecular electronic, rotational, or vibrational quantum states, followed by radiative emission and de-excitation. 

In this framework, several molecular tracers of star formation have been exploited. CO rotational lines significantly contribute to the cooling process of the molecular clouds, with a spectral line energy distribution responding to the local gas density and kinetic temperature. Non-local thermodynamic equilibrium (non-LTE) models including multi-component molecular gas are usually required to constrain the CO spectral line energy distribution (SLED) of high-redshift galaxies. Nevertheless, observations of individual high-$z$ galaxies show that the low-J CO lines, with their minimal excitation requirements, are mostly collisionally excited by an extended cool molecular component, thus tracing the bulk of the H$_{2}$ gas.
 In particular, the CO(1–0) emission
is dominated by the dense gas component; at higher redshifts (n(H$_2$) $\sim 10^{5}$ cm$^{-3}$), galaxies have a higher gas fraction and a denser gas than at low redshifts,  which implies that the CO-to-H$_2$ conversion factor should be higher than for local galaxies (\citealt{Weiss2007}). 
Furthermore, this conversion factor is also dependent on metallicities, 
making it more unreliable especially at high-$z$ (\citealt{Bolatto2013}). 

On the other hand, the mid/high-J CO emission is responsive to a warm molecular component, with a volume density of $10^3-10^4$ cm$^{-3}$, thus probing the molecular gas that fuels the star formation activity (\citealt{Weiss2007}, \citealt{Ivison2010}, \citealt{Danielson2011}). A caveat is that high-J CO transitions are plausibly also excited in X-ray dominated regions associated with AGNs (\citealt{Carilli}, \citealt{Kirkpatrick2019}, \citealt{Pozzi17}, \citealt{Mingozzi18}), which can bias the determination of the Star Formation Rate (SFR) in all those cases in which a mild and/or obscured AGN cooperates with the star formation to the gas heating and excitation. 

Another interstellar molecule of key importance in the astrophysical environments of star-forming galaxies is H$_2$O.  Milky Way observations (\citealt{Caselli2010}) suggest that water abundance in the gas phase, defined as $ \left[ \rm{H_20} \right] / \left[\rm{H}_2 \right] = \rm{X(H_20)} $, is quite low in cold molecular clouds, where it is strongly depleted on dust grains to levels $\rm{X(H_{2}O)} < 10^{-9}$. However, water  becomes the third most abundant species (after H$_2$ and CO) in dense warm regions, where star formation raises the dust temperature above the ice evaporation limit, or in shock-heated regions such as those dominated by AGN-driven winds (\citealt{Bergin2003}; \citealt{Gonzalez2013A}, \citealt{Cernicharo2006},  \citealt{Pensabene2022}).
This makes water a unique and powerful tracer of the gas component involved in the highly energetic processes associated with compact nuclear starbursts or with the extreme environments of AGNs, potentially even more reliable than other molecular gas tracers (such as CO and HCN) traditionally used to probe the densest ISM regions. In fact, the H$_2$O lines offer diagnostics of warm gas regions which are usually deeply buried in dust and in star-forming regions and they can be almost as intense as CO lines and more prominent than HCN, as observed in the ULIRG Mkr 231 (\citealt{Feruglio2015}, \citealt{Aalto}, \citealt{vanderWerf2011}). All in all, we can consider H$_2$O as a 'beacon' signaling the molecular clouds energy repository (\citealt{vanDishoeck2021}, \citealt{Liu2017}, hereafter L17). 

Because of its high dipolar moment, extremely rich spectrum, and high-level spacing (compared to other molecules with low-level transitions in the millimetric range), H$_2$O is strongly coupled with the FIR radiation in compact and warm star-forming regions; in this vein, \cite{Omont2013} and \cite{Yang2013} found a strong correlation $L_{\rm{H_2O}_{submm}} \propto \rm{L_{IR}}^{\alpha}$, with $\alpha \sim 0.9-1.2$, extending from local to high-z sources. In addition, the excitation of water lines is also sensitive to collisions with Hydrogen molecules. Since the lowest vibrational band of H$_2$O lies at $\sim 6.3\, \mu$m  (shorter than collisionally excited lines, e.g. HCN at $\sim$14.7$\mu$m), the continuum radiation is too weak to excite H$_2$O vibrational states: the relevant water transitions involve rotational states at the ground vibrational level. Pumping through \emph{pure} rotational transitions, over the lowest vibrational level, is particularly important for an asymmetric top molecule like H$_2$O, since it makes the energy level structure more complex than the simple rotational ladder typical of linear or spinless molecules like HCN or CO. 

The importance of radiative pumping in water vapor rotational transitions has been extensively discussed by \cite{Gonzalez2004, Gonzalez2008}, \cite{Weiss2010}, \cite{Gonzalez2010, Gonzalez2014, Gonzalez2021, Gonzalez2022}, \cite{Pereira-Santaella2017} and L17. Absorption of FIR photons can populate mid/high-energy (E$_{up}>150$ K) rotational levels, which decay through a cascade process in which sub-mm photons are emitted; such energy levels are then radiatively excited even when collisions alone are ineffective due to their inadequate kinetic energy. 
However, the relative importance of collisions and IR pumping on  water excitation in extragalactic sources turns out to be strongly dependent on the ambient ISM conditions. We aim to investigate this issue
by taking advantage of the high spectral and spatial resolution offered by the Atacama Large Millimeter/submillimeter Array (ALMA) on a strongly-lensed Optical/NIR-dark galaxy HATLASJ113526.2-01460 (hereafter J1135) at redshift z $\sim$ 3.1. The lensed system was originally discovered in the Gama $12^{th}$ field of the $Herschel$-ATLAS survey and is found to be remarkably peculiar, given the strong faintness of the background source and the foreground deflector in the optical/near-IR(NIR) regime.  \cite{Giulietti2022b} analyzed  its continuum, the CO(8-7), and [CII] spectral line emissions detected from ALMA Archival Data. The detailed lens modeling of these components, revealed that J1135 is magnified by a factor $\mu \sim 6-13$. Thanks to gravitational lensing, that stretches the angular sizes by factors $\sqrt{\mu}\sim 3$, it is possible to locate the SFR region in a compact core of $\lesssim $ 0.5 kpc, with no clear evidence of gas rotation or ongoing merging events.  

In this paper, we present the modeling, analysis, and thermal interpretation of three molecular water line transitions, namely p-H$_2$0 2$_{02}-1_{11}$, o-H$_2$0 3$_{21}-3_{12}$ and  p-H$_2$0 $4_22-4_{13}$ made available from the high resolution ($\lesssim$0.3") ALMA observations in different bands of J1135, and we provide a calibration of the SFR-L$_{H_20}$ relation. The paper is organized as follows: in Sect. \ref{sec:data} we describe the ALMA dataset used in this analysis; in Sect. \ref{sec:waterex}, we introduce water excitation models from literature, outlining the main intervening physical processes in the water spectra. In Sect. \ref{sec:imaging} and Sect. \ref{section:lineshapes} we discuss the imaging  of water emission and the distribution of their fluxes with respect to the continuum emission in the mapping, and provide a  model to fit the corresponding line shapes.  In Sect. \ref{sec:analysis} we interpret our results in light of the most updated radiative transfer models existing in literature. In Sect. \ref{sec:SFR}, we discuss the reliability of water lines as SFR calibrators. In Sec. \ref{sec:conclusions} we finally draw our summary and conclusions. Throughout this work, we adopt a flat $\Lambda$CDM cosmology (\citealt{PlanckCollaboration2020}) with $h=0.67$, $\Omega_m=0.3$ and $\Omega_{\Lambda}=0.7$, and a Salpeter initial mass function (IMF).

%%%%%%%%%%%%%%% TABLE OF LINES PARAMETERS %%%%%%%%%%%%%%%%%
\begin{table*}
\begin{center}
\begin{tabular} {c c c c c c c c c c c}
\hline    
 \hline
Transition  & $\nu _{rest}$  & E$_{upper}$  & Log A$_{ij}$ &  ALMA& $\nu _{obs}$ & Resolution & P.A. & pixel scale & rms \\ 
           &   [GHz]  & [K] & [s$^{-1}$] &  Band &  [GHz] & [arcsec$^{2}$] & [deg] & [arcsec] & [mJy beam$^{-1}$ km s$^{-1}$] \\ 
 \hline
 % $p$-H$_2$O $2_{02}$-$1_{11}$ &2& 3& 4& 5 \\
   $p$-H$_2$O $2_{02}$-$1_{11}$& 987.927  & 100.8 & -2.23 &  6 & 239.376 &0.305 $\times$ 0.270 & -86.74 & 0.048 & 0.44 \\ 
  $o$-H$_2$O $3_{21}$-$3_{12}$&  1162.912 & 305.2 & -1.64 &  7 & 281.766 &0.259 $\times$ 0.226 & -77.51 & 0.039 & 0.26 \\
 $p$-H$_2$O $4_{22}$-$4_{13}$&  1207.639  & 454.3 & -1.55 &  7 & 292.621 &0.251 $\times$ 0.220 & -78.19 & 0.039 & 0.24 \\ 
    CO(J=8-7) & 921.800 & 199.1 & -4.29 & 6 & 223.583 & 0.328 $\times$ 0.289 & -85.21 &  0.048 & 0.35 \\
%    $[$C II$]$  & 1900.548 & 91.25 & -5.6 & 8 & 460.510&\\
 [1ex]
 \hline
\end{tabular}
\caption{Rest-frame properties of the water and CO(J=8-7) lines (frequency, upper-level energy, critical density) and information of the relative ALMA images (band, spectral window central frequency, resolution, position angle, pixel scale, and rms) analyzed in this work. For comparison we also report the line properties values for the [CII] fine structure transition $^2 P_{3/2} \rightarrow ^2 P_{1/2}$.}
\label{table:linesparams}
\end{center}
\end{table*}
%%%%%%%%%%%%%%%%%%%%%%%%%%%%%%%%%%%%%%%%%%%%%%%%%%%%%%%%%%%%%%%%%%%
\section{ALMA data}\label{sec:data}

J113526 has been the target of several ALMA observations, including continuum between 250 and 450 GHz, [CII], CO, and water lines \citep{Giulietti2022b}. 
In this paper, we analyze the observations of our target included in a Cycle 6 project (2018.1.00861.S, PI: Yang), publicly available on the ALMA Science Archive. These observations
aimed at tracing H$_2$O and CO (J=8$-$7) lines in candidate lensed galaxies at high redshift ($z\sim 2-4$); they are performed by exploiting ALMA band 6 and 7 data with a maximum baseline of 1397m, corresponding to a $\sim0.2$ arcsec resolution in both bands.
Band 6 targets the spectral line emission of H$_2$O(J=2$_{\rm0,2}$-1$_{\rm 1,1}$) and CO(J=8-7) with two spectral windows centered at 239.376 GHz and 223.583 GHz respectively, while other two windows centered at 235.940 and 221.705 GHz are dedicated to continuum observations.
The H$_2$O(J=3$_{\rm2,1}$-3$_{\rm 1,2}$) and H$_2$O(J=4$_{\rm2,2}$-4$_{\rm 1,3}$) spectral lines are targeted in band 7 with two spectral windows, centered at 281.766 and 292.621 GHz respectively. The remaining two windows centered at 280.314 and 294.266 GHz observe the continuum emission. 
In both the bands, each spectral window has 1.875 GHz bandwidth and 240$\times$7.8 MHz channels. 

Calibration was performed by running the available pipeline scripts in the Common Astronomy Software Applications (CASA, \citealt{McMullin2007}) package version 5.4.0-68.
The continuum subtraction was manually done using the task \texttt{uvcontsub}. Imaging was performed manually by adopting a Briggs weighting scheme with a robustness parameter equal to 0.5 and an Hogbom deconvolution algorithm. Images cubes are generated with a spectral resolution of 10.6 km s$^{-1}$ per channel. 
Details of the beam size and observing frequencies are collected in Table \ref{table:linesparams}, together with the rest frame properties of the lines (rest frequency, upper energy value of the transition, Einstein coefficient).

%%%%%%%%%%%%%%%%%% FIGURE Radiative pumping %%%%%%%%%%%%%%%
\begin{figure}
\centering
    \includegraphics[width=8cm]{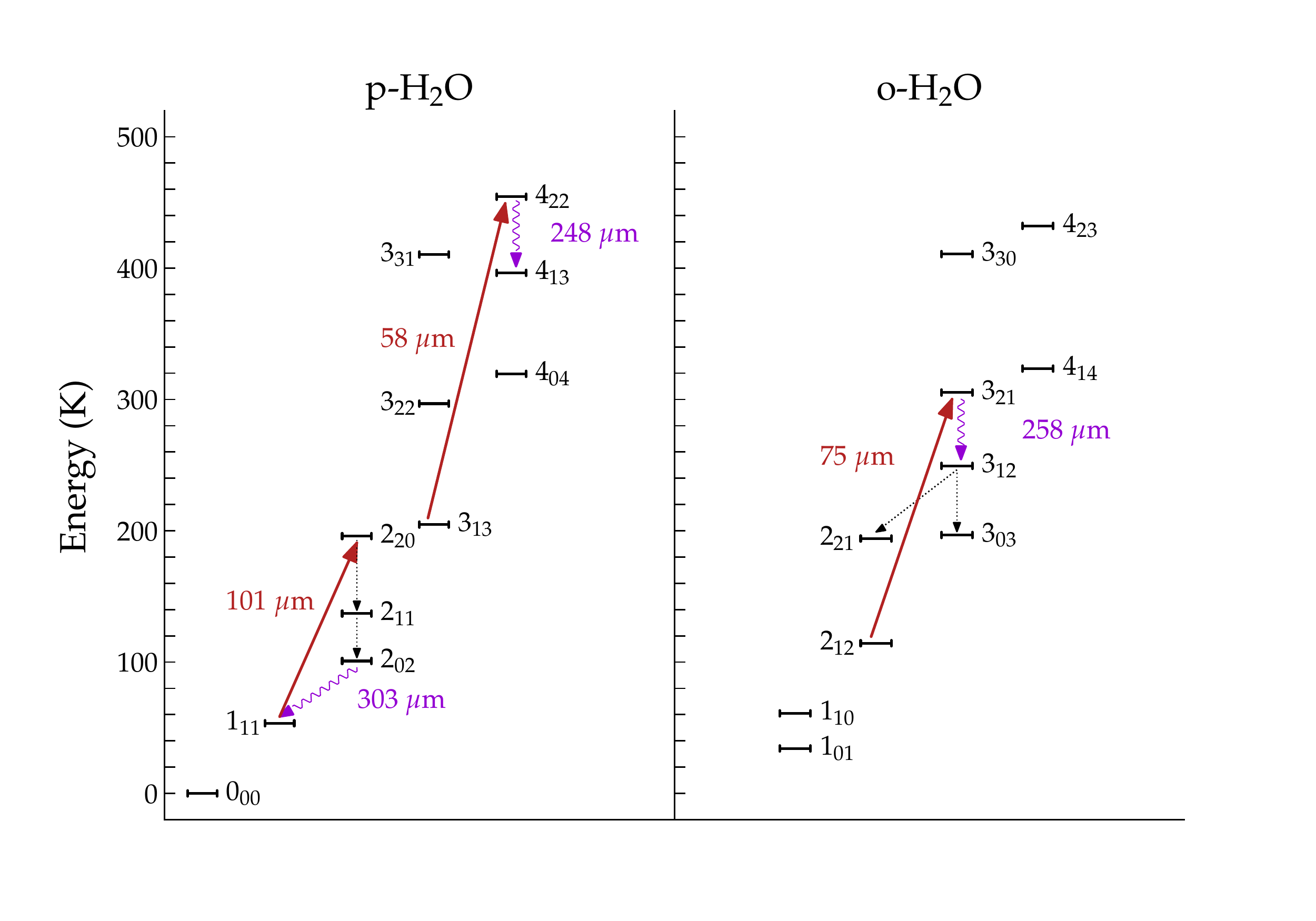}
    \caption{Energy levels of rotational water transitions. The red arrows indicate the FIR pumping from dust photons; the corresponding wavelengths are also indicated. The downward arrows indicate the radiative cascade of de-excitations in the pumping cycles. The three purple downward arrows highlight the three emission lines which are the subject of this work. Collisions may partially contribute to the population of the backbone levels. The energy levels are taken from LAMDA database, \citealt{LAMDA}.}
    \label{figure:pumping}
\end{figure}
%%%%%%%%%%%%%%%%%%%%%%%%%%%%%%%%%%%%%%%%%%%%%%%%%%%%%%%%%%%%%%%%%%%%
\section{Water excitation} 
\label{sec:waterex}
H$_2$O lines  are among the strongest molecular lines  in high-$z$ ultraluminous starburst galaxies. They are a powerful probe of the physical, geometrical, and dynamical structure of the ISM under the peculiar and extreme conditions of star-forming molecular clouds. However, retrieving such key information is not straightforward, both because most observed water lines have high optical depths, which prevent us from directly inferring the column densities from the measured line intensities, and because the water spectrum is less trivial than other gas tracers, such as CO, due to the fuddled and complex interplay between collisional and radiative excitation processes. 

H$_2$O can be an efficient coolant of the dense, warm ISM, as long as it is excited mainly collisionally below the critical density; in this case, the interactions with a colliding partner (typically, H$_2$ or He) are able to bring kinetic energy off the gas through spontaneous decay since each collision results in the emission of a  photon (\citealt{Draine2011}) and the level de-excitation. 
However, due to its high dipolar moment (1.85 Debye) and its strong coupling with the background radiation field, the water molecule is also subject to an important radiative excitation mechanism channeled by the FIR continuum. FIR photons radiatively excite the water molecules from low-lying "base" levels; water molecule responds to this pumping through a cascade process back to the starting level, thus triggering a pumping cycle that fuels the observed submillimeter lines emission. 
In other words, both collisions and photon absorption may, in principle, contribute to populate an energy level, while the de-excitation mechanism generating the observed lines is radiative and requires subcritical density. 
In Figure \ref{figure:pumping} we highlight with red downward arrows the three transitions targeted by ALMA observations of J1135 which are the subject of this paper: namely, a "low excitation" (E$_{up}<$250 K ) line,  p-H$_2$0 $(2_{02}-1_{11})$, a "medium excitation" (150$<$E$_{up}<$350 K) line, o-H$_2$0 $(3_{21}-3_{12})$, and a "high excitation" (E$_{up}>$350 K) line,  p-H$_2$0 $(4_{22}-4_{13})$.  
They are pumped in the corresponding cycles by dust-emitted photons at 101, 75, and  58 $\mu$m, respectively.

Radiative excitation is not independent on collisions, though: the latter may significantly contribute to populate the lower backbone level of a cycle, boosting the FIR pumping itself. 
In particular, collisional excitation of the $1_{11}$ and $2_{12}$ rotational energy levels can significantly affect the pumping cycles corresponding to the absorption of dust-emitted photons at 101 $\mu$m and 75 $\mu$m, enhancing also the emission of two of the lines under study, namely  the p-H$_2$0 2$_{02}-1_{11}$ and  the o-H$_2$0 3$_{21}-3_{12}$. 
Furthermore, the emission of the p-H$_2$0 2$_{02}-1_{11}$ line may be enhanced in regions of low continuum opacity but warm and dense gas, where the population of the upper level 
p-H$_2$0  2$_{02}$ may get a collisional contribution in addition to that coming from the decay of $2_{11}$ in the FIR pumping cycle (\citealt{Gonzalez2014}).  In order to quantify the relative importance of FIR pumping and collisions, a radiative transfer model must include the parameters of the dust radiation field as well as those describing the physical properties of the gas component. In particular, the "low-level" transition analyzed in this work will reflect the before-mentioned interplay between collisional and radiative excitation mechanisms.
In general (\citealt{vanDishoeck2021}, \citealt{Yang2013}), lines connected with the ground state levels of o- H${_2}$0 and p-H${_2}$0 are prime diagnostics of the cold gas; they usually show strong self-absorption or can even be purely in absorption.  The 'medium level' lines probe the warm gas and are less affected (if not at all) by absorption. The "high-level" excited lines originate from energy levels only populated in high-temperature gas and strong shocks. 

The interpretation of the observed line intensity ratios would require a radiative transfer  model for the ISM in LTE, which includes the interplay between collisions and FIR pumping. Building such a model is beyond the purpose of this paper.  Nevertheless, such a physical model (describing mainly in-situ processes) does not depend on the source redshift, and we can take advantage of radiative transfer models existing for local star-forming galaxies, as the one described by L17, to extract some general diagnostics about J1135. 
 This model was used to analyze a survey of multiple velocity-resolved water vapor FIR spectra in a sample of local galaxies with different nuclear environments, from pure nuclear starbursts to starburst nuclei hosting an AGN. To assess the relative importance of collisions and FIR pumping on the line emission, they modeled each ISM component as an ensemble of clumps with identical physical properties, varying the parameters characterizing the radiation field  as well as those of the gas in each clump model. 
Then, water excitation in a multiphase ISM is obtained using an extended 3D escape probability method to solve non-LTE radiative transfer that includes the dust emission, with the aim of probing the physical and chemical conditions in the nuclei of actively star-forming galaxies. 
Water excitation is calculated for three typical ISM components:
a cold extended component (with gas and dust temperatures of $20-30$K, the density of the order $\sim 10 ^4$ cm$^{-3}$ and column density  $\sim 10^{23}$ cm$^{-2}$); a warm component with gas and dust temperature between 40 and 70 K, typical densities of the order $\sim 10^5-10^6$ cm$^{-3}$, and column density of $\sim 10^{24}$ cm$^{-2}$;
and a third, hot and dense component,  needed to explain the high excitation transitions (such as those observed in Arp 220 and Mrk 231, which are commonly thought to host an AGN in  their nuclei, see e.g, \citealt{Soifer1999}, \citealt{Downes2007}, \citealt{Aalto2009}, \citealt{Gonzalez2014}, \citealt{Fischer2010}, \citealt{vanderWerf2011}, \citealt{Rangwala2011}): this hot component 
has gas and dust temperatures $\sim 100-200$ K at densities n(H)$\geq 10^{6}$, with high column densities $N_H \geq 5 \times 10^{24}$ cm$^{-2}$.  Figure 3 of L17 illustrates the excitation model for these ISM components, which will be used as a reference model to interpret ALMA water lines observations of J1135. 
%%%%%%%%%%%%%%%%%%%%%%%%%%%%%%%%%%%%%%%%%%%%%%%%%%%%%%%%%%
\section{Imaging analysis} 
\label{sec:imaging}
The $\lesssim 0.3$ arcsec resolution ALMA images presented here provide the highest angular resolution of water emission ever reached in observations of high-$z$ star-forming galaxies. 
This high angular resolution, combined with the stretching of angular scales provided by the intervening gravitational lens, allows us to dissect the (lensed) image plane of the observations, and to map it into a detailed reconstruction of the unlensed galaxy in the source plane. 
We take advantage of the gravitational amplification of angular scales to  circumscribe the molecular clouds from which the observed water spectra arise, and whose physical properties will be discussed in Section \ref{sec:analysis}.  

\begin{figure*}
    \includegraphics[width=0.8\textheight]
    {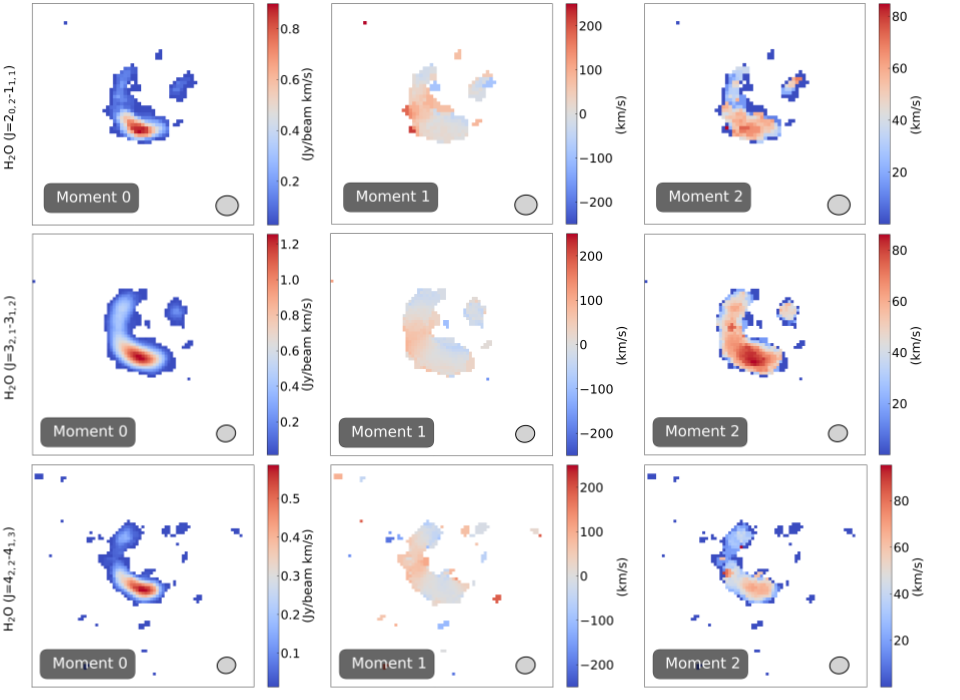}
    \caption{Zero, first, and second momenta for the three H$_2$O line transitions analyzed in this work, corresponding to the integrated brightness, and
   velocity distributions and the velocity dispersion, integrated over the velocity range v$_p \pm$ FWHM. Moments 1 and 2 include pixels $\gtrsim 5 \sigma$. }
    \label{fig:moments}
\end{figure*}

In Figure \ref{fig:moments} we report image plane moment maps of J115326 from ALMA observations, showing  zero, first, and second momenta for the three H$_2$O line transitions analyzed in this work, respectively the integrated brightness, velocity distributions, and velocity dispersion. Momenta are computed considering the velocity range v$_p$ $\pm$ Full Width at Half Maximum (FWHM) reported in Table \ref{table:linesvalues} (see Section \ref{section:lineshapes}) and including only pixels above a 5$\sigma$ threshold, where $\sigma$ is the rms of the map.
Besides random pixels of no statistical significance, the three transitions seem to arise from the same physical region, a nucleus stretched into an Einstein ring up to a size of a few kpcs in the image plane. The observed emissions are consistent with a single molecular component, due to the strikingly similar kinematic properties; in addition, average velocities and velocity dispersions show no indication of merging nor rotation of the emitting cloud.

\begin{figure*}
\centering
\includegraphics[width=\textwidth]{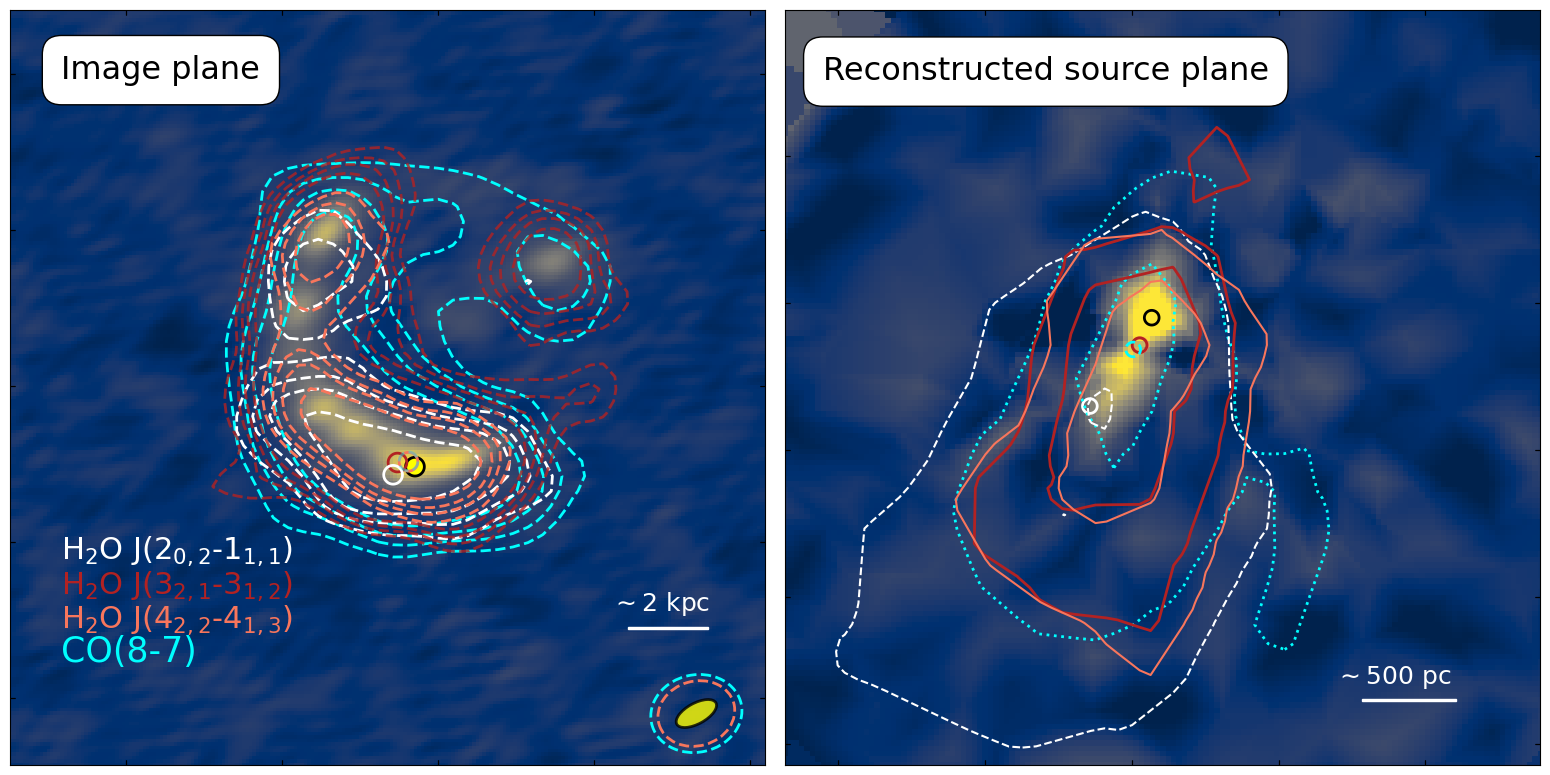}
\caption{Left panel: Lensed image of J1135 ALMA Band 8 continuum at 640 $\mu$m. The superimposed contours of water lines and CO(8-7) moment 0 maps are shown at 5, 7, 9,12 $\sigma$ levels. Circles represent the position of the respective peak emissions. Right panel: reconstructed (de-lensed) source and line emissions. Contours are shown at 3 and 11 $\sigma$ levels}.
\label{fig:contours_all}
\end{figure*}

The left panel of Figure \ref{fig:contours_all} shows the contour plots of the same lines at 5, 6, 9, and 12 $\sigma$ and the peak of the corresponding emissions, superimposed to the continuum emission at 640 
$\mu$m. For completeness, we also show the emission peak and contours of the CO (8-7) emission described in \cite{Giulietti2022b}. 
The lens modeling of J1135 is then performed through the open source Python 3.6+ code \texttt{PyAutoLens} (\citealt{Nightingale2018, Nightingale2021}), which implements the Regularized Semi-Linear Inversion (SLI) Method described in \cite{Warren2003} together with the adaptive source plane pixelization scheme described in \cite{Nightingale2015} adapted to interferometric data. Further details on this method are available e.g. in \cite{Dye2018,Dye2022}, \cite{Enia2018}, \cite{Massardi2017}, \cite{Maresca2022}. 
We modeled the three water spectral line emissions adopting the best-fit lens model obtained in \cite{Giulietti2022b}, described as a Singular Isothermal Ellipsoid (SIE; \citealt{Kormann1994}) i.e. an elliptical power-law density distribution which goes as $\rho \propto r^{-\gamma}$, with $r$ being the elliptical radius and with a fixed slope value $\gamma=2$. The parameters describing the lens model are the Einstein radius $\theta_{\rm E}$, the lens centroid positions $x_c$, $y_c$, the ratio between the semi-major and semi-minor axis q, and the positional lens angle $\phi$. The corresponding best-fit values are $\theta_{\rm Ein}^{\rm{mass}} = 0.4241 \pm {0.0005}$, $y_{c} = -0.2329 \pm {0.0005}$ , $x_c = 0.1494 \pm {0.001}$, $q = 0.637 \pm {0.001}$ and $\phi =-35.31\pm 0.04$. 

Details on the modeling pipeline and the computation of the physical parameters are described in \cite{Giulietti2022b}. The pipeline consists of two main steps. First, we determined the best-fit lens model via a non-linear parametric fitting performed by \texttt{PyAutoLens} through the nested sampling algorithm \texttt{Dynesty} (\citealt{Speagle2020}), sampling the parameter space and computing the posterior probability distributions for the parameters of a given lens model.
We assumed a Sérsic light profile for the source and simultaneously fit the three ALMA continuum bands (including spectral lines). In the second step, we kept fixed the best-fit lens model parameters described above and reconstructed the source emission for all three water spectral lines through the regularized SLI inversion method.

The fit is performed on a number of pixels delimited by a circular mask, where the radius changes according to the resolution of the cleaned ALMA image, in order to obtain a satisfactory fit without exceeding in terms of computational cost.
The resulting reconstructed source contains only pixels excluded from the masked lensed image area.

Magnification factors are computed as $\mu = A_{\rm IP}/A_{\rm SP}$, where $A_{\rm IP}$ and $A_{\rm SP}$ are the areas enclosing significant (i.e. $>3 \sigma$ and $>5 \sigma$) pixels in the reconstructed image plane (IP) and the reconstructed source plane (SP) respectively. The noise is estimated as the rms in the reconstructed source map. From the area enclosing all the pixels with signal-to-noise ratio $\gtrsim 3$ and $\gtrsim 5$ in the reconstructed source plane, the effective radius can be computed as $r_{\rm eff}=(A_{\rm SP}/ \pi)^{0.5} $. 
Uncertainties on magnifications and effective radii are computed from a set of samples provided by the non-linear search performed during the inversion. Each sample corresponds to a set of inversion parameters (i.e. the regularization coefficient and the pixelization's shape) that were evaluated and accepted by the non-linear search. 
Uncertainties are then retrieved as the 16th and 84th quantiles of the parameter distribution drawn from $\sim 200$ accepted samples (see Table \ref{tab:lens_parameters}). 

\begin{figure*}
\includegraphics[width=\textwidth]{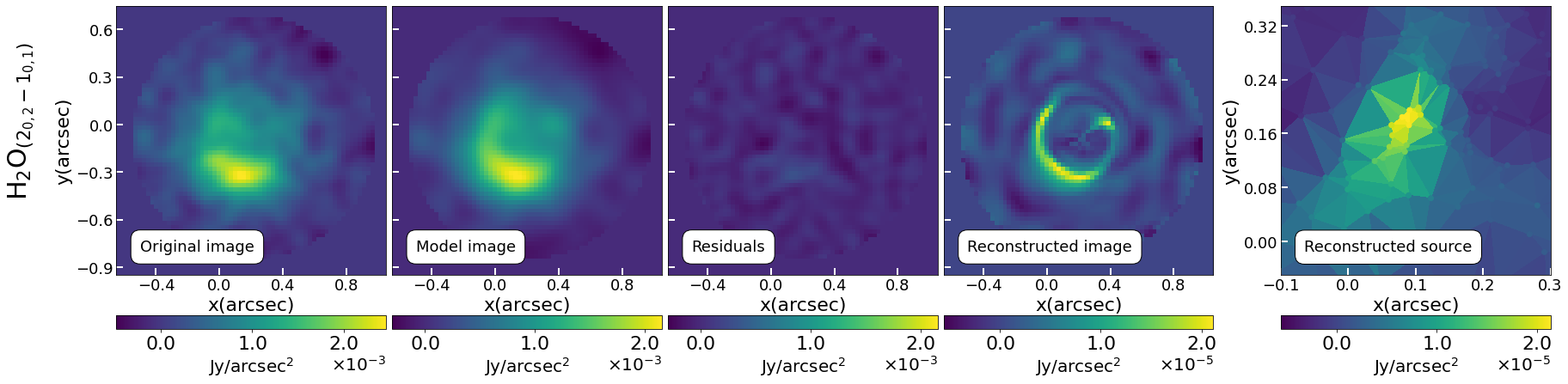}
\includegraphics[width=\textwidth] 
{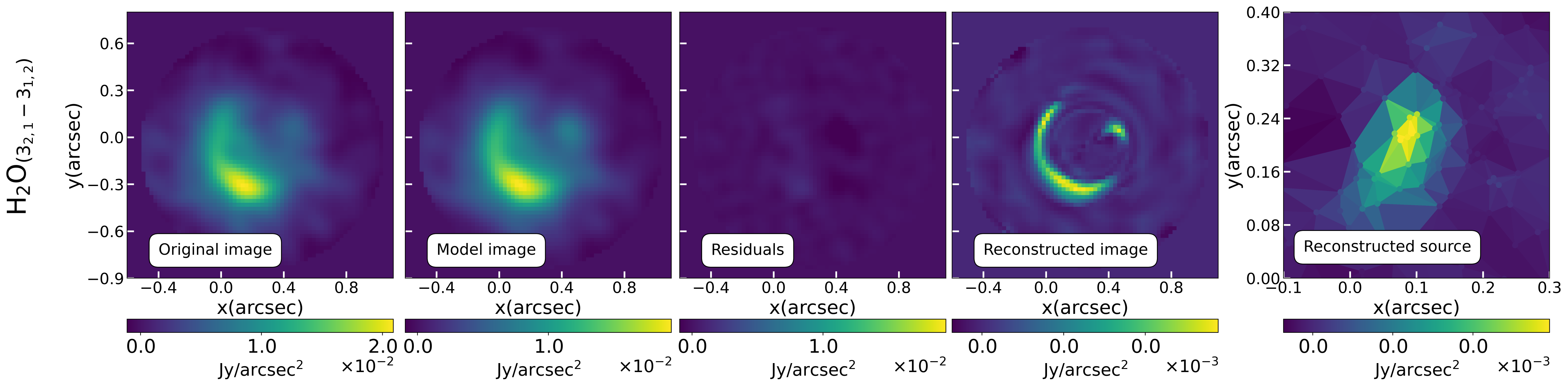}
\includegraphics[width=\textwidth]{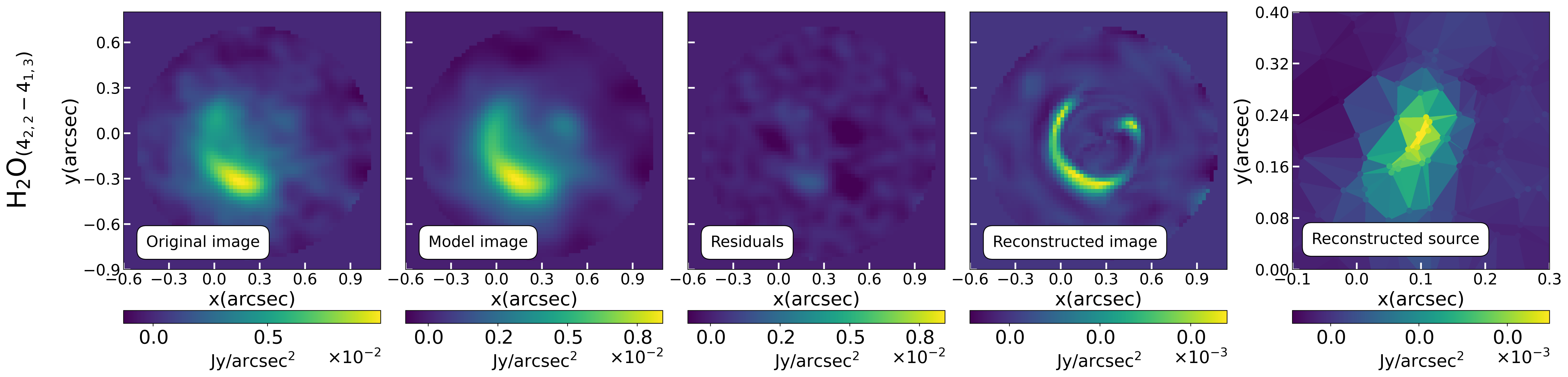}
  \caption{Results of the lens modeling and source reconstruction procedure for J1135 water lines emission. From the first column to the right: the ALMA dirty image, the best-fit lensed model dirty image, the residuals, the image plane's model, and the reconstructed source plane. Please note that surface brightness values of the former are de-magnified. The color bar indicates the surface brightness in units of Jy arcsec$^{-2}$.  }
   \label{fig:water_reconstruction}
\end{figure*}

In Figure \ref{fig:water_reconstruction}, we report the results for the lens modeling of the water lines  $p-{\rm{H_2O}} (2_{02}-1_{11})$, $o-{\rm{H_2O}} (3_{21}-3_{12})$, and $p-{\rm{H_2O}} (4_{22}-4_{13})$.

In the right panel of Figure \ref{fig:contours_all}, the  $3 \sigma$ and $11 \sigma$ contours of the same lines in the source plane are superimposed to the continuum in Band 8. The reconstructed image reveals the H$_2$O with an unprecedented resolution of $<500$ pc scale.
Notice that the comparison is made between observations at different ALMA resolutions: the low-level line is observed in Band 6, where the resolution is lower with respect to the other lines. However, we can still locate a central nucleus, showing a clumpy structure corresponding to a region of less than 500 pc, where all the water lines have their peak of emission, superimposed to the CO(8-7) emission peak and adjacent to the 640 $\mu$m continuum peak. 
The reconstructed low-level transition peak is slightly ($\lesssim$ 500 pc) displaced from the peak of the other two lines, but  still located in the central region . The $3 \sigma$ contours level of this line  shows a tail  of few hundred parsecs stretching from the nucleus, resulting in a more extended region in which the low-level line emission is appreciably detectable, with respect to the medium and high-level lines. Noticeably, the peaks of CO(8-7) and $p-{\rm{H_2O}}(4_{22}-4_{13})$ turn out to be completely superimposed to each other on both  observed and reconstructed maps: the absence of differential lensing between these two emissions is another confirmation that they stem from the very same physical region. In order to interpret the observed emissions in terms of the underlying gas and dust physical properties, we analyze the corresponding line profiles in the next section. 

%%%%%%%%%%%%%%%   TABELLA DEI PARAMETRI DI LENSING %%%%%%%%
\begin{table}
\begin{center}
\begin{tabular}{lcccc}
\hline
  \multicolumn{1}{c}{} &
 \multicolumn{1}{c}{$\mu_{3\sigma}$ }  & \multicolumn{1}{c}{$\mu_{5\sigma}$ } & \multicolumn{1}{c}{R$_{\rm eff, 3\sigma}$} & \multicolumn{1}{c}{R$_{\rm eff, 5\sigma}$}  \\
 &   & (pc) & (pc) 
 \\
 \hline
p-H$_2$O ($2_{02}$-$1_{11}$) & 7.67$^{+0.68}_{-0.31}$ &  6.73$^{+0.99}_{-0.78}$ & 1043$_{-56}^{+55}$  & 912$^{+67}_{-65}$  \\
o-H$_2$O ($3_{21}$-$3_{12}$) &  12.74$^{+0.27}_{-0.66}$ & 12.23$^{+0.31}_{-0.47}$ & 739$^{+57}_{-64}$ &  672$^{+33}_{-31}$ \\
p-H$_2$O ($4_{22}$-$4_{13}$) &  12.04$^{+1.49}_{-1.82}$ & 10.83$^{+1.06}_{-1.01}$ & 782$_{89}^{+76}$ &  745$^{+62}_{-75}$\\

\hline
\end{tabular}
\end{center}
\caption{Output properties of the lens-modeling and source reconstruction analysis. From the left: magnification factors and effective radii computed at 3$\sigma$ and 5$\sigma$.  }
\label{tab:lens_parameters}
\end{table}
%%%%%%%%%%%%%%%%%%%%%%%%%FIGURA  MOMENTI %%%%%%%%%%%%%%%%%%%%%%%%
%%%%%%%%% FIGURE IMAGING continuum and line contours %%%%%%%
%%%%%%%%%%%%%%%%%%%%%%%%%%%%%%%%%%%%%%%%%%%%%%%%%%%%%%%%%%%%%%
%%%%%%%%%%%%%%%%%% FIGURE Water maps oberved and delensed %%%%%%%%%%%

\section{Lines reconstruction}
\label{section:lineshapes}
\begin{figure*}[htbp]
\gridline{
  \fig{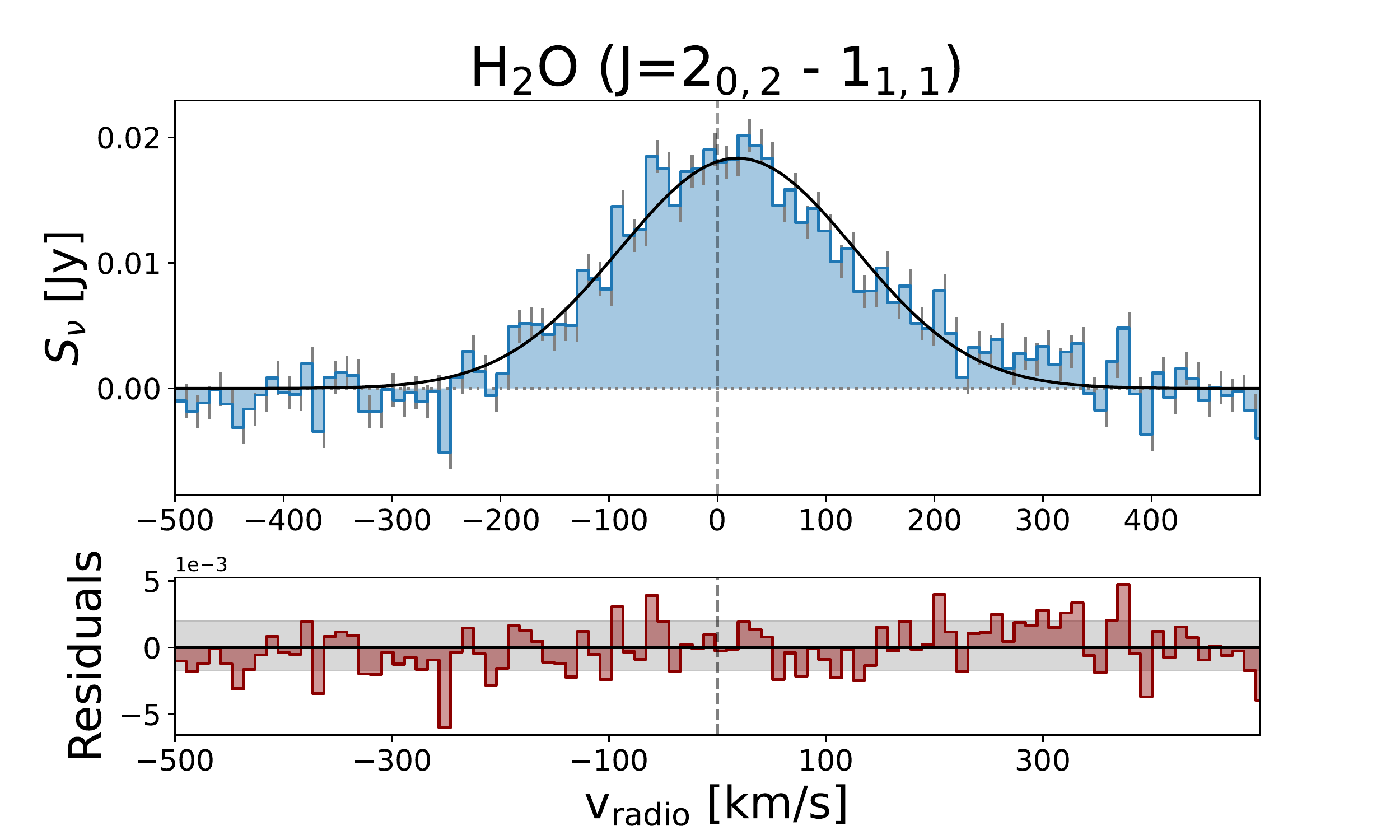}{0.5\textwidth}{}
  \fig{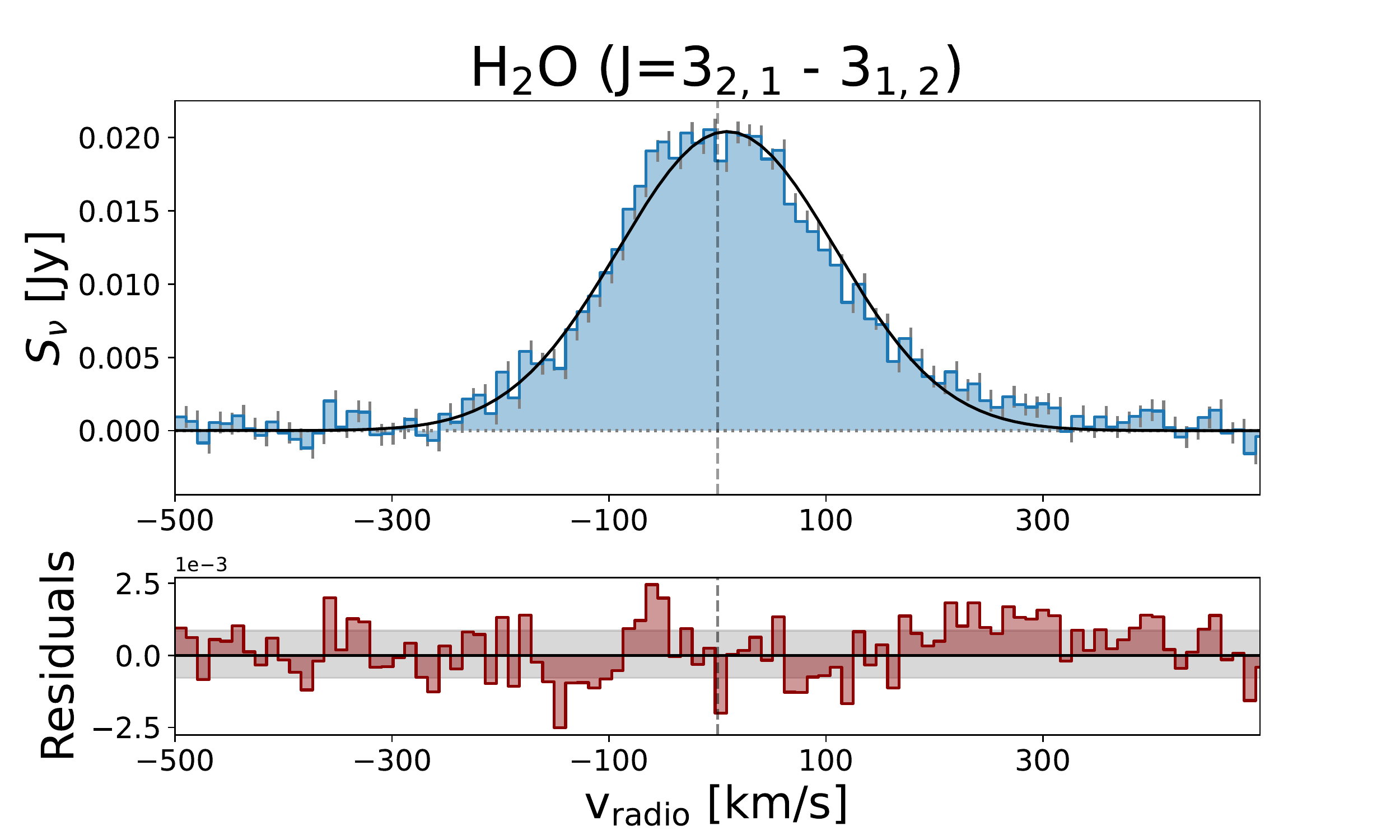}{0.5\textwidth}{}
}
\gridline{
  \fig{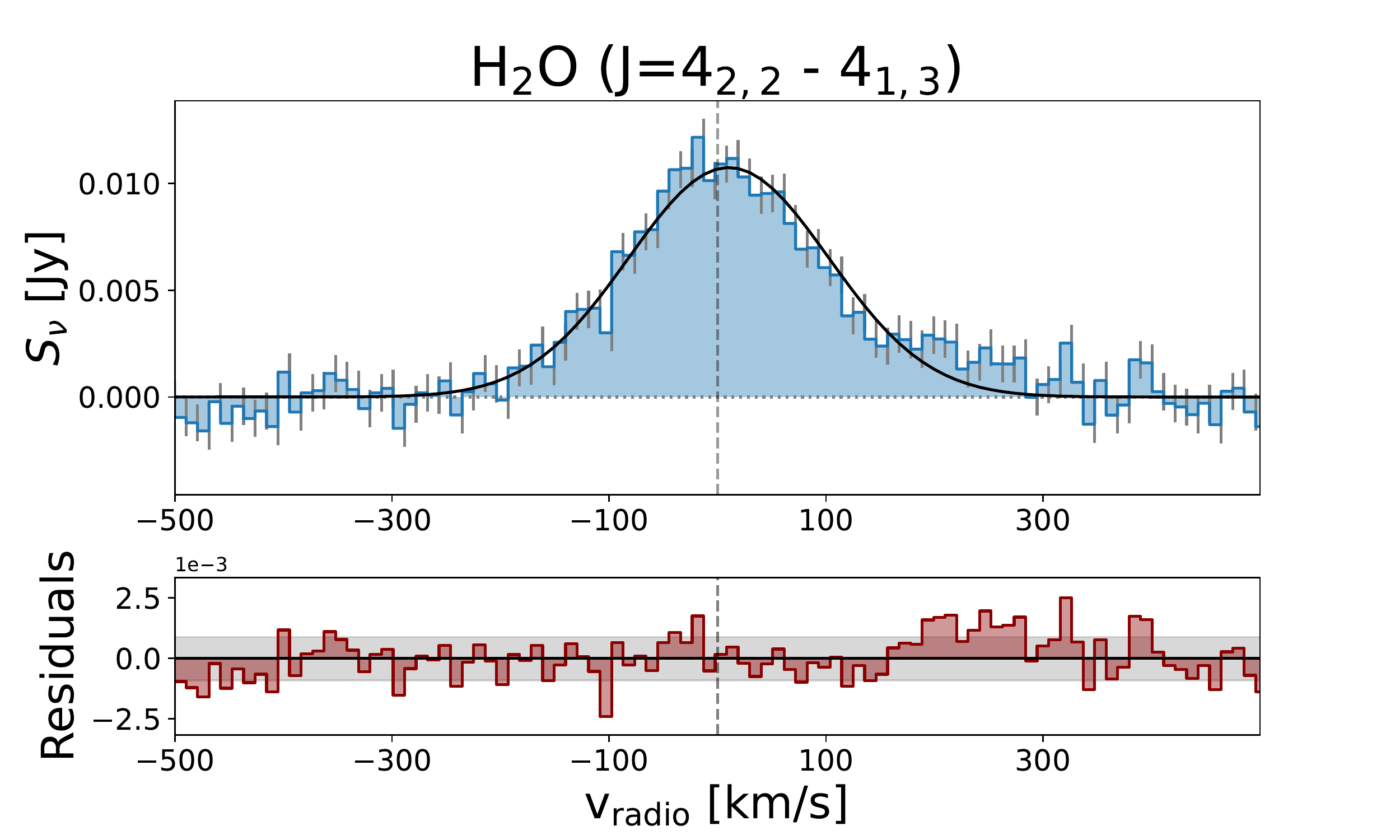}{0.5\textwidth}{}
  \fig{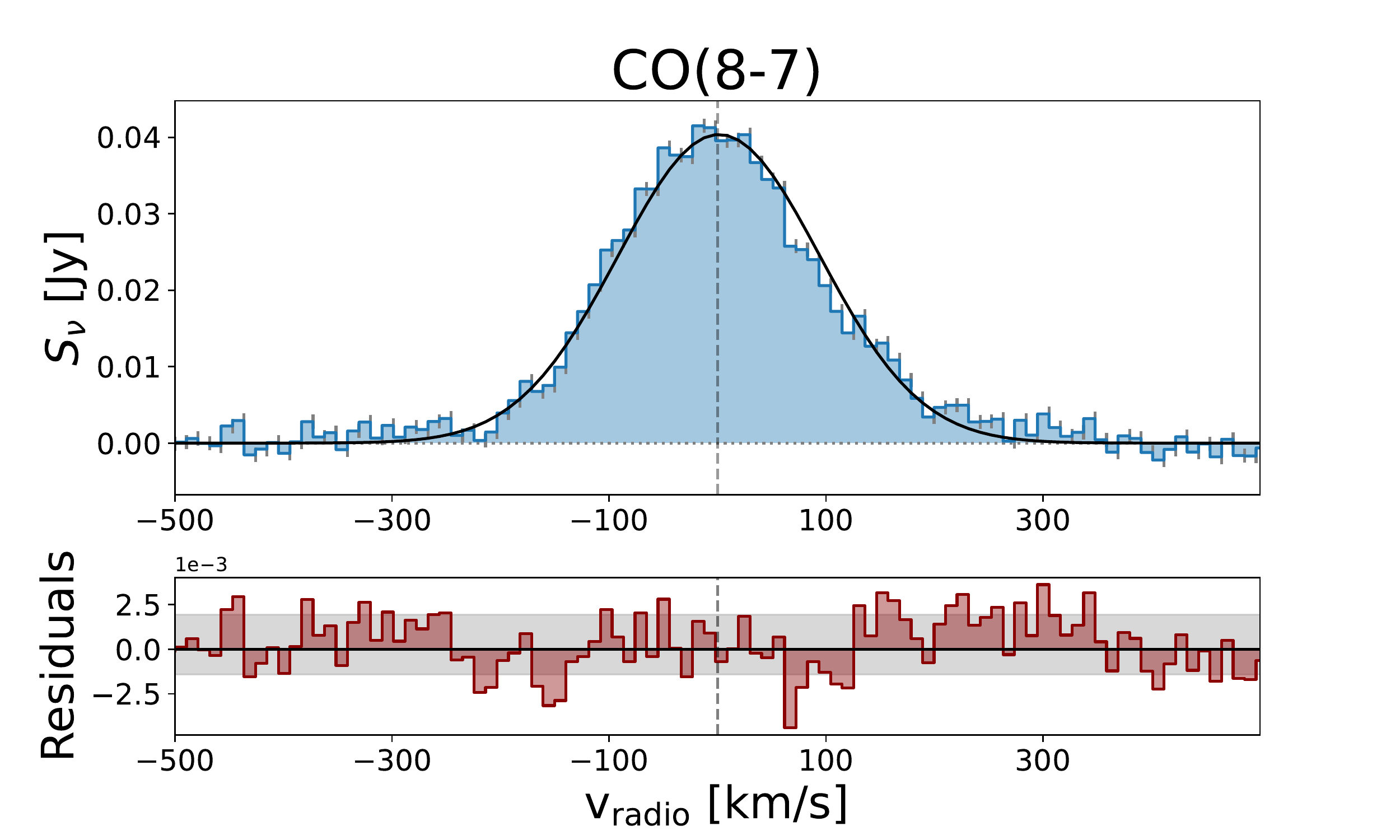}{0.5\textwidth}{}
}
\caption{Targeted water lines as observed in ALMA Bands 6 and 7, and CO(8-7) line observed in ALMA Band 6, from \cite{Giulietti2022b}. Lines are extracted within a region enclosing pixels above 5$\sigma$ drawn from the moment-0 map. The zero velocity corresponds to the rest-frame frequency of the spectral line. Data are represented by the blue step function along with the associated errors, i.e., the gray bars. The fit of the spectral lines is represented by the black solid line. In the bottom boxes, we have shown the magnitude of the residuals (red step function) relative to each fit together with their mean value (black solid line) and their standard deviation (spanning the gray filled area).}
\label{fig:profili}
\end{figure*}

\begin{figure}
\begin{minipage}{0.46\textwidth}
\centering
\includegraphics[width=.9\textwidth]
{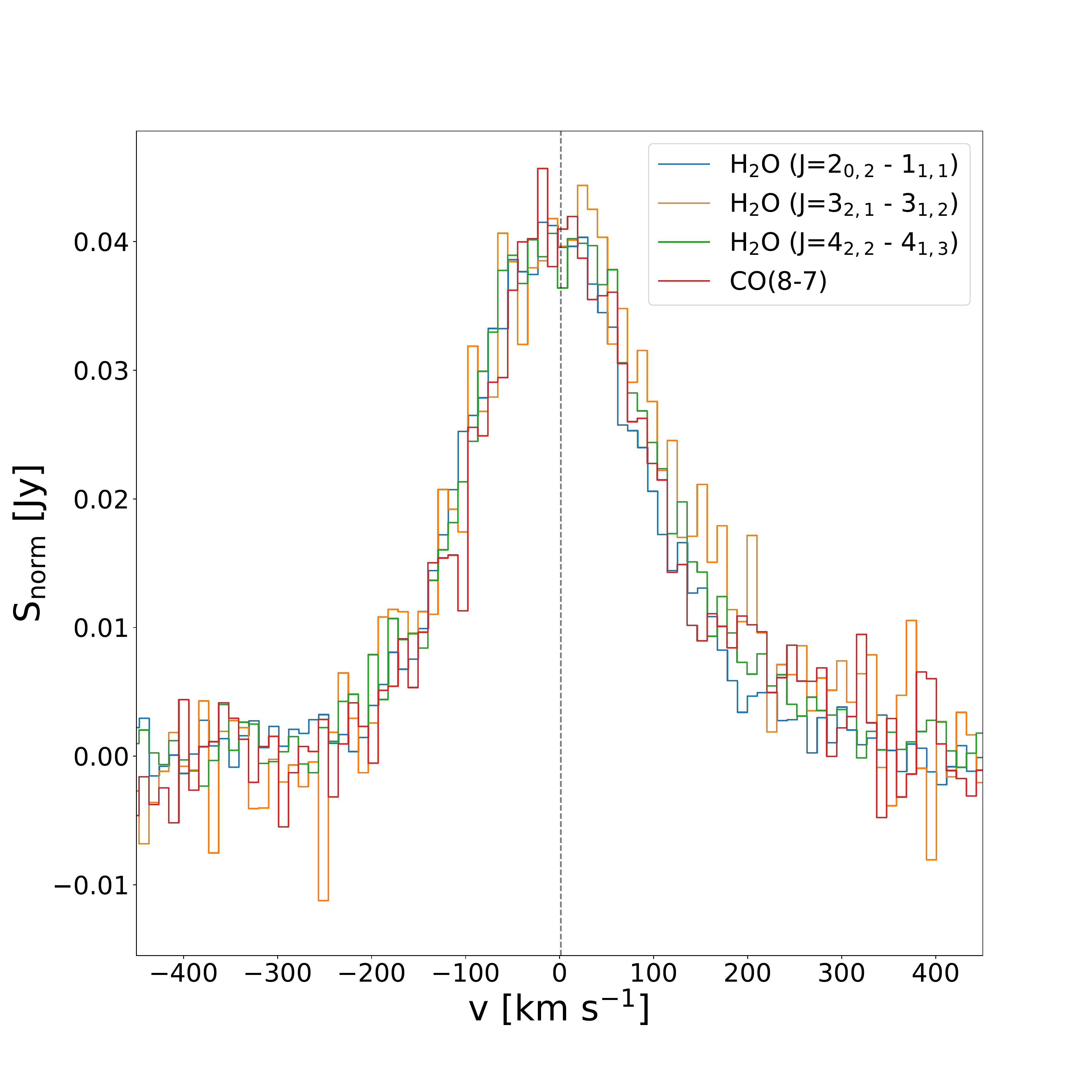}
    \caption{Water emission line profiles superimposed to the CO(J=8-7) line and scaled to the CO peak. The velocity scale is relative to the systemic velocity of J1135. }
\label{fig:sovrapposti}
\end{minipage}
\end{figure}

Figure \ref{fig:profili} reports the observed (magnified) $p$-H$_2$0 $(2_{02}-1_{11})$, $o$-H$_2$0 $(3_{21}-3_{12})$, and $p$-H$_2$0 $(4_{22}-4_{13})$ lines of J1135. For comparison, the line profile of CO(8-7) of the source, detailed in \cite{Giulietti2022b}, is also shown.  The intensity of the water lines is smaller (by a factor 2-4)  but comparable to that of CO, with H$_2$O covering the same velocity range as CO. The water emission line profiles, scaled to the peak value of the CO profile for better visualization of the line profiles, are superimposed to the CO(8-7) line in Figure \ref{fig:sovrapposti}. The CO(8-7) and the  water vapor lines show consistent velocity dispersion, reinforcing the idea that they plausibly stem from the very same physical region inside the galaxy. The extended emission tail of the low-level transition in Figure \ref{fig:contours_all} is only  weakly widening the velocity profile of the line. 

%%%%%%%%%%%%%%%%%% TABLE of lines Gaussian fits %%%%%%%%%%%%
\begin{table*}
\begin{center}
\begin{tabular} {c c c c c c c c c}  
 \hline
    Line & S$_{\mathrm{p}}$ & v$_{\mathrm{p}}$ & $\sigma_\mathrm{v}$ & $\mathrm{FWHM}$ & S$_{\mathrm{p}} \cdot \mathrm{FWHM}$ & $\langle \mathrm{S}_{\mathrm{r}} \rangle$ & $\sigma_{\mathrm{r}}$ & $\mu$L \\  
    & (10$^{-4}$ Jy) & (km s$^{-1}$) & (km s$^{-1}$) & (km s$^{-1}$) & (Jy km s$^{-1}$) & (Jy) & (Jy) & ($10^8$ L$_{\odot}$)\\
 \hline
  $p$-H$_2$O $2_{02}$-$1_{11}$& $183 \pm 4$ & 18 $\pm$ 3 & $108 \pm 3$ & $255 \pm 7$ & $4.7 \pm 0.2$ & $1.5 \cdot 10^{-4}$ & $1.9 \cdot 10^{-3}$ & $8.22 \pm 0.35$ \\ 
  $o$-H$_2$O $3_{21}$-$3_{12}$& $204 \pm 4$ & 9 $\pm$ 2 & $100 \pm 2$ & $235 \pm 4$ & $4.8 \pm 0.2$ & $4.3 \cdot 10^{-5}$ & $8.2 \cdot 10^{-4}$ & $9.88 \pm 0.41$\\
  $p$-H$_2$O $4_{22}$-$4_{13}$& $107 \pm 3$ & $10 \pm 3$ & $92 \pm 3$ & $218 \pm 8$ & $2.3 \pm 0.2$ & $-1.7 \cdot 10^{-5}$ & $8.9 \cdot 10^{-4}$ & $4.92\pm 0.43$ \\
  CO(8-7) & $404 \pm 5$ & $1 \pm 1$ & $93.0 \pm 0.8$ & $219 \pm 2$ & $8.8 \pm 0.2$ & $2.7 \cdot 10^{-4}$ & $1.7 \cdot 10^{-3}$ & $14.4 \pm 0.03$
\\ [1ex]
 \hline
\end{tabular}
\caption{Observed emission lines parameters derived from the Bayesian MCMC Gaussian fit. Here S$_{\mathrm{p}}$ is the Gaussian's peak, $v_{\mathrm{p}}$ is the peak position on the velocity axis, $\sigma_\mathrm{v}$ is the standard deviation (or the Gaussian RMS width) and FWHM represents the full width at half maximum of the Gaussian, computed as $\mathrm{FWHM}=2 \sqrt{2 \ln 2} \sigma_\mathrm{v}$. The product $\mathrm{S}_{\mathrm{p}} \cdot \mathrm{FWHM}$ represents the line flux. $\langle \mathrm{S}_{\mathrm{r}} \rangle$ and $\sigma_{\mathrm{r}}$ respectively represent the mean fit residuals and their standard deviation. The last column reports the inferred magnified luminosity. The fit refers to the lensing-magnified line profiles in Figures \ref{fig:profili} .}
\label{table:linesvalues}
\end{center}
\end{table*}
%%%%%%%%%%%%%%%%%%%%%%%%%%%%%%%%%%%%%%%%%%%%%%%%%%%%%%%%%%
This result is confirmed by the mapping represented in Figure \ref{fig:contours_all},  
evidencing a central, compact nucleus of $<$500 pc where CO (8-7) and the water lines have their maximum.  As for the low-level line, the reconstructed emission of $p$-H$_2$O $(2_{02}-1_{11})$ (right panel of Figure \ref{fig:contours_all}) is spread over a larger zone, approximately delimited by its 3$\sigma$ contour extending for about 1 kpc.
We reconstructed the shape of each spectral line exploiting a Bayesian Monte Carlo Markov Chain (MCMC) framework, implemented numerically via the Python package \texttt{emcee} (\citealt{ForemanMackey2013}; further details on the fit procedure and the resulting contour plots can be found in Appendix A). 
We fit each spectral line with a single Gaussian function, described by the parameter vector $\theta=\{ \mathrm{S}_{\mathrm{p}}, \mathrm{v}_{\mathrm{p}}, \sigma_\mathrm{v} \}$
where $\mathrm{S}_{\mathrm{p}}$ stands for the peak flux, $\mathrm{v}_{\mathrm{p}}$ represents the position of the peak in the velocity axis, and $\sigma_\mathrm{v}$ is the standard deviation of the Gaussian (also called Gaussian RMS width).
The main parameters of the spectral line fit are reported in Table \ref{table:linesvalues}, along with estimates of the FWHM of the fitting functions, line flux, the mean, and the standard deviation of the residuals of the fits. The water line luminosities are obtained from the velocity-integrated line flux from the relation given in \cite{Solomon2005}:
\begin{equation}
{\rm{L}_{\rm{ H_2O}}  }= (1.04 \times 10^{-3} ) \rm{I}_{\rm{H_2O}} \nu_{rest} D_L^2(1+z)^{-1}
\end{equation}
In order to identify possible signatures of rotations or outflows, we also performed a fit of the profiles using two (and even three) Gaussian components, without finding any striking statistical evidence in favor of any of these supplementary components. 

The joint  water line and maps analysis are thus pointing toward a single, homogeneous, and compact ($<$ 500 pc) ISM component in the core of J1135, in which  water transitions are ignited by FIR pumping and collisions, together with collisional excitation of CO and, partially, of the low-level water line. The latter is also weakly excited in a more extended region of the ISM. In Section \ref{sec:analysis} we will discuss these findings comparing them with radiative transfer models predictions existing in literature. 

%%%%%%%%%%%%%%%%   FIGURE Water lines profiles %%%%%%%%%%%%%%

%%%%%%%%%%%%%%%%%%Figura profili di linea sovrapposti%%%%%%%

%%%%%%%%%%%%%% FIGURE: CONTINUUM EMISSION OF THE TARGET %%%%%
\begin{figure}\centering
\begin{minipage}{0.49\textwidth}
\centering
\includegraphics[width=1.05\textwidth,height=0.27\textheight]{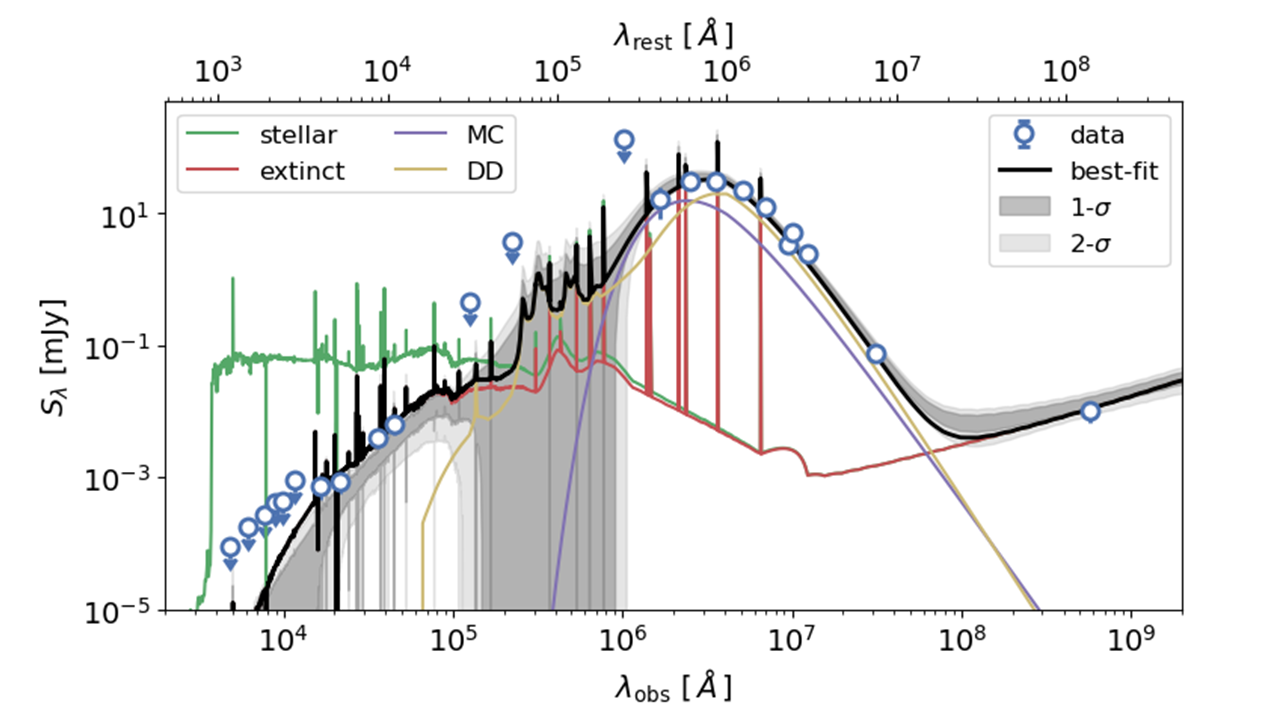}
\caption{ Best-fit model (black line) and 1- and 2-$\sigma$ confidence intervals around the mean of samples (dark and light grey shaded areas, respectively) of the UV-to-radio SED of J1135 performed with GalaPy (Ronconi et al. 2023, in preparation). The total best-fitting model is decomposed into the different emission components: green and red solid line show respectively the un-attenuated and dust-attenuated stellar emission (including line and free-free emission from nebular regions as well as the supernova synchrotron continuum), the yellow and purple solid lines mark the grey-body emitted by the two assumed dust components with temperatures of about $\sim 40$ K and $\sim 70$ K respectively. Round markers with error bars represent the detected fluxes while round markers with downward arrows represent 1$\sigma$ upper limits.}
\label{fig:continuum}
\end{minipage}
\end{figure}

%%%%%%%%%%%%%%%%%%%%%%%%%%%%%%%%%%%%%%%%%%%%%%%%%%%%%%%%%%%%%%%%%%
\section{Qualitative  thermal diagnostics}
\label{sec:analysis}

The high angular resolution observations of J1135 presented in the previous Section reveal that our target galaxy has a central, compact zone where low, medium, and high E$_{up}$ water transitions are ignited, as well as emission from CO(8-7) transition. Though the rigorous diagnostic from these observations would require a full Radiative Transfer model, some general considerations can be done by sticking to the detailed models of L17 and  \citealt{Gonzalez2010}, \citealt{Gonzalez2014},
\citealt{Gonzalez2022}.
 The L17 model was specifically developed in order to produce a diagnostic tool for the ISM in the nuclei of star-forming galaxies. Several nuclear environments are covered, from pure nuclear starbursts to starburst nuclei hosting an AGN. Each galaxy is modeled with different ISM components, where each component is an ensemble of molecular clumps with identical physical properties. The excitation temperature and level population of the gas molecules  in a cloud is determined under different physical conditions of the dust and of the gas itself. The SED modeling of J1135 by Ronconi et al. 2023 (in preparation), shown in Figure \ref{fig:continuum}, successfully reproduces the continuum emission with  two dust components: one is associated with the molecular clouds hosting star formation, where the young stellar populations, still embedded in their dense envelope,  are heating the dust up to  T$_{dust}$ $\sim 70$ K; the second  component is associated to the diffuse, optically thinner ISM: here, the diffuse dust is heated by the radiation field of the old stars populations and by the photons which could escape the molecular clouds sites of star formation, making a diffuse cold component at T$_{dust}$  $\sim 40$ K. In the starburst core, the molecular clouds hosting young stellar populations are assumed to be embedded in the diffuse component.
We take these fiducial values of T$_{dust}$ together with the L17 model as starting points to interpret the thermal status of the ISM associated with the observed water excitation.  For each of the two dust components, we calculate the total flux (integrated over the whole galaxy) expected from the SED modeling at the water-pumping frequencies we are interested in; the corresponding values are reported in Table 
\ref{table:IRratios}, normalized to the flux at $101$ $ \mu$m. 
%%%%%%%%%%%%%%%%%%%%%%%%%%%%%%%%%%%%%%%%%%%%%%%%%%%%%%%%%%%%%%%%%%%%
%%%%%%%%%%%%%%%%%%%%%%%%%%%%%%%%%%%%%%%%%%%%%%%%%%
The L17 model analyzes, in particular,  the water excitation, in terms of levels population, in a single clump having characteristic parameters of a typical warm molecular component, with gas kinetic temperature T$_{K}$=50 K, number density n(H)=$10^{5} $ cm$^{-3}$ and water  abundance X(H${_2}$0)=$10^{-5}$, for increasing dust temperatures. 
Ignoring the effect of pumping (i.e., setting T$_{dust}$=0 K), collisions alone excite $p$-H$_2$O  ($o$-H$_2$O ) up to levels with upper energy 250 K (350 K). This scheme is almost unchanged until the dust temperature increases to a value of 40-50 K, when FIR pumping starts to populate levels with  $250-350 {\rm K} < {\rm E}_{up}/{\rm K}_B  < 500-700 {\rm K}$. The line strength of transitions  is sketched in Figure 5 of L17:
 the intensity of transitions with $E_{up}>$ 250 K (350 K) (as the $p$-H$_2$O$(4_{22}-4_{13})$ line) increases with dust temperature, while the  transitions with  $E_{up}<$ 250 K (350 K)  (as the $o$-H$_2$O$(3_{21}-3_{12})$ line) weakly depend on T$_{dust}$,
 and some of the low level lines with $E_{up}<$ 200 K  disappear for increasing T$_{dust}$, because of the depopulation by the dust photons.
In this framework, line $p$-H$_2$O$(2_{02}-1_{11})$ in this dense, warm gas (T$_{gas}\sim 50 $ K)  component, and for dust temperatures $\gtrsim$ 50 k, feels the combined effect of collisions (populating levels 1$_{11}$ and 2$_{02}$) and the weak effect of pumping, which  
populates level 2$_{20}$ through the transition 1$_{11}$-2$_{20}$ while depopulating the same level 2$_{20}$ through the transition 
2$_{20}$-3$_{31}$. The last two effects balance, in such a way that the pumping effect on the transition  $p$-H$_2$O$(2_{02}-1_{11})$ is subdominant with respect to the collisional excitation. This explains the weak dependence on the dust temperature of this line at these  relatively high density, and thus we can infer with a certain level of confidence that this line is mainly collisionally excited in the ISM component associated with the "molecular" warm dust. 
 The L17 analysis of the level population fractions shows that the collisional excitation drives the p-H$_2$O (o-H$_2$O) lines with $E_{upper}<$  100 (200) K toward a Boltzmann distribution at the gas kinetic temperature (thermalizing these levels)  and dominates the population of levels with $E_{upper}<$ 250 (350) K, almost independently on the dust temperature.
Therefore, line  $p$-H$_2$O$(2_{02}-1_{11})$ is collisionally excited, as long as the gas density is as large as  n(H)=$10^{5} $ cm$^{-3}$. 

 This thermalization can occur also at gas densities much smaller than the critical density of these water transitions ($n_{crit} \sim 10^8-10^9 cm^{-3}$, reflecting the large values of the Einstein coefficients for water rotational transitions), due to the small escape probability of the emitted photons and the consequent large optical depth and important radiative trapping which lowers the effective density for thermalization.   
For $E_{upper}>$ 250-350 K, the excitation is almost completely determined by the FIR pumping alone. In this range, the population of the levels is driven towards a Boltzmann distribution at the dust temperature. 
At the typical gas density of dense molecular clouds, $p$-H$_2$O$(4_{22}-4_{13})$ line is excited by pumping from dust photons (58 $\mu$m) only for dust temperatures above 50 K, and  the level population of $4_{22}$ with respect to $3_{21}$ increases for increasing T$_{dust}$.
We can infer  that also the intensity of line $p$-H$_2$O$(4_{22}-4_{13})$ with respect to  $o$-H$_2$O$(3_{21}-3_{12})$ is increasing with T$_{dust}$, although their ratio depends on line optical depths, and the solution of the radiative transfer equation is necessary to diagnostic the dust temperature from the line ratios. 

Considering the nuclear region of J1135 depicted in Figure \ref{fig:contours_all}, roughly delimited by the 3$\sigma$ contours of the medium and high-level water lines, we can compare the observed excitation  of the lines with the trend arising from the thermal model of  L17. We will analyze this nucleus separately from the extended tail of line $p$-H$_2$O$(2_{02}-1_{11})$ later in this Section. 

Adopting the continuum modeling of J1135 previously described, the observations of the water line in the central core are consistent with a $\sim$500 pc extended ensemble of molecular clouds hosting bursts of star formation activity, featuring  a warm  "molecular" dust component at a temperature of  $T_{dust} \sim $ 70 K or higher, embedded in a diffuse dust component,  and warm gas with density n(H)$\sim 10^{5} $ cm$^{-3}$ and $T_{gas} \sim$ 50 K.

The contribution of the colder, diffuse dust component at $\sim$40 K (in which the warm component is embedded), adds up to contribute FIR photons, in the central starburst region, although with fractional fluxes different than in the "molecular" warm component (see Table \ref{table:IRphotons}). This means that the FIR photons from the diffuse dust component are most effective in exciting the low-level line, with 101 $ \mu m$ photons than the medium and high-level, which require  continuum injection of 75 $\mu m $ and 58 $\mu m $ photons respectively. 

Summarizing, the water emission in the core region of the starburst J113625 witnesses the effect of radiative pumping on all three water lines, due to FIR photons injected by both the molecular (warm) and diffuse (colder) dust component, as well as the effect of collisions on the low-level (and, partially, on the medium level) line.  In the denser ISM, the p-H$_2$O $(2_{02}-1_{11})$ line is mostly excited by collisions, while the medium and high-level lines are mostly excited by FIR pumping. In the diffuse, colder ISM, in which warm molecular clouds are embedded, FIR pumping is mostly exciting the low-level line. In the core of the galaxy, we assume these components are mixed.

It is important to stress here  that  the relative balance between FIR pumping and collisional excitation of the low-level lines is strongly dependent on the gas and dust environment  parameters, which present a plethora of conditions in a different context (see, e.g. \citealt{Gonzalez2021}).

While the forementioned L17 model refers to a dense and warm gas component, we can tentatively interpret the more extended emission of the low-level line  (Figure \ref{fig:contours_all}) as arising far from the dense clouds cocooning young stellar populations, i.e. in a region where only the "diffuse dust" component is contributing to the local radiation field. Here, the effect of collisions is reduced with respect to the central nucleus,  due to the lower gas density, though they are still efficient in populating level 1$_{11}$, basis of the $101 \mu$m pumping. 
The continuum from the diffuse dust component is then appreciably populating the low energy water levels by absorption of $101 \mu$m photons, but it is ineffective to excite the medium and high-level lines, due to a reduced  injection of photons with the proper pumping frequencies (see the second row of Table \ref{table:IRratios}). The abundance of H$_2$O molecules may also be reduced here since the lower temperature of the radiation field is lowering the fraction of water evaporation from the dust grains ice mantles  into the gas component. The low-level line in this extended tail is then plausibly FIR pumped by the 101 $\mu$m photons of the colder, diffuse dust component, in a less dense and more extended region where the other lines do not emit, owing to the weaker FIR continuum at 75 and 50 $\mu m$. 

Translating level populations into line intensities is not trivial, since the observed emission lines are differently affected by radiative trapping and, ultimately, by their line and continuum optical depths. Usually, while the low-level lines in the water spectrum feature high optical depths ($\tau_{line} > 1$)  (and lower escape probabilities, which allows them to trace densities much lower than the line critical density), the medium-high lines are found to be optically thinner ($\tau_{line} \sim 1$) because of the higher energy required for excitation. Because of the different optical depths of those lines, the interpretation of their flux ratios requires the solution of the full radiative transfer problem, which will be the subject of a forthcoming study (Perrotta et al. 2023, in preparation). 

Finally, we notice that we are assuming an {\it a priori} value of T$_{dust} \sim 70$K, as derived from the SED fitting,  but the observed excitation of the high-level water line is also consistent with a higher dust temperature T$_{dust} \sim 100$K (the "hot" component in the L17 model). Nevertheless, the possible contribution of this hot dust component to the integrated continuum  emission is expected to be subdominant, as it  might be compact and deeply buried. The continuum emission  by this component would  be attenuated by the dense surrounding ISM, making J1135 strongly obscured at optical/NIR wavelengths. Due to the lack of mid-IR detections, this possible hot dust  component is still largely unconstrained although it  could still be included consistently with the upper limits in the mid-IR.

\begin{figure}
\begin{minipage}{0.46\textwidth}
\centering
\includegraphics[width=1.05\textwidth]
{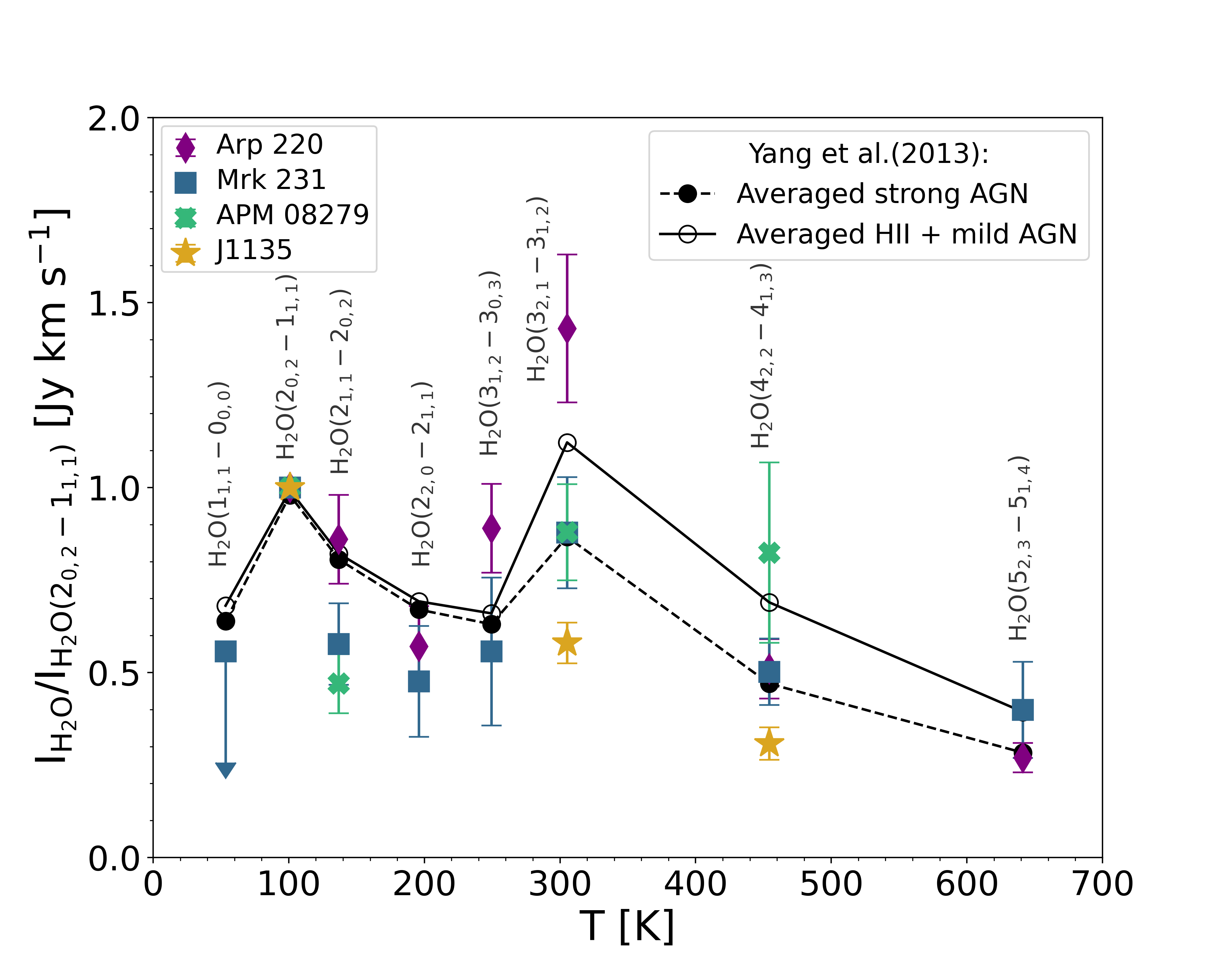}
    \caption{Integrated fluxes of the three J1135 de-magnified water lines in Jy km s$^{-1}$ (this work), compared with the sample analyzed in \cite{Yang2013} and with Mrk231 (\citealt{Gonzalez2010}), APM 08279 (\citealt{vanderWerf2011}), Arp 220 (\citealt{Rangwala2011}).  The fluxes are normalized to  the intensity of the  p-H$_2$O $(2_{02}-1_{11})$ line. }
\label{fig:waterSLED}
\end{minipage}
\end{figure}

The ratios of the line intensities, after correcting for lensing magnification, are represented in Figure  \ref{fig:waterSLED}, where they are compared with the results of \cite{Yang2013} for two samples taken from the NASA/IPAC Extragalactic Database (NED),  with Mrk 231 (\citealt{Gonzalez2010}), with Arp 220 (\citealt{Rangwala2011}), and with the lensed QSO APM08279+5255 (\citealt{vanderWerf2011}). One of the NED samples has  optically identified, strong, AGN-dominated sources (Seyfert types 1 and 2), and the other sample has star-forming-dominated galaxies, possibly with mild AGNs (classes HII, composite, and LINER of \cite{Kewley2006}, “HII+mild-AGN”).
The two groups show similar ratios in H$_2$O emission, indicating that a strong AGN may have little impact on water excitation. Noticeably, the ratio of the high to the medium water lines of J1135 (which are the lines excited mostly by FIR photons from dust) is analogous to that of the sample containing a mild AGN.  In each sample, as well as in our target galaxy, collisions alone cannot explain the relevant excitation of the medium and high-level transition. One could  formulate a dubitative hypothesis since, up to date,  there is no striking evidence for an obscured AGN (the radio luminosity of J1135 at 6 cm from available EVLA observations is consistent with the star-formation activity so that no significant contribution from a central AGN is emerging, see \citealt{Vishwas2018}).  Our results seem to point to a hot nucleus possibly powered by a starburst and maybe hosting a mild AGN in J1135.  

To further explore this possibility, we analyzed the HCN/HCO+ ratio (data taken from ALMA Science Archive, project 2017.1.01694.S, P.I. Oteo), which we will discuss in the next Subsection.

%%%%%%%%%%%%%%%%%% TABLE Flux ratios of the dust components %%%%%%%%%%%%
\begin{table}
\begin{center}
\begin{tabular} {c c c }  
\hline
Dust component  &  F$_{58}$/F$_{101}$   &  F$_{75}$/F$_{101}$  \\  
\hline
Dense molecular clouds & 1.66 & 1.40  \\

  Diffuse dust & 0.66  &  0.95  \\
 \hline
\end{tabular}
\caption{Flux ratios of J1135 FIR pumping photons from dust associated to the dense molecular clouds (warm dust component) and from the diffuse (cold) dust component, having dust temperature, respectively, of $\sim 70$K and  $\sim 40$K   (from the SED modeling by Ronconi et al 2023, represented in Figure \ref{fig:continuum}). F$_{58}$, F$_{75}$, F$_{101}$ are the fluxes at 58, 75, 101 $\mu$m, respectively, integrated over all the source.   } 
\label{table:IRphotons}
\end{center}
\end{table}

%%%%%%%%%%%%%%%%%%%%%%%%%%%%%%%%%%%%%%%%%%%%%%%%%%%%%%%%%%%
\subsection{Is J1135 hosting an AGN?}

%It is commonly accepted that local (U)LIRGs nuclei are %surrounded by a high 
%concentration of molecular gas and dust \cite{Sanders}, making %the 
%optical detection of AGN signatures very difficult: a 
%putative compact AGNs may indeed easily be buried (i.e., %obscured 
%in virtually all directions). The issue of identifying a %possible AGN is even more challenging in high-$z$ dusty star %forming galaxies. 
Co-evolutionary models between supermassive black holes (BHs) and their host galaxies predict  that processes in the early stages of galaxy formation, such as nuclear activity and star formation, are strictly related and coordinated in time (\citealt{Lapi2006},  \citealt{Lilly2013},
\citealt{Mancuso2017}, \citealt{Pantoni2019}). 
In this (in-situ) scenario, the star-formation and BH accretion
are triggered by the fast collapse of baryons toward a gravitational center,
during which most of the mass of the system is
accreted and fuels the central region of the galaxies with gas (\citealt{Lapi2018}). The intense
star formation activity is accompanied by an exponential growth of the active nucleus, whose feedback will
eventually sweep away the interstellar medium. Observing a star-forming galaxy hosting an AGN would thus
substantiate this scenario, fortunately catching the shortly-lasting time period in which star formation coexists with 
the central AGN accretion. 
The star formation would indeed  be eventually  quenched on a relatively short timescale, while the nucleus will shine as an optical quasar. 
%The vigorous star formation requires a large amount
%of gas, suggesting that the host galaxy’s ISM significantly %contributes to the obscuration of the ultraviolet (UV) and  X-ray emissions originated in the central nuclear region (\citealt{Hickox2018}), up to the Compton-thick regime
%(N(H) $>$ 10$^{24}$ cm$^{-2}$), in addition to pc-scale torus %(\citealt{Damato2020}, \citealt{Peca2021}, \citealt{Gilli2022}). %The fraction of obscured AGNs is found to be higher at earlier %epochs ($3 < z < 5$) up to $\sim 80 \%$ (e.g. \citealt{Vito2018})
%with respect to that in the local Universe %(\citealt{Burlon2011}), likely driven by the increase of the gas %content in distant
%galaxies (e.g., \citealt{Carilli}).

From the galactic evolutionary point of view, then, 
a mostly important issue is to determine whether
the obscured compact cores in high-$z$ star-forming galaxies are powered by very compact starbursts
and/or AGNs, from which strong emission is produced in very compact regions around the central accreting BH. 

The question is if our target galaxy J1135 is confirming this scenario. Our analysis of the
water excitation is pointing to 
the presence of a high excitation status in the ISM associated to the central region, but how could we 
distinguish between a purely starburst excitation and a (buried) starburst+AGN mechanism? 
Being J1135 obscured in the optical and 
IR, and due to the lack of data in the X band, the question of whether 
it hosts a buried AGN is still open.
The lack of a good sampling of the MIR
part of the SED and of X-ray data for J1135 prevents the characterisation of the central BH properties.
The CO high-J transition line (J=8-7) in J1135 reveals the presence of warm
high-density molecular gas, which can be either associated to the far-UV photons produced by star formation
processes or to the X-ray emission originated by the AGN (e.g. \citealt{Vallini2018}). Indeed, the inner regions of the galaxy
($\sim$ 0.5 kpc) are the most affected by the photons coming from the AGN, which penetrate deeply in the
giant molecular clouds generating X-ray dominated regions (XDRs, \citealt{Maloney1996}). 

\begin{figure}
\begin{minipage}{0.46\textwidth}
\centering
\includegraphics[width=1.0\textwidth]{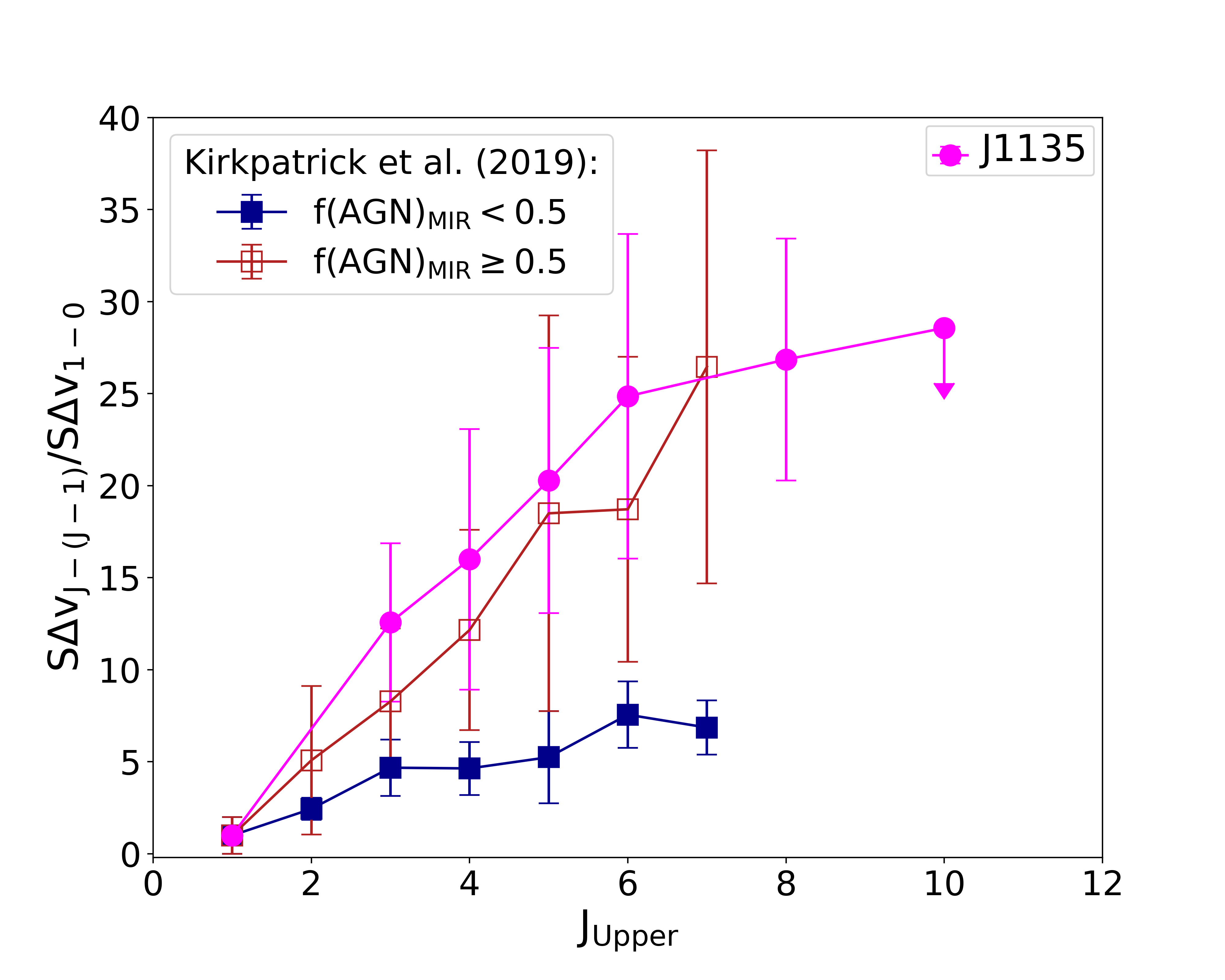}
\caption{Median CO SLEDs and line ratio uncertainties for selected sources with mid-IR AGN fractions $\geq  50 \% $(red open squares/thick lines) and mid-IR AGN fractions $ < 50 \%$ (dark blue filled squares/thick lines), from \cite{Kirkpatrick2019}. The magenta line refers to the CO SLED of HATLAS-J113625 (\citealt{Harris2012}, \citealt{Giulietti2022b}, \citealt{Yang2017}).}
\label{figure:cosled} 
\end{minipage}
\end{figure}

Generally, the
extreme compact nature of high-$z$ galaxies hinders establishing the extension of the XDRs even with high-resolution observations performed with ALMA (e.g. \citealt{Schleicher2010}), and the effects of the AGN could be visible
on the global CO spectral line energy distribution (SLED).  Figure \ref{figure:cosled} compares the results
of \cite{Kirkpatrick2019} to the  CO-SLED of J1135, which includes data from the IRAM-30 m (\citealt{Yang2017}), from ALMA (\citealt{Giulietti2022b}), and from the Green Bank telescope (\citealt{Harris2012}) . From this comparison, we found that our target is consistent with the CO-SLED of high-
redshift sources with mid-IR AGN fraction $>0.5$.
 However, Figure \ref{figure:cosled} should not be considered as  a diagnostic, as one cannot find from the analysis of \cite{Kirkpatrick2019}  a robust
statistically significant difference between the CO excitation of star-formation dominated galaxies and AGN  using the line flux alone.

Thanks to the strong lensing effect, \cite{Giulietti2022b} reconstructed and resolved the CO(8-7) emission extension for J1135, inferring a quite large effective radius of
R$_{eff} \sim 1.2$ kpc. This result is strongly hinting toward the presence of an heavily obscured AGN activity
in J1135 at the stage of its accretion or, alternatively, of a feedback associated to the central star formation activity.

Besides the indications from mid and high-water excitations and from the (8-7) CO transition,  previously discussed,
one of the keys to  discern between a 
starburst and a putative AGN, or the ensemble of both,  relies in 
the different mechanisms which rule their energetic mechanisms. Specifically,  the main powering source in starbusts is nuclear fusion, while in AGNs is accretion  into a supermassive BH. As a consequence,  UV is  the predominant energetic radiation in a starburst, while an AGN strongly  emits in the X band in addition to the UV. However, this has consequences on  the physical properties  and on the excitation state
of the dust and molecules surrounding the nuclei. 

In particular,  a peculiar and sensitive probe of the excitation mechanism resides in sub-millimeter and  millimeter wavelength  emission lines (weakly affected by dust extinction) from those molecules  which mostly characterize these very dense regions where star formation takes place, i.e. HCN and HCO$^{+}$, and in their abundance ratio. Both HCN and HCO$^{+}$ are powerful tracers of the densest (n$_{H_2}> 10 ^4$ cm$^{-3}$ )  molecular gas, thanks to their high dipole moment, so their intensity ratio is insensitive to the fraction of dense molecular gas, relative to diffuse.  While their typical ratio in our galactic molecular  cloud is $ \sim 1$ (\citealt{Blake}; \citealt{Pratap}; \citealt{Dickens}), an overabundance of HCN relative to HCO$^{+}$ is predicted for dense molecular gas when illuminated by an X-ray emitting source (\citealt{Meijerink}, \citealt{Krips2008}).  Observations of the  HCN and HCO$^{+}$ (J=1-0), (J=2-1), (J=3-2) and (J=4-3)  transitions  initially  confirmed this trend since it was found that the HCN emission is systematically stronger than the HCO$^{+}$ in  AGN-dominated galaxy nuclei than in starburst galaxies. The trend of a strong ($>1$) HCN ($1-0$)
to HCO$^{+}$ (1–0) emission ratio was further confirmed in luminous buried AGN candidates
(\citealt{Krips2008}, \citealt{Bussmann2008}, \citealt{Gracia}, \citealt{Juneau}, \citealt{Gao}, 
\citealt{Riechers}, \citealt{Kohno2005},
\citealt{Imanishi2004}, \citealt{Imanishi2006}, 
 \citealt{ImanishiNakanishi}, \citealt{Imanishi2007}, \citealt{Imanishi2009},  \citealt{Imanishi2010}, \citealt{Aalto2015}, \citealt{Imanishi2016}, \citealt{Oteo2017}).
 For this reason, enhanced emission from the dense gas tracer HCN (relative to HCO+) has been proposed as a signature of active galactic nuclei (AGN). 
However, this diagnostic was later questioned by 
\cite{Privon2020}, following the hard X-ray observations of a sample of four galaxies with HCN/HCO$^{+}$ (1-0) intensity ratios consistent with those of many AGN. No X ray evidence of an obscured AGN was found, indicating that HCN/HCO$^{+}$+ intensity ratios are not driven by the energetic dominance of AGN, nor are they reliable indicators of ongoing supermassive BH accretion (\citealt{Costagliola2011}, \citealt{Snell2011}). Low HCN/HCO$^{+}$(1-0) intensity ratios were also found in AGNs (\citealt{Sani2012}).  
This inconsistency could be due to spectral contamination from a coexisting starburst  activity that dilutes emission from an AGN. Furthermore, the reason for the HCN /HCO$^{+}$ enhancement in AGNs is still not totally clear: it may reside in a high temperature-driven chemistry (\citealt{Izumi2013}), a non-collisional excitation such as an IR pumping through the reradiation from UV/X-ray heated dust (\citealt{Sakamoto}, \citealt{Aalto}, \citealt{Garcia2015}),  or  in shocked regions  at a few hundred parsecs from the supermassive black holes due to outflowing material (\citealt{Martin}). Different heating  mechanisms, higher gas opacities, densities, and temperatures,  abundance variations, can contribute to this enhancement. 

The HCN and HCO$^{+}$ (J=1–0) transitions have rest wavelengths $\lambda_{rest}=3.385$ mm and $\lambda_{rest}=3.364$ mm respectively, 
thus suffering the limited resolutions of current observations.
\cite{Izumi2013} proposed a possibly more reliable diagnostic based on the so-called "submillimeter HCN-enhancement", making use of the HCN/HCO$^{+}$ (J=4-3) ratio, whose integrated intensity seems again to
be higher in AGNs than in starburst galaxies. These higher-J lines have submillimeter rest frame wavelengths so that higher angular resolution is easily achievable compared to (J=1–0) transitions; this is especially important to exclude contaminations from starburst activity to the line emission from AGN-heated gas. Moreover, such submillimeter lines can be covered by ALMA up to a redshift of $\sim$ 3-4. Since this feature could potentially be an extinction-free energy diagnostic tool of nuclear regions of galaxies (\citealt{Izumi2015, Izumi2016}), 
we did a preliminary analysis of the ratio HCN(4–3)/HCO$^{+}$ (J=4–3) in J1135 (purely rotational transition in the fundamental vibration state, rest frequencies $\nu_{rest} = 354.5$ GHz for HCN (J=4-3)  and $\nu_{rest} = 354.5$ GHz for HCO$^{+}$ (J=4-3), using archive ALMA data from 2017.1.01694.S, P.I. Oteo). 
The HCN/HCO$^{+}$ ratio turns out to be $\sim 2.2$
(Perrotta et al. 2023 in preparation); 
this observed  value of the HCN-to-HCO$^{+}$ (J=4-3) is suggestive of possible AGN coexisting with a starburst-dominated region.
Further observations are required, though, to confirm this result, which would be an important piece of information for the galaxy evolution scenarios. 

%%%%%%%%%%%%%%%%%%%%%%%%%%%%%%%%%%%%%%%%%%%%%%%%%%%%%%%%%%%

%%%%%%%%%%%%%%%  FIGURE CO SLED %%%%%%%%%%%%%%%%%%%%%%%%%%%%

%%%%%%%%%%%%%%%%%%%%%%%%%%%%%%%%%%%%%%%%%%%%%%%%%%%%%%%%%
\section{The \texorpdfstring{L$_{\rm H_20}$-L$_{\rm IR}$}{Lg} relation and SFR calibration} 
\label{sec:SFR} 
Multi-wavelength observations, ranging from UV to radio frequencies, have greatly improved our understanding of the SFR calibration. Direct tracers, like UV emission from newborn stars, recombination lines such as  H$_{\alpha}$ or FIR cooling lines like $[$CII$]$ at  158 $\mu$m  are, however, largely affected by dust attenuation (\citealt{Casey2014}). In contrast, the total IR luminosity L$_{\rm{IR}}$ (IR luminosity integrated into the rest frame 8-1000 $\ mu$m band)  is a promising SFR indicator precisely because of dust: in the limit of high obscuration, essentially all of the UV-optical light from young stars is absorbed and re-emitted into the IR. Thus, the IR luminosity should be one of the best SFR indicators in those situations, such as for starburst galaxies. The SFR-L$_{\rm{IR}}$ correlation is much more robust than it is for the [CII] fine structure line or for the CO(J=1-0) line, which are sublinearly correlated with L$_{\rm{FIR}}$  even when only pure star-forming galaxies are considered.
 (\citealt{DiazSantos2013}, \citealt{Solomon1997} , \citealt{Aravena2016}).
The total IR luminosity  of galaxies has been used  to infer the SFR of galaxies for decades (see \citealt{Kennicutt1998} and \citealt{Kennicutt2012} for reviews). The SFR is generally estimated from the scaling relations of  \cite{Kennicutt2012}, 
that assumes a Salpeter IMF in the mass range 0.1-100 M$_{\Sun}$, 
solar metallicity and for continuous bursts of age 
10 - 100 Myr: 
\begin{equation}
\label{eq:kennicutt}
\rm{SFR} \left[\rm{M}_\Sun yr ^{-1} \right] \approx 1.47 \times 10^{-10} \rm{L_{\rm{IR}}}\, \left[\rm{L}_{\Sun}\right]
\end{equation}

A caveat to take into account is that L$_{\rm{IR}}$  may  overestimate the SFR in all those cases in which dust heating is enhanced by physical mechanisms other than star formation, for example, extra UV emission or X-rays, as arising in the light of evolved stars or AGNs (\citealt{Kennicutt2009}; \citealt{Murphy2011}; \citealt{Hayward2014}). In all those cases,  converting the IR luminosity into an SFR using a standard calibration will overestimate the true SFR.
Besides this caveat,  retrieving L$_{\rm{IR}}$ requires a good sampling or modeling of the source SED, which is not always available. A way to encompass this possible lack of detailed information about the SED is to use tracers of L$_{\rm{IR}}$  itself. 
L$_{\rm{H_2O}}$ was found to have a strong dependence on the IR luminosity, varying as L$_{\rm{H_2O}}$ $\sim$  L${_{\rm IR}}^{1.2}$ (\citealt{Omont2013}, \citealt{Yang2013}, \citealt{Yang2016}), slightly steeper than linear,  and equivalent to a linear relation in log–log space with a slope very close to unity. It extends over four orders of magnitude of the luminosity range, regardless of whether or not a strong AGN signature is present.
This relation is indicative of the fundamental role of radiative IR excitation of the water lines and implies that high-z galaxies with L$_{\rm IR} $ $\gtrsim 10^{13}$  L$_{\Sun}$ tend to be very strong emitters in H$_2$O, that have no equivalent in the local universe. 

Our modeling of J1135 SED with a warm dust component and a cold, diffuse one, allows us to infer for the total IR luminosity a value of L$_{\rm{IR}} \approx 1.21 \times 10^{13}$ L$_\sun$
and a FIR luminosity, integrated into the (42.5-122.5  $\mu$m) band, L$_{\rm{FIR}} \approx 7.05 \times 10^{12}$ L$_\sun$, after correcting for lensing magnification.   Equation  (\ref{eq:kennicutt}) then provides a SFR of $\sim 1.77 \times 10^3\, M_\odot$ yr$^{-1}$. 
After the demagnification of the water line intensities given in Table \ref{tab:lens_parameters}, we  calculate for each transition the corresponding ratio to the total IR luminosity and the steepness $\alpha$ for the relation L$_{H_2O}$= L$_{\rm IR}^{\alpha}$, 
where, for each water emission line, the luminosity L$_{H_2O}$ is in units of $10^{7} $ L$_{\Sun}$ and L$_{\rm IR}$ in units of $10^{12} $ L$_{\Sun}$. Our results are reported in Table \ref{table:IRratios} and compared with the best-fit values for the sample by \cite{Yang2013}, initially performed on a sample of 45 water-emitting galaxies spanning a luminosity range (1-300 $\times 10^{10} $ L$_\sun$) and on high-$z$ dusty star-forming galaxies, and later also including lensed hyper-luminous IR galaxies at z $\sim 2-4$ (\citealt{Yang2016}, \citealt{Yang2020}). 

Comparing our ratios with \cite{Yang2013}, 
we find a consistent value (within $3 \sigma$) for
all the transitions. In particular, this confirms the trend, already found in \cite{Yang2013}, that the inclusion of high-z dusty star-forming galaxies at the high L$_{\rm IR}$ end of the L$_{\rm H_2O}$-L$_{\rm IR}$ relation has the effect of slightly increasing L${_{\rm H_2O}}$/L$_{\rm IR}$ when looking at the ${2_{02}-1_{11}}$ transition. 
%On the contrary, for the two transitions ${3_{21}-3_{12}}$ and  ${4_{22}-4_{13}}$ we obtain $\alpha$ values smaller by a factor 25$\%$ and  $38\%$, respectively,  with respect to the best fit of \cite{Yang2013}. 
We explain this effect by noticing that,  according to our interpretation of the J1135 water emission described in  Section \ref{sec:analysis}, the 988 GHz line is enhanced by a substantial contribution from collisions; since collisions are not a direct tracer of the warm/hot phase of the star-forming ISM, the use of this low-level line may bias the  L${_{\rm H_2O}}$-L$_{\rm IR}$ correlation. Thus, the use of the H$_2$O $ {2_{02}-1_{11}}$  line as a tracer of IR emission may result in overestimating the SFR. This mechanism could be acting in high-z DSFGs, explaining the observed enhancement of the steepness in the L$_{\rm H_2O}$-L$_{\rm IR}$ relation for this  particular class of sources, in which J1135 is included. 

In the same manner, our detailed imaging analysis joined with  our model for the J1135 dust components, suggests that considering the total IR luminosity may not give precise results, as the flux between 8 and 1000 $\mu$m may also include small contributions from old stellar populations and possibly relevant AGNs. In this way, the L$_{\rm IR}$ would not be a direct tracer of the dust emission  strictly related to star formation. Thus, it would  overestimate the SFR if used as a direct SF tracer, or underestimate it  if we use  water lines only excited by IR pumping (such as the H$_2$O ${4_{22}4_{13}}$ or higher $E_{up}$)  as calibrators since they would be divided  by L$_{\rm IR}$ instead than by the effective warm/hot dust component emission (smaller than L$_{\rm IR}$).

For this reason, we find  more reliable to relate any SFR indicators to the FIR luminosity rather than to the total IR.
As outlined in \cite{Kennicutt1998}, 
the efficacy of the FIR luminosity as a SFR tracer depends on the contribution of young stars to the heating of the dust, and on the optical depth of the dust in the star-forming regions. The simplest physical situation is one in which newly born stars dominate the UV–visible radiation field  and the dust opacity is high everywhere, in which case the FIR luminosity measures the bolometric luminosity of the starburst. In such a limiting case the FIR luminosity is the ultimate SFR tracer. For this reason, at the high optical depths, typical of starbursts,  L$_{\rm FIR}$, integrated over the rest frame wavelength range 42.5-122.5 $\mu$m turns out to be a reliable 
tracer of SF, possibly more than L$_{\rm IR}$.
As long as our analysis is limited to high FIR luminosities typical of starbursts (i.e. L$_{\rm{FIR}} \geq 10^{12}$L$_{\Sun}$),  the luminosity of the spectral water line $p$-H$_2$O$(2_{02}-1_{1,1})$, corresponding to the rest-frame frequency of 988 GHz, does linearly correlate with L$_{\rm{FIR}}$,
 suggesting that the luminosity of this particular water line, L$_{(2_{02}-1{1,1})}$  may be conveniently used as a SFR calibrator of high redshift starbursts. The crucial point here is that,
 as discussed in Section \ref{sec:analysis},  this correlation could be easily explained by the  partial fuelling of this transition by  101 $\mu$m photons from the diffuse dust component in which the dense component is embedded (\citealt{Jarugula2019}, \citealt{vanderWerf2011}, \citealt{Yang2013, Yang2016}).  We have to notice, however, that the excitation fraction due to pumping  depends on
 the gas and dust parameters, since, as shown in the L17 model and discussed in Section \ref{sec:analysis}, FIR pumping of this line in a dense warm ISM component starts to contribute for T$_{dust}>40$K weakly, and it is more important in the less dense ISM component, while collisions in a dense ISM are contributing to excite this line even without a FIR radiation field.
  Noticeably, \cite{Jarugula2019}  found that the 988 GHz line is correlated with  L$_{\rm{FIR}}$ not only on global scales but also on resolved kiloparsec scales within starbursts and AGNs, in source regions with uniform dust temperature and opacity.  On this respect, high-resolution observations could select the water emission arising from a selected zone around the peaks of the line luminosities with high signal-to-noise ratio, and plausibly uniform physical conditions. For the $p$-H$_2$O$(2_{02}-1_{1,1})$ transition, considering the whole area of this line emission, we have a lower limit, because we are also counting the line flux coming from a region not directly associated with the molecular clouds hosting star formation: 
L$_{\rm H_2O}$/L$_{\rm FIR}$ 1.63 $\pm$ 0.03 $\times 10^{-6}$, in good agreement with the results from \cite{Jarugula2019} for a sample of gravitationally lensed dusty star-forming galaxies at z$\sim$3.
This reinforces the hypothesis that $p$-H$_2$O$(2_{02}-1_{1,1})$ in such galaxies is correlated with the FIR luminosity of the starburst.  
However, we suggest that a more reliable diagnostic of the SFR in dusty starburst galaxies based on the thermal dust emission, should better
rely on the high-level H$_2$O transitions (like the ${4_{22}-4_{13}}$ analyzed in this paper). Indeed, in this case, the ratio L$_{\rm{H_2O}}$ / L$_{\rm FIR} $ is dominated by IR pumping (actually, almost purely pumping induced), thus it is expected to increase for increasing dust temperatures and SFR, unlike the low-level transition  at 988 GHz, whose upper energy level  tends to depopulate at increasing temperatures.
For completeness, then, we report in the last column of Table \ref{table:IRratios} our  J1135 calibrations for the SFR/ L$_{\rm{H_2O}} $ obtained from  eq. \ref{eq:kennicutt}, from the measured water line luminosities and from the  L$_{\rm IR}$/L$_{\rm FIR}$= 1.72  value obtained by Ronconi et al. 2023 for our target galaxy.

%%%%%%%%%%%%%%%%%% TABLE of lines Gaussian fits %%%%%%%%%%%%
\begin{table*}
\begin{center}
\begin{tabular} {c c c c c c c}    
 \hline
    Line & L$_{\rm{H_2O}}$  & (L$_{\rm{H_2O}}$/L$_{\rm IR}) \times 10^6$ & $({\rm L}_{\rm{H_2O}}/{\rm L}_{\rm IR} ) ^{(a)}\times 10^6 $   & $\alpha$  & ${\alpha}^{(a)}$ & SFR/ ${\rm L}_{\rm{H_2O}}$ \\  
     & $(10^7$ L$_{\Sun})$ & &  &  &  &  M$_{\Sun} {\rm yr}^{-1}$ L${_{\Sun}^{-1}}$  \\
 \hline
 $p$-H$_2$O $2_{02}$-$1_{11}$ & 11.4 $\pm$ 0.2 & 9.4 $\pm$ 1.6 & 7.58 & 0.97 $\pm$ 0.06 & 1.12 $\pm$ 0.04 &  1.56 $\times 10^{-5}$   \\
 $o$-H$_2$O $3_{21}-3_{12}$ & 7.9 $\pm$ 0.6 & 6.54 $\pm$ 0.55 & 10.70 & 0.83 $\pm$ 0.03 & 1.11 $\pm$ 0.05 &  2.25 $\times 10^{-5}$ \\
  $p$-H$_2$O $4_{22}$-$4_{13}$ & 4.3 $\pm$ 0.83 & 3.56  $\pm$ 0.65 & 5.71 & 0.58 $\pm$ 0.07 & 0.94 $\pm$ 0.12 &  4.13 $\times 10^{-5}$  \\
[1ex]
 \hline
\end{tabular}
\caption{De-magnified luminosities of the water lines analyzed in this paper; ratios of water luminosities to the IR total luminosity of J1135, where the index (a) refers to the results reported in \cite{Yang2013};  $\alpha$ coefficient for the relation L$_{H_2O}$= L$_{\rm IR}^{\alpha}$; SFR vs. water line luminosity. The SFRs were obtained from  eq. \ref{eq:kennicutt} and linked to the luminosities of the water transitions.}
\label{table:IRratios}
\end{center}
\end{table*}

%%%%%%%%%%%%%%%%%%%%%%%%%%%%%%%%%%%%%%%%%%%%%%%%%%%%%%
\section{Summary and Conclusions} 
\label{sec:conclusions}
We  presented the high-resolution ALMA images of the water emission lines in the strongly lensed highly optical/NIR obscured galaxy J1135 at $z\sim 3.1$. We analyzed the distribution of the water and of CO(8-7) emission lines with respect to the continuum emission,  modeled through the data SED fitting to include a diffuse dust component and a warmer component associated with molecular clouds hosting young stellar populations.  

We have been able to directly associate, already in the visual imaging, a central nucleus (the possible starburst region) to the peak of those lines which are more sensitive to the FIR pumping of photons from a warm ($T_{dust} \sim 70 $K) dust component (namely,  {\it{o}}-{H$_2$O} ${3_{21}-3_{12}}$ and {\it{p}}-{H$_2$O} $({4_{22}-4_{13}}$). 
The low-level line {\it{p}}-{H$_2$O} ${2_{02}-1_{11}}$ peaks  in the same central nucleus, despite a small displacement, and presents a more extended emission; its excitation, on the core and on the extended tail,  is probably due to  different combinations of collisions and pumping of $101 \mu $m photons from the denser, warmer and from the diffuse, colder dust component. 

An accurate fitting of the line profiles showed no evidence for outflows nor rotation of the central nucleus from which the water lines arise, confirming previous results obtained by \cite{Giulietti2022b} on the basis of the [CII] emission. A single-Gaussian fit turned out to reproduce the water and the CO(8-7) emission lines at a high confidence level, indicating that the mid- and high-level water excitation region encompasses the same dense gas region where CO is collisionally excited.  We compared our imaging and line analysis to the existing radiative transfer models which account for the effect of dust-emitted photons in the excitation of ISM molecules  as well as of the collisional excitations. This comparison is  suggestive of a warm, or even hot  ($\gtrsim $ 70 K), dust component in the central nucleus, which, together with a colder, diffuse component, could explain the observed excitations.
%The maps and lines analysis of Sections \ref{sec:imaging} and %\ref{section:lineshapes}, together with the SED modeling,  is thus indicating that the %target galaxy exhibits a  warm/hot nuclear region (core) with size R $\sim$ 500 pc, a %diffuse ISM component at T$_{dust} \sim 40$K mixed with a warm/hot component with $70 %< $ T$_{dust} < $ 100 K, optically thick at optical/NIR wavelengths, where the low %level transition, for a fiducial value of n(H)=$10^{5} $ cm$^{-3}$,  is mainly excited %by collisions, and the pumping contribution increases for increasing E$_{upper}$ and %increasing  T$_{dust}$.
The lack of  X-ray observations and  MIR detections of the continuum are currently preventing us from discerning the nature of this central hot nucleus.

We discussed the importance, for the in-situ evolutionary scenario,  of detecting an accreting AGN in a star forming galaxy, outlining the spectroscopic observations which 
are, to date,  suggestive of that case, though further observations are needed to confirm the case.  

Finally, we exploited  the robustness of  the 
water lines as possible SFR calibrators, comparing water luminosity-IR luminosity correlation with the values obtained in literature for samples of local and high-z sources and for strongly lensed dusty star-forming galaxies. 
Our physically motivated analysis, based on the high-resolution imaging of the targeted source, suggested that the better SFR indicators should be the high-level water transitions, since they are exclusively excited by FIR pumping from dust, with negligible contribution from collision, and  direct tracers of the warm and hot dust typically powered in star-forming environments. The sensitivity of the high-level water line to FIR pumping in starburst can thus be a faithful SFR calibrator, bypassing the knowledge of the full IR or FIR galaxy spectrum. We provided the corresponding calibrations for J1135, but a similar analysis on additional galaxies will be needed to obtain statistically significant confirmation of our results. 

\begin{acknowledgments}
This work is partially supported by the PRIN MIUR 2017 prot. 20173ML3WW, `Opening the ALMA window on the cosmic evolution of gas, stars and supermassive black holes'. AL acknowledges funding from the EU H2020-MSCA-ITN-2019 Project 860744 `BiD4BESt: Big Data applications for black hole Evolution STudies'. F.P. acknowledges P. Marsiaj for the technical support, L. Silva and  M. Rybak for useful discussions. 

\bigskip

This work has been developed by the GOThA (Galaxy Observational and Theoretical Astrophysics; see \texttt{https://gotha-1.jimdosite.com/}) inter-disciplinary and inter-institutional team. 

\bigskip

This paper makes use of the following ALMA data: ADS/JAO.ALMA\#2018.1.00861.S, ADS/JAO.ALMA\#2017.1.01694.. ALMA is a partnership of ESO (representing its member states), NSF (USA) and NINS (Japan), together with NRC (Canada), MOST and ASIAA (Taiwan), and KASI (Republic of Korea), in cooperation with the Republic of Chile. The Joint ALMA Observatory is operated by ESO, AUI/NRAO and NAOJ. For the exploitation of the ALMA Science Archive, the paper made use of the products of the Additional Representative Images for Legacy \citep[ARI-L,][]{Massardi2021} project.
\end{acknowledgments}

\software{astropy \citep{2013A&A...558A..33A,2018AJ....156..123A},  
CASA (v4.7.2; \citealt{McMullin2007}), PyNUFFT (\citealt{Lin2018}), PyLops (\citealt{Ravasi2020}), PyAutoLens (\citealt{Nightingale2018,Nightingale2021}), emcee (\citealt{ForemanMackey2013}), getdist (\citealt{Lewis2019}). }

\appendix
\section{Bayesian lines shape reconstruction}
\begin{figure*}
\centering
    \resizebox{8.5 cm}{!}{\includegraphics[width=\textwidth]{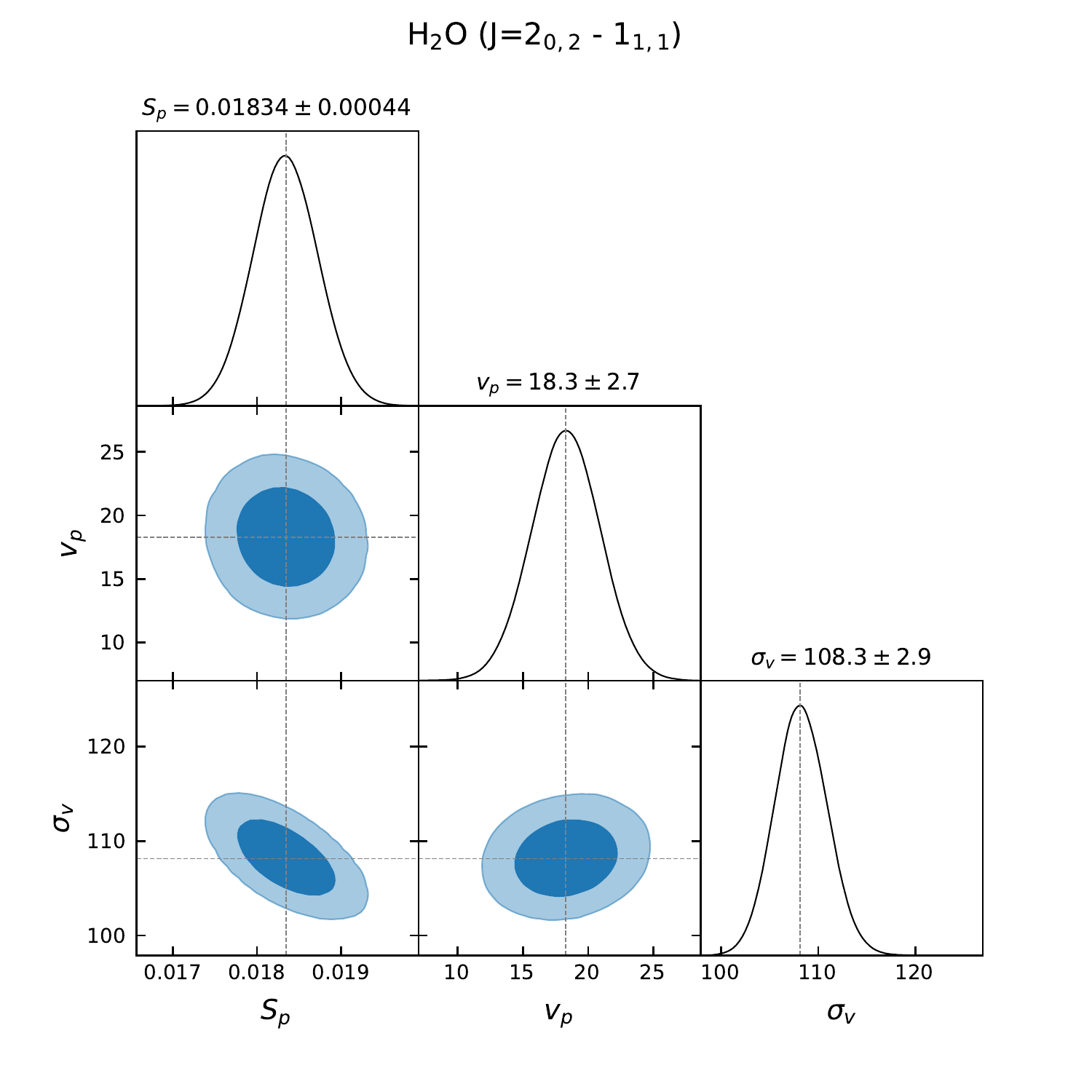}}
    \resizebox{8.5 cm}{!}{\includegraphics[width=\textwidth]{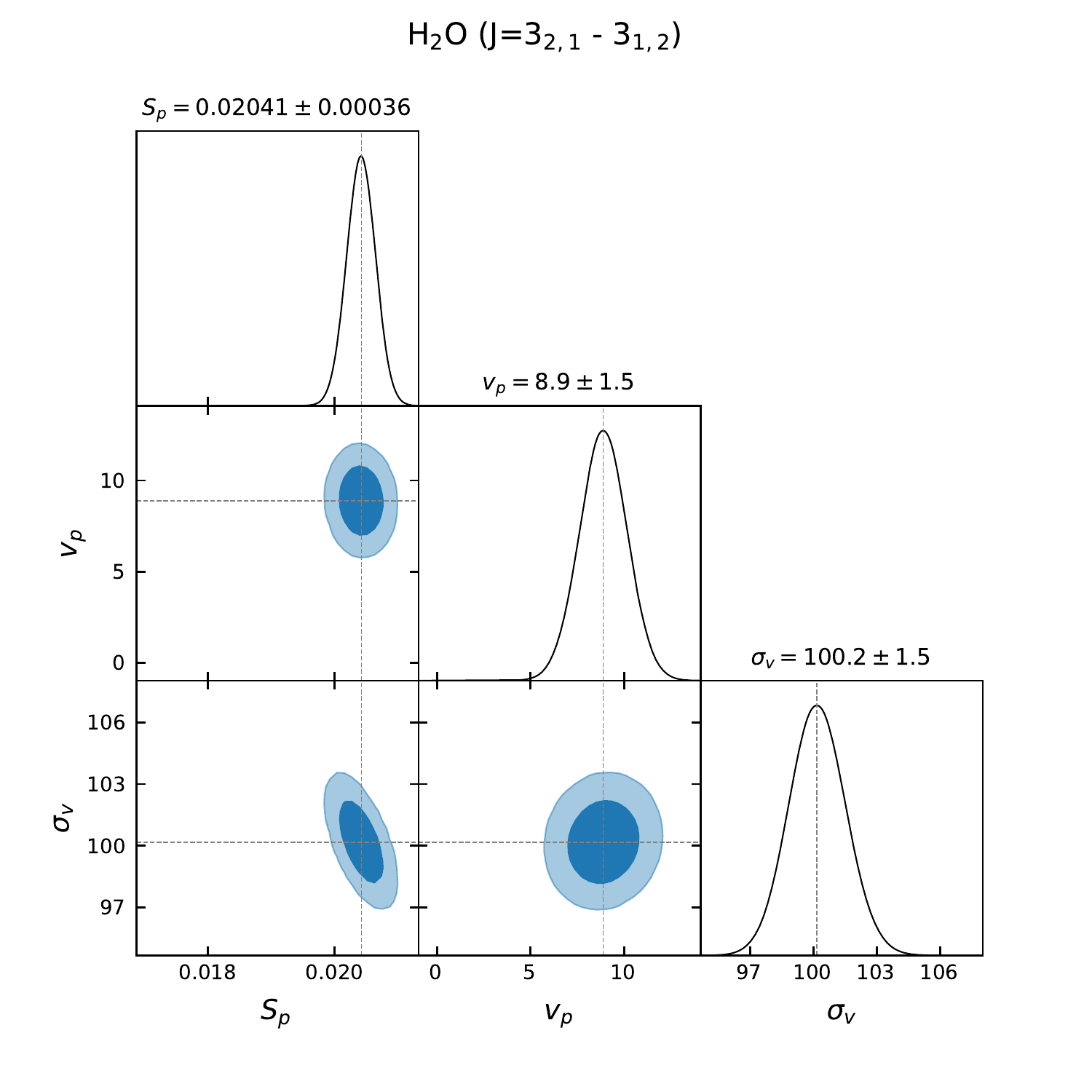}}
    \resizebox{8.5cm}{!}{\includegraphics[width=\textwidth]{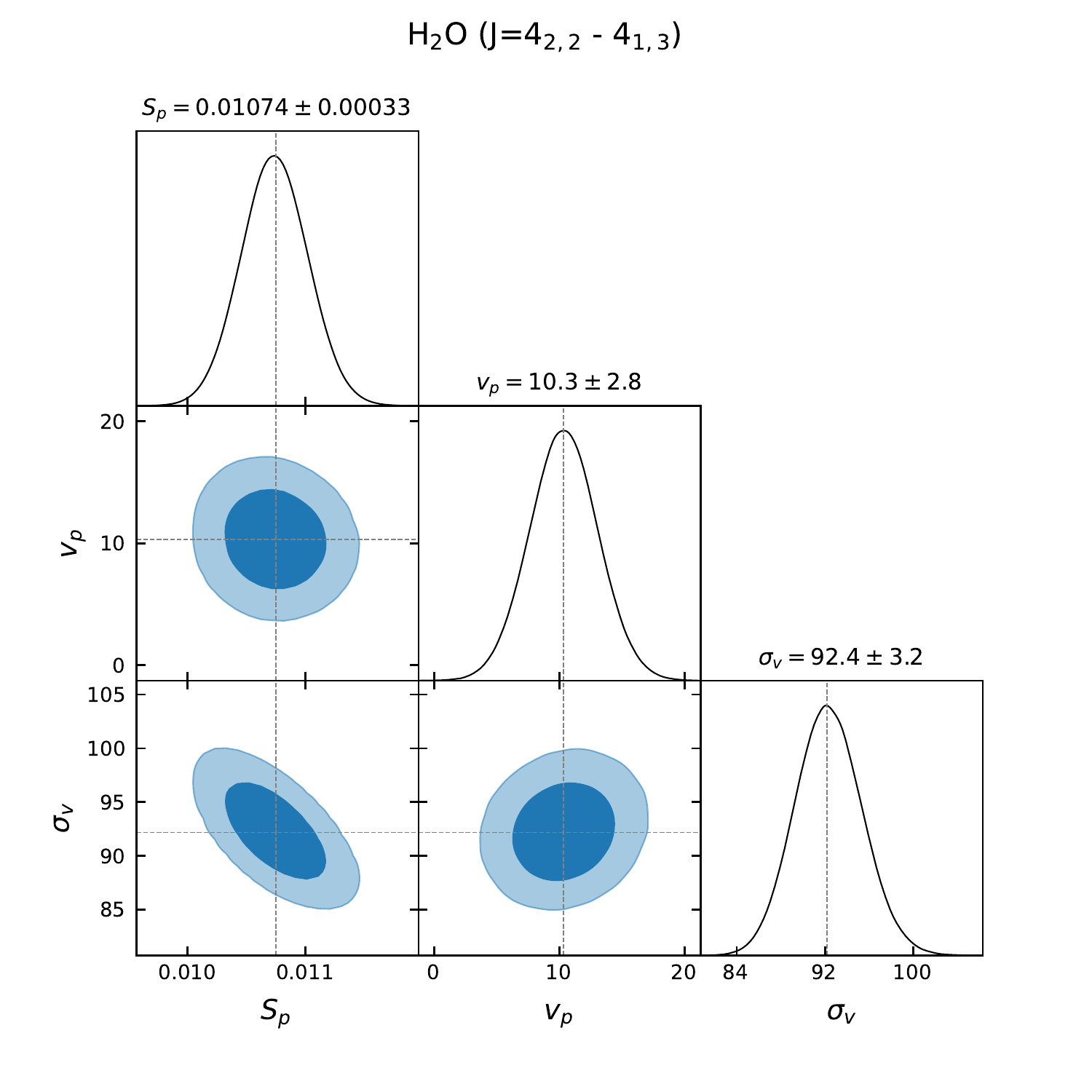}}
    \resizebox{8.5cm}{!}{\includegraphics[width=\textwidth]{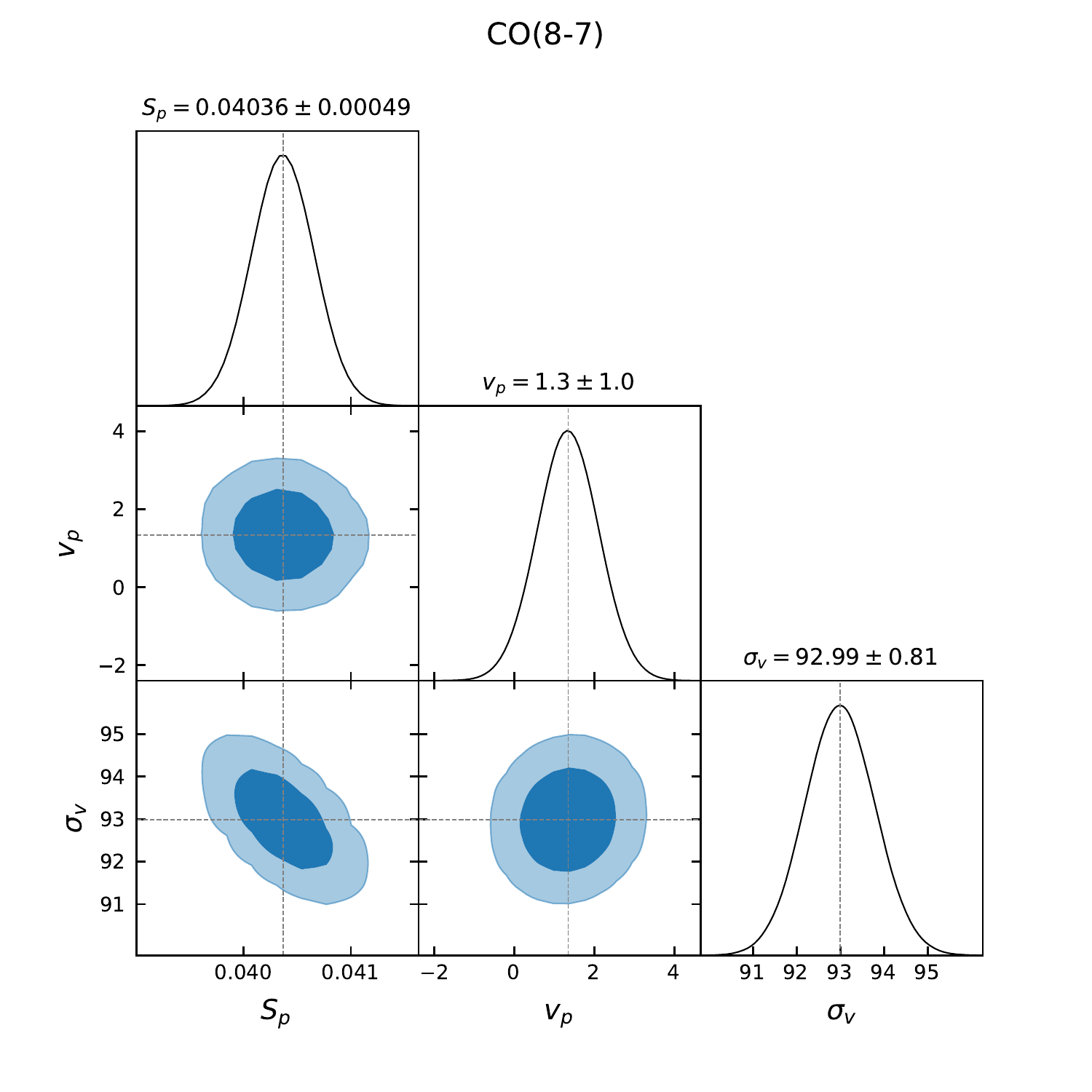}}
       \caption{Contour plots for our Bayesian lines shape reconstruction obtained with the \texttt{getdist} Python package.}
       \label{fig:contour_plots}
\end{figure*}

{This appendix summarizes relevant information about the technique of spectral line fitting. Our MCMC fit maximizes the likelihood $\mathcal{L}(\theta) \equiv - \chi^2(\theta) / 2$, where $\chi^2=[\mathcal{M}(\theta)-\mathcal{D}]^2 / \sigma_{\mathcal{D}}^2$ is obtained by comparing the expectations of our empirical model $\mathcal{M}(\theta)$ with the data $\mathcal{D}$ having uncertainties $\sigma_{\mathcal{D}}^2$. We adopted flat prior $\pi(\theta)$ on the fit parameters, sampling the posterior distribution $\mathcal{P}(\theta) \propto \mathcal{L}(\theta) \pi(\theta)$ by running \texttt{emcee} with $10^4$ iterations and 300 walkers. We initialized each walker with a random position uniformly sampled from the (flat) priors and discarded a fraction of the initial iterations of the MCMC to allow the chain to reach statistical equilibrium. The fraction of the chain to discard was determined from the Markov Chain's autocorrelation time computed for each parameter thanks to the Gelmen-Rubin criterion (\citealt{GelmanRubin1992}). In our case, the fraction of rejected chain is about 18$\%$ of the MCMC on average for each fitted spectral line. In Figure \ref{fig:contour_plots} we show the contour plots obtained from the chains via the Python package \texttt{getdist} (\citealt{Lewis2019}).}
%%%%%%%%%%%%%%%%%%%%%%%%%%%%%%%%%%%%%%%%%%%%%%%%%%%%%%%%%%%
%%%%%%%%%%%%%%%%%%%%%%%%%%%%%%%%%%%%%%%%%%%%%%%%

\bibliography{main}{}

\begin{thebibliography}{}
\expandafter\ifx\csname natexlab\endcsname\relax\def\natexlab#1{#1}\fi
\providecommand{\url}[1]{\href{#1}{#1}}
\providecommand{\dodoi}[1]{doi:~\href{http://doi.org/#1}{\nolinkurl{#1}}}
\providecommand{\doeprint}[1]{\href{http://ascl.net/#1}{\nolinkurl{http://ascl.net/#1}}}
\providecommand{\doarXiv}[1]{\href{https://arxiv.org/abs/#1}{\nolinkurl{https://arxiv.org/abs/#1}}}

\bibitem[{{Aalto} {et~al.}(1995){Aalto}, {Booth}, {Black}, \&
  {Johansson}}]{Aalto}
{Aalto}, S., {Booth}, R.~S., {Black}, J.~H., \& {Johansson}, L.~E.~B. 1995,
  \aap, 300, 369

\bibitem[{{Aalto} {et~al.}(2009){Aalto}, {Wilner}, {Spaans}, {Wiedner},
  {Sakamoto}, {Black}, \& {Caldas}}]{Aalto2009}
{Aalto}, S., {Wilner}, D., {Spaans}, M., {et~al.} 2009, VizieR Online Data
  Catalog, J/A+A/493/481

\bibitem[{{Aalto} {et~al.}(2015){Aalto}, {Garcia-Burillo}, {Muller}, {Winters},
  {Gonzalez-Alfonso}, {van der Werf}, {Henkel}, {Costagliola}, \&
  {Neri}}]{Aalto2015}
{Aalto}, S., {Garcia-Burillo}, S., {Muller}, S., {et~al.} 2015, \aap, 574, A85,
  \dodoi{10.1051/0004-6361/201423987}

\bibitem[{{Aravena} {et~al.}(2016){Aravena}, {Spilker}, {Bethermin},
  {Bothwell}, {Chapman}, {de Breuck}, {Furstenau}, {G{\'o}nzalez-L{\'o}pez},
  {Greve}, {Litke}, {Ma}, {Malkan}, {Marrone}, {Murphy}, {Stark}, {Strandet},
  {Vieira}, {Weiss}, {Welikala}, {Wong}, \& {Collier}}]{Aravena2016}
{Aravena}, M., {Spilker}, J.~S., {Bethermin}, M., {et~al.} 2016, \mnras, 457,
  4406, \dodoi{10.1093/mnras/stw275}

\bibitem[{{Astropy Collaboration} {et~al.}(2013){Astropy Collaboration},
  {Robitaille}, {Tollerud}, {Greenfield}, {Droettboom}, {Bray}, {Aldcroft},
  {Davis}, {Ginsburg}, {Price-Whelan}, {Kerzendorf}, {Conley}, {Crighton},
  {Barbary}, {Muna}, {Ferguson}, {Grollier}, {Parikh}, {Nair}, {Unther},
  {Deil}, {Woillez}, {Conseil}, {Kramer}, {Turner}, {Singer}, {Fox}, {Weaver},
  {Zabalza}, {Edwards}, {Azalee Bostroem}, {Burke}, {Casey}, {Crawford},
  {Dencheva}, {Ely}, {Jenness}, {Labrie}, {Lim}, {Pierfederici}, {Pontzen},
  {Ptak}, {Refsdal}, {Servillat}, \& {Streicher}}]{2013A&A...558A..33A}
{Astropy Collaboration}, {Robitaille}, T.~P., {Tollerud}, E.~J., {et~al.} 2013,
  \aap, 558, A33, \dodoi{10.1051/0004-6361/201322068}

\bibitem[{{Astropy Collaboration} {et~al.}(2018){Astropy Collaboration},
  {Price-Whelan}, {Sip{\H{o}}cz}, {G{\"u}nther}, {Lim}, {Crawford}, {Conseil},
  {Shupe}, {Craig}, {Dencheva}, {Ginsburg}, {VanderPlas}, {Bradley},
  {P{\'e}rez-Su{\'a}rez}, {de Val-Borro}, {Aldcroft}, {Cruz}, {Robitaille},
  {Tollerud}, {Ardelean}, {Babej}, {Bach}, {Bachetti}, {Bakanov}, {Bamford},
  {Barentsen}, {Barmby}, {Baumbach}, {Berry}, {Biscani}, {Boquien}, {Bostroem},
  {Bouma}, {Brammer}, {Bray}, {Breytenbach}, {Buddelmeijer}, {Burke},
  {Calderone}, {Cano Rodr{\'\i}guez}, {Cara}, {Cardoso}, {Cheedella}, {Copin},
  {Corrales}, {Crichton}, {D'Avella}, {Deil}, {Depagne}, {Dietrich}, {Donath},
  {Droettboom}, {Earl}, {Erben}, {Fabbro}, {Ferreira}, {Finethy}, {Fox},
  {Garrison}, {Gibbons}, {Goldstein}, {Gommers}, {Greco}, {Greenfield},
  {Groener}, {Grollier}, {Hagen}, {Hirst}, {Homeier}, {Horton}, {Hosseinzadeh},
  {Hu}, {Hunkeler}, {Ivezi{\'c}}, {Jain}, {Jenness}, {Kanarek}, {Kendrew},
  {Kern}, {Kerzendorf}, {Khvalko}, {King}, {Kirkby}, {Kulkarni}, {Kumar},
  {Lee}, {Lenz}, {Littlefair}, {Ma}, {Macleod}, {Mastropietro}, {McCully},
  {Montagnac}, {Morris}, {Mueller}, {Mumford}, {Muna}, {Murphy}, {Nelson},
  {Nguyen}, {Ninan}, {N{\"o}the}, {Ogaz}, {Oh}, {Parejko}, {Parley}, {Pascual},
  {Patil}, {Patil}, {Plunkett}, {Prochaska}, {Rastogi}, {Reddy Janga},
  {Sabater}, {Sakurikar}, {Seifert}, {Sherbert}, {Sherwood-Taylor}, {Shih},
  {Sick}, {Silbiger}, {Singanamalla}, {Singer}, {Sladen}, {Sooley},
  {Sornarajah}, {Streicher}, {Teuben}, {Thomas}, {Tremblay}, {Turner},
  {Terr{\'o}n}, {van Kerkwijk}, {de la Vega}, {Watkins}, {Weaver}, {Whitmore},
  {Woillez}, {Zabalza}, \& {Astropy Contributors}}]{2018AJ....156..123A}
{Astropy Collaboration}, {Price-Whelan}, A.~M., {Sip{\H{o}}cz}, B.~M., {et~al.}
  2018, \aj, 156, 123, \dodoi{10.3847/1538-3881/aabc4f}

\bibitem[{{Bergin} {et~al.}(2003){Bergin}, {Kaufman}, {Melnick}, {Snell}, \&
  {Howe}}]{Bergin2003}
{Bergin}, E.~A., {Kaufman}, M.~J., {Melnick}, G.~J., {Snell}, R.~L., \& {Howe},
  J.~E. 2003, \apj, 582, 830, \dodoi{10.1086/344674}

\bibitem[{{Blain}(1996)}]{Blain1996}
{Blain}, A.~W. 1996, \mnras, 283, 1340, \dodoi{10.1093/mnras/283.4.1340}

\bibitem[{{Blain} {et~al.}(2002){Blain}, {Smail}, {Ivison}, {Kneib}, \&
  {Frayer}}]{Blain2002}
{Blain}, A.~W., {Smail}, I., {Ivison}, R.~J., {Kneib}, J.~P., \& {Frayer},
  D.~T. 2002, \physrep, 369, 111, \dodoi{10.1016/S0370-1573(02)00134-5}

\bibitem[{{Blake} {et~al.}(1987){Blake}, {Sutton}, {Masson}, \&
  {Phillips}}]{Blake}
{Blake}, G.~A., {Sutton}, E.~C., {Masson}, C.~R., \& {Phillips}, T.~G. 1987,
  \apj, 315, 621, \dodoi{10.1086/165165}

\bibitem[{{Bolatto} {et~al.}(2013){Bolatto}, {Wolfire}, \&
  {Leroy}}]{Bolatto2013}
{Bolatto}, A.~D., {Wolfire}, M., \& {Leroy}, A.~K. 2013, \araa, 51, 207,
  \dodoi{10.1146/annurev-astro-082812-140944}

\bibitem[{{Bussmann} {et~al.}(2008){Bussmann}, {Narayanan}, {Shirley},
  {Juneau}, {Wu}, {Solomon}, {Vanden Bout}, {Moustakas}, \&
  {Walker}}]{Bussmann2008}
{Bussmann}, R.~S., {Narayanan}, D., {Shirley}, Y.~L., {et~al.} 2008, \apjl,
  681, L73, \dodoi{10.1086/590181}

\bibitem[{{Carilli} \& {Walter}(2013)}]{Carilli}
{Carilli}, C.~L., \& {Walter}, F. 2013, \araa, 51, 105,
  \dodoi{10.1146/annurev-astro-082812-140953}

\bibitem[{{Caselli} {et~al.}(2010){Caselli}, {Keto}, {Pagani}, {Aikawa},
  {Y{\i}ld{\i}z}, {van der Tak}, {Tafalla}, {Bergin}, {Nisini}, {Codella}, {van
  Dishoeck}, {Bachiller}, {Baudry}, {Benedettini}, {Benz}, {Bjerkeli}, {Blake},
  {Bontemps}, {Braine}, {Bruderer}, {Cernicharo}, {Daniel}, {di Giorgio},
  {Dominik}, {Doty}, {Encrenaz}, {Fich}, {Fuente}, {Gaier}, {Giannini},
  {Goicoechea}, {de Graauw}, {Helmich}, {Herczeg}, {Herpin}, {Hogerheijde},
  {Jackson}, {Jacq}, {Javadi}, {Johnstone}, {J{\o}rgensen}, {Kester},
  {Kristensen}, {Laauwen}, {Larsson}, {Lis}, {Liseau}, {Luinge}, {Marseille},
  {McCoey}, {Megej}, {Melnick}, {Neufeld}, {Olberg}, {Parise}, {Pearson},
  {Plume}, {Risacher}, {Santiago-Garc{\'\i}a}, {Saraceno}, {Shipman}, {Siegel},
  {van Kempen}, {Visser}, {Wampfler}, \& {Wyrowski}}]{Caselli2010}
{Caselli}, P., {Keto}, E., {Pagani}, L., {et~al.} 2010, \aap, 521, L29,
  \dodoi{10.1051/0004-6361/201015097}

\bibitem[{{Casey} {et~al.}(2014){Casey}, {Narayanan}, \& {Cooray}}]{Casey2014}
{Casey}, C.~M., {Narayanan}, D., \& {Cooray}, A. 2014, \physrep, 541, 45,
  \dodoi{10.1016/j.physrep.2014.02.009}

\bibitem[{{Cernicharo} {et~al.}(2006){Cernicharo}, {Pardo}, \&
  {Weiss}}]{Cernicharo2006}
{Cernicharo}, J., {Pardo}, J.~R., \& {Weiss}, A. 2006, \apjl, 646, L49,
  \dodoi{10.1086/506473}

\bibitem[{{Costagliola} {et~al.}(2011){Costagliola}, {Aalto}, {Rodriguez},
  {Muller}, {Spoon}, {Mart{\'\i}n}, {Per{\'e}z-Torres}, {Alberdi}, {Lindberg},
  {Batejat}, {J{\"u}tte}, {van der Werf}, \& {Lahuis}}]{Costagliola2011}
{Costagliola}, F., {Aalto}, S., {Rodriguez}, M.~I., {et~al.} 2011, in EAS
  Publications Series, Vol.~52, EAS Publications Series, ed. M.~{R{\"o}llig},
  R.~{Simon}, V.~{Ossenkopf}, \& J.~{Stutzki}, 285--286,
  \dodoi{10.1051/eas/1152049}

\bibitem[{{Danielson} {et~al.}(2011){Danielson}, {Swinbank}, {Smail}, {Cox},
  {Edge}, {Weiss}, {Harris}, {Baker}, {De Breuck}, {Geach}, {Ivison}, {Krips},
  {Lundgren}, {Longmore}, {Neri}, \& {Flaquer}}]{Danielson2011}
{Danielson}, A.~L.~R., {Swinbank}, A.~M., {Smail}, I., {et~al.} 2011, \mnras,
  410, 1687, \dodoi{10.1111/j.1365-2966.2010.17549.x}

\bibitem[{{D{\'\i}az-Santos} {et~al.}(2013){D{\'\i}az-Santos}, {Armus},
  {Charmandaris}, {Stierwalt}, {Murphy}, {Haan}, {Inami}, {Malhotra},
  {Meijerink}, {Stacey}, {Petric}, {Evans}, {Veilleux}, {van der Werf}, {Lord},
  {Lu}, {Howell}, {Appleton}, {Mazzarella}, {Surace}, {Xu}, {Schulz},
  {Sanders}, {Bridge}, {Chan}, {Frayer}, {Iwasawa}, {Melbourne}, \&
  {Sturm}}]{DiazSantos2013}
{D{\'\i}az-Santos}, T., {Armus}, L., {Charmandaris}, V., {et~al.} 2013, \apj,
  774, 68, \dodoi{10.1088/0004-637X/774/1/68}

\bibitem[{{Dickens} {et~al.}(2000){Dickens}, {Irvine}, {Snell}, {Bergin},
  {Schloerb}, {Pratap}, \& {Miralles}}]{Dickens}
{Dickens}, J.~E., {Irvine}, W.~M., {Snell}, R.~L., {et~al.} 2000, \apj, 542,
  870, \dodoi{10.1086/317040}

\bibitem[{{Downes} \& {Eckart}(2007)}]{Downes2007}
{Downes}, D., \& {Eckart}, A. 2007, \aap, 468, L57,
  \dodoi{10.1051/0004-6361:20077301}

\bibitem[{{Draine}(2011)}]{Draine2011}
{Draine}, B.~T. 2011, {Physics of the Interstellar and Intergalactic Medium}
  (Princeton, NJ: Princeton Univ. Press)

\bibitem[{Dye {et~al.}(2018)Dye, Furlanetto, Dunne, Eales, Negrello, Nayyeri,
  van~der Werf, Serjeant, Farrah, Michalowski, Baes, Marchetti, Cooray,
  Riechers, \& Amvrosiadis}]{Dye2018}
Dye, S., Furlanetto, C., Dunne, L., {et~al.} 2018, \mnras, 476, 4383,
  \dodoi{10.1093/mnras/sty513}

\bibitem[{Dye {et~al.}(2022)Dye, Eales, Gomez, Jones, Smith, Borsato, Moss,
  Dunne, Maresca, Amvrosiadis, Negrello, Marchetti, Corsini, Ivison, Bendo,
  Bakx, Cooray, Cox, Dannerbauer, Serjeant, Riechers, Temi, \&
  Vlahakis}]{Dye2022}
Dye, S., Eales, S.~A., Gomez, H.~L., {et~al.} 2022, \mnras, 510, 3734,
  \dodoi{10.1093/mnras/stab3569}

\bibitem[{Enia {et~al.}(2018)Enia, Negrello, Gurwell, Dye, Rodighiero,
  Massardi, De~Zotti, Franceschini, Cooray, van~der Werf, Birkinshaw,
  Micha{\l}owski, \& Oteo}]{Enia2018}
Enia, A., Negrello, M., Gurwell, M., {et~al.} 2018, \mnras, 475, 3467,
  \dodoi{10.1093/mnras/sty021}

\bibitem[{{Feruglio} {et~al.}(2015){Feruglio}, {Fiore}, {Carniani},
  {Piconcelli}, {Zappacosta}, {Bongiorno}, {Cicone}, {Maiolino}, {Marconi},
  {Menci}, {Puccetti}, \& {Veilleux}}]{Feruglio2015}
{Feruglio}, C., {Fiore}, F., {Carniani}, S., {et~al.} 2015, \aap, 583, A99,
  \dodoi{10.1051/0004-6361/201526020}

\bibitem[{{Fischer} {et~al.}(2010){Fischer}, {Sturm}, {Gonz{\'a}lez-Alfonso},
  {Graci{\'a}-Carpio}, {Hailey-Dunsheath}, {Poglitsch}, {Contursi}, {Lutz},
  {Genzel}, {Sternberg}, {Verma}, \& {Tacconi}}]{Fischer2010}
{Fischer}, J., {Sturm}, E., {Gonz{\'a}lez-Alfonso}, E., {et~al.} 2010, \aap,
  518, L41, \dodoi{10.1051/0004-6361/201014676}

\bibitem[{{Foreman-Mackey} {et~al.}(2013){Foreman-Mackey}, {Hogg}, {Lang}, \&
  {Goodman}}]{ForemanMackey2013}
{Foreman-Mackey}, D., {Hogg}, D.~W., {Lang}, D., \& {Goodman}, J. 2013,
  Publications of the Astronomical Society of the Pacific, 125.
\newblock \doarXiv{1202.3665}

\bibitem[{{Gao} {et~al.}(2007){Gao}, {Carilli}, {Solomon}, \& {Vanden
  Bout}}]{Gao}
{Gao}, Y., {Carilli}, C.~L., {Solomon}, P.~M., \& {Vanden Bout}, P.~A. 2007,
  \apjl, 660, L93, \dodoi{10.1086/518244}

\bibitem[{{Garc{\'\i}a-Burillo} {et~al.}(2015){Garc{\'\i}a-Burillo}, {Combes},
  {Usero}, {Aalto}, {Colina}, {Alonso-Herrero}, {Hunt}, {Arribas},
  {Costagliola}, {Labiano}, {Neri}, {Pereira-Santaella}, {Tacconi}, \& {van der
  Werf}}]{Garcia2015}
{Garc{\'\i}a-Burillo}, S., {Combes}, F., {Usero}, A., {et~al.} 2015, \aap, 580,
  A35, \dodoi{10.1051/0004-6361/201526133}

\bibitem[{Gelman \& Rubin(1992)}]{GelmanRubin1992}
Gelman, A., \& Rubin, D.~B. 1992, Statistical Science, 7, 457

\bibitem[{{Giulietti} {et~al.}(2023){Giulietti}, {Lapi}, {Massardi}, {Behiri},
  {Torsello}, {D'Amato}, {Ronconi}, {Perrotta}, \& {Bressan}}]{Giulietti2022b}
{Giulietti}, M., {Lapi}, A., {Massardi}, M., {et~al.} 2023, \apj, 943, 151,
  \dodoi{10.3847/1538-4357/aca53f}

\bibitem[{{Gonz{\'a}lez-Alfonso} {et~al.}(2014){Gonz{\'a}lez-Alfonso},
  {Fischer}, {Aalto}, \& {Falstad}}]{Gonzalez2014}
{Gonz{\'a}lez-Alfonso}, E., {Fischer}, J., {Aalto}, S., \& {Falstad}, N. 2014,
  \aap, 567, A91, \dodoi{10.1051/0004-6361/201423980}

\bibitem[{{Gonz{\'a}lez-Alfonso} {et~al.}(2022){Gonz{\'a}lez-Alfonso},
  {Fischer}, {Goicoechea}, {Yang}, {Pereira-Santaella}, \&
  {Stewart}}]{Gonzalez2022}
{Gonz{\'a}lez-Alfonso}, E., {Fischer}, J., {Goicoechea}, J.~R., {et~al.} 2022,
  \aap, 666, L3, \dodoi{10.1051/0004-6361/202244700}

\bibitem[{{Gonz{\'a}lez-Alfonso} {et~al.}(2008){Gonz{\'a}lez-Alfonso}, {Smith},
  {Ashby}, {Fischer}, {Spinoglio}, \& {Grundy}}]{Gonzalez2008}
{Gonz{\'a}lez-Alfonso}, E., {Smith}, H.~A., {Ashby}, M. L.~N., {et~al.} 2008,
  \apj, 675, 303, \dodoi{10.1086/527292}

\bibitem[{{Gonz{\'a}lez-Alfonso} {et~al.}(2004){Gonz{\'a}lez-Alfonso}, {Smith},
  {Fischer}, \& {Cernicharo}}]{Gonzalez2004}
{Gonz{\'a}lez-Alfonso}, E., {Smith}, H.~A., {Fischer}, J., \& {Cernicharo}, J.
  2004, \apj, 613, 247, \dodoi{10.1086/422868}

\bibitem[{{Gonz{\'a}lez-Alfonso} {et~al.}(2010){Gonz{\'a}lez-Alfonso},
  {Fischer}, {Isaak}, {Rykala}, {Savini}, {Spaans}, {van der Werf},
  {Meijerink}, {Israel}, {Loenen}, {Vlahakis}, {Smith}, {Charmandaris},
  {Aalto}, {Henkel}, {Wei{\ss}}, {Walter}, {Greve}, {Mart{\'\i}n-Pintado},
  {Naylor}, {Spinoglio}, {Veilleux}, {Harris}, {Armus}, {Lord}, {Mazzarella},
  {Xilouris}, {Sanders}, {Dasyra}, {Wiedner}, {Kramer}, {Papadopoulos},
  {Stacey}, {Evans}, \& {Gao}}]{Gonzalez2010}
{Gonz{\'a}lez-Alfonso}, E., {Fischer}, J., {Isaak}, K., {et~al.} 2010, \aap,
  518, L43, \dodoi{10.1051/0004-6361/201014664}

\bibitem[{{Gonz{\'a}lez-Alfonso} {et~al.}(2013){Gonz{\'a}lez-Alfonso},
  {Fischer}, {Bruderer}, {M{\"u}ller}, {Graci{\'a}-Carpio}, {Sturm}, {Lutz},
  {Poglitsch}, {Feuchtgruber}, {Veilleux}, {Contursi}, {Sternberg},
  {Hailey-Dunsheath}, {Verma}, {Christopher}, {Davies}, {Genzel}, \&
  {Tacconi}}]{Gonzalez2013A}
{Gonz{\'a}lez-Alfonso}, E., {Fischer}, J., {Bruderer}, S., {et~al.} 2013, \aap,
  550, A25, \dodoi{10.1051/0004-6361/201220466}

\bibitem[{{Gonz{\'a}lez-Alfonso} {et~al.}(2021){Gonz{\'a}lez-Alfonso},
  {Pereira-Santaella}, {Fischer}, {Garc{\'\i}a-Burillo}, {Yang},
  {Alonso-Herrero}, {Colina}, {Ashby}, {Smith}, {Rico-Villas},
  {Mart{\'\i}n-Pintado}, {Cazzoli}, \& {Stewart}}]{Gonzalez2021}
{Gonz{\'a}lez-Alfonso}, E., {Pereira-Santaella}, M., {Fischer}, J., {et~al.}
  2021, \aap, 645, A49, \dodoi{10.1051/0004-6361/202039047}

\bibitem[{{Graci{\'a}-Carpio} {et~al.}(2008){Graci{\'a}-Carpio},
  {Garc{\'\i}a-Burillo}, {Planesas}, {Fuente}, \& {Usero}}]{Gracia}
{Graci{\'a}-Carpio}, J., {Garc{\'\i}a-Burillo}, S., {Planesas}, P., {Fuente},
  A., \& {Usero}, A. 2008, \aap, 479, 703, \dodoi{10.1051/0004-6361:20078223}

\bibitem[{{Harris} {et~al.}(2012){Harris}, {Baker}, {Frayer}, {Smail},
  {Swinbank}, {Riechers}, {van der Werf}, {Auld}, {Baes}, {Bussmann},
  {Buttiglione}, {Cava}, {Clements}, {Cooray}, {Dannerbauer}, {Dariush}, {De
  Zotti}, {Dunne}, {Dye}, {Eales}, {Fritz}, {Gonz{\'a}lez-Nuevo}, {Hopwood},
  {Ibar}, {Ivison}, {Jarvis}, {Maddox}, {Negrello}, {Rigby}, {Smith}, {Temi},
  \& {Wardlow}}]{Harris2012}
{Harris}, A.~I., {Baker}, A.~J., {Frayer}, D.~T., {et~al.} 2012, \apj, 752,
  152, \dodoi{10.1088/0004-637X/752/2/152}

\bibitem[{{Hayward} {et~al.}(2014){Hayward}, {Lanz}, {Ashby}, {Fazio},
  {Hernquist}, {Mart{\'\i}nez-Galarza}, {Noeske}, {Smith}, {Wuyts}, \&
  {Zezas}}]{Hayward2014}
{Hayward}, C.~C., {Lanz}, L., {Ashby}, M. L.~N., {et~al.} 2014, \mnras, 445,
  1598, \dodoi{10.1093/mnras/stu1843}

\bibitem[{{Imanishi} \& {Nakanishi}(2006)}]{ImanishiNakanishi}
{Imanishi}, M., \& {Nakanishi}, K. 2006, \pasj, 58, 813,
  \dodoi{10.1093/pasj/58.5.813}

\bibitem[{{Imanishi} {et~al.}(2016){Imanishi}, {Nakanishi}, \&
  {Izumi}}]{Imanishi2016}
{Imanishi}, M., {Nakanishi}, K., \& {Izumi}, T. 2016, \aj, 152, 218,
  \dodoi{10.3847/0004-6256/152/6/218}

\bibitem[{{Imanishi} {et~al.}(2006){Imanishi}, {Nakanishi}, \&
  {Kohno}}]{Imanishi2006}
{Imanishi}, M., {Nakanishi}, K., \& {Kohno}, K. 2006, \aj, 131, 2888,
  \dodoi{10.1086/503527}

\bibitem[{{Imanishi} {et~al.}(2004){Imanishi}, {Nakanishi}, {Kuno}, \&
  {Kohno}}]{Imanishi2004}
{Imanishi}, M., {Nakanishi}, K., {Kuno}, N., \& {Kohno}, K. 2004, \aj, 128,
  2037, \dodoi{10.1086/424620}

\bibitem[{{Imanishi} {et~al.}(2007){Imanishi}, {Nakanishi}, {Tamura}, {Oi}, \&
  {Kohno}}]{Imanishi2007}
{Imanishi}, M., {Nakanishi}, K., {Tamura}, Y., {Oi}, N., \& {Kohno}, K. 2007,
  \aj, 134, 2366, \dodoi{10.1086/523598}

\bibitem[{{Imanishi} {et~al.}(2009){Imanishi}, {Nakanishi}, {Tamura}, \&
  {Peng}}]{Imanishi2009}
{Imanishi}, M., {Nakanishi}, K., {Tamura}, Y., \& {Peng}, C.-H. 2009, \aj, 137,
  3581, \dodoi{10.1088/0004-6256/137/3/3581}

\bibitem[{{Imanishi} {et~al.}(2010){Imanishi}, {Nakanishi}, {Yamada}, {Tamura},
  \& {Kohno}}]{Imanishi2010}
{Imanishi}, M., {Nakanishi}, K., {Yamada}, M., {Tamura}, Y., \& {Kohno}, K.
  2010, \pasj, 62, 201, \dodoi{10.1093/pasj/62.1.201}

\bibitem[{{Ivison} {et~al.}(2010){Ivison}, {Smail}, {Papadopoulos}, {Wold},
  {Richard}, {Swinbank}, {Kneib}, \& {Owen}}]{Ivison2010}
{Ivison}, R.~J., {Smail}, I., {Papadopoulos}, P.~P., {et~al.} 2010, \mnras,
  404, 198, \dodoi{10.1111/j.1365-2966.2010.16322.x}

\bibitem[{{Izumi} {et~al.}(2013){Izumi}, {Kohno}, {Mart{\'\i}n}, {Sheth},
  {Matsushita}, {Espada}, \& {NGC 1097 collaborators}}]{Izumi2013}
{Izumi}, T., {Kohno}, K., {Mart{\'\i}n}, S., {et~al.} 2013, in Astronomical
  Society of the Pacific Conference Series, Vol. 476, New Trends in Radio
  Astronomy in the ALMA Era: The 30th Anniversary of Nobeyama Radio
  Observatory, ed. R.~{Kawabe}, N.~{Kuno}, \& S.~{Yamamoto}, 309

\bibitem[{{Izumi} {et~al.}(2015){Izumi}, {Kohno}, {Aalto}, {Doi}, {Espada},
  {Fathi}, {Harada}, {Hatsukade}, {Hattori}, {Hsieh}, {Ikarashi}, {Imanishi},
  {Iono}, {Ishizuki}, {Krips}, {Mart{\'\i}n}, {Matsushita}, {Meier}, {Nagai},
  {Nakai}, {Nakajima}, {Nakanishi}, {Nomura}, {Regan}, {Schinnerer}, {Sheth},
  {Takano}, {Tamura}, {Terashima}, {Tosaki}, {Turner}, {Umehata}, \&
  {Wiklind}}]{Izumi2015}
{Izumi}, T., {Kohno}, K., {Aalto}, S., {et~al.} 2015, \apj, 811, 39,
  \dodoi{10.1088/0004-637X/811/1/39}

\bibitem[{{Izumi} {et~al.}(2016){Izumi}, {Kohno}, {Aalto}, {Espada}, {Fathi},
  {Harada}, {Hatsukade}, {Hsieh}, {Imanishi}, {Krips}, {Mart{\'\i}n},
  {Matsushita}, {Meier}, {Nakai}, {Nakanishi}, {Schinnerer}, {Sheth},
  {Terashima}, \& {Turner}}]{Izumi2016}
---. 2016, \apj, 818, 42, \dodoi{10.3847/0004-637X/818/1/42}

\bibitem[{{Jarugula} {et~al.}(2019){Jarugula}, {Vieira}, {Spilker},
  {Apostolovski}, {Aravena}, {B{\'e}thermin}, {de Breuck}, {Chen},
  {Cunningham}, {Dong}, {Greve}, {Hayward}, {Hezaveh}, {Litke}, {Mangian},
  {Narayanan}, {Phadke}, {Reuter}, {Van der Werf}, \& {Weiss}}]{Jarugula2019}
{Jarugula}, S., {Vieira}, J.~D., {Spilker}, J.~S., {et~al.} 2019, \apj, 880,
  92, \dodoi{10.3847/1538-4357/ab290d}

\bibitem[{{Juneau} {et~al.}(2009){Juneau}, {Narayanan}, {Moustakas}, {Shirley},
  {Bussmann}, {Kennicutt}, \& {Vanden Bout}}]{Juneau}
{Juneau}, S., {Narayanan}, D.~T., {Moustakas}, J., {et~al.} 2009, \apj, 707,
  1217, \dodoi{10.1088/0004-637X/707/2/1217}

\bibitem[{{Kennicutt}(1998)}]{Kennicutt1998}
{Kennicutt}, Robert~C., J. 1998, \araa, 36, 189,
  \dodoi{10.1146/annurev.astro.36.1.189}

\bibitem[{{Kennicutt} {et~al.}(2009){Kennicutt}, {Hao}, {Calzetti},
  {Moustakas}, {Dale}, {Bendo}, {Engelbracht}, {Johnson}, \&
  {Lee}}]{Kennicutt2009}
{Kennicutt}, Robert~C., J., {Hao}, C.-N., {Calzetti}, D., {et~al.} 2009, \apj,
  703, 1672, \dodoi{10.1088/0004-637X/703/2/1672}

\bibitem[{{Kennicutt} \& {Evans}(2012)}]{Kennicutt2012}
{Kennicutt}, R.~C., \& {Evans}, N.~J. 2012, \araa, 50, 531,
  \dodoi{10.1146/annurev-astro-081811-125610}

\bibitem[{{Kewley} {et~al.}(2006){Kewley}, {Groves}, {Kauffmann}, \&
  {Heckman}}]{Kewley2006}
{Kewley}, L.~J., {Groves}, B., {Kauffmann}, G., \& {Heckman}, T. 2006, \mnras,
  372, 961, \dodoi{10.1111/j.1365-2966.2006.10859.x}

\bibitem[{{Kirkpatrick} {et~al.}(2019){Kirkpatrick}, {Sharon}, {Keller}, \&
  {Pope}}]{Kirkpatrick2019}
{Kirkpatrick}, A., {Sharon}, C., {Keller}, E., \& {Pope}, A. 2019, \apj, 879,
  41, \dodoi{10.3847/1538-4357/ab223a}

\bibitem[{{Kohno}(2005)}]{Kohno2005}
{Kohno}, K. 2005, in American Institute of Physics Conference Series, Vol. 783,
  The Evolution of Starbursts, ed. S.~{H{\"u}ttmeister}, E.~{Manthey},
  D.~{Bomans}, \& K.~{Weis}, 203--208, \dodoi{10.1063/1.2034987}

\bibitem[{Kormann {et~al.}(1994)Kormann, Schneider, \&
  Bartelmann}]{Kormann1994}
Kormann, R., Schneider, P., \& Bartelmann, M. 1994, \aap, 284, 285.
\newblock \url{https://ui.adsabs.harvard.edu/abs/1994A&A...284..285K}

\bibitem[{{Krips} {et~al.}(2008){Krips}, {Neri}, {Garc{\'\i}a-Burillo},
  {Mart{\'\i}n}, {Combes}, {Graci{\'a}-Carpio}, \& {Eckart}}]{Krips2008}
{Krips}, M., {Neri}, R., {Garc{\'\i}a-Burillo}, S., {et~al.} 2008, \apj, 677,
  262, \dodoi{10.1086/527367}

\bibitem[{{Lapi} {et~al.}(2006){Lapi}, {Shankar}, {Mao}, {Granato}, {Silva},
  {De Zotti}, \& {Danese}}]{Lapi2006}
{Lapi}, A., {Shankar}, F., {Mao}, J., {et~al.} 2006, \apj, 650, 42,
  \dodoi{10.1086/507122}

\bibitem[{{Lapi} {et~al.}(2018){Lapi}, {Pantoni}, {Zanisi}, {Shi}, {Mancuso},
  {Massardi}, {Shankar}, {Bressan}, \& {Danese}}]{Lapi2018}
{Lapi}, A., {Pantoni}, L., {Zanisi}, L., {et~al.} 2018, \apj, 857, 22,
  \dodoi{10.3847/1538-4357/aab6af}

\bibitem[{Lewis(2019)}]{Lewis2019}
Lewis, A. 2019, {GetDist: MCMC sample analysis, plotting and debugging tools},
  Zenodo, \dodoi{10.5281/zenodo.3376730}

\bibitem[{{Lilly} {et~al.}(2013){Lilly}, {Carollo}, {Pipino}, {Renzini}, \&
  {Peng}}]{Lilly2013}
{Lilly}, S.~J., {Carollo}, C.~M., {Pipino}, A., {Renzini}, A., \& {Peng}, Y.
  2013, \apj, 772, 119, \dodoi{10.1088/0004-637X/772/2/119}

\bibitem[{Lin(2018)}]{Lin2018}
Lin, J.-M. 2018, Journal of Imaging, 4, \dodoi{10.3390/jimaging4030051}

\bibitem[{{Liu} {et~al.}(2017){Liu}, {Wei{\ss}}, {Perez-Beaupuits},
  {G{\"u}sten}, {Liu}, {Gao}, {Menten}, {van der Werf}, {Israel}, {Harris},
  {Martin-Pintado}, {Requena-Torres}, \& {Stutzki}}]{Liu2017}
{Liu}, L., {Wei{\ss}}, A., {Perez-Beaupuits}, J.~P., {et~al.} 2017, \apj, 846,
  5, \dodoi{10.3847/1538-4357/aa81b4}

\bibitem[{{Maloney} {et~al.}(1996){Maloney}, {Hollenbach}, \&
  {Tielens}}]{Maloney1996}
{Maloney}, P.~R., {Hollenbach}, D.~J., \& {Tielens}, A.~G.~G.~M. 1996, \apj,
  466, 561, \dodoi{10.1086/177532}

\bibitem[{{Mancuso} {et~al.}(2017){Mancuso}, {Lapi}, {Prandoni}, {Obi},
  {Gonzalez-Nuevo}, {Perrotta}, {Bressan}, {Celotti}, \&
  {Danese}}]{Mancuso2017}
{Mancuso}, C., {Lapi}, A., {Prandoni}, I., {et~al.} 2017, \apj, 842, 95,
  \dodoi{10.3847/1538-4357/aa745d}

\bibitem[{Maresca {et~al.}(2022)Maresca, Dye, Amvrosiadis, Bendo, Cooray,
  De~Zotti, Dunne, Eales, Furlanetto, Gonz{\'a}lez-Nuevo, Greener, Ivison,
  Lapi, Negrello, Riechers, Serjeant, Tergolina, \& Wardlow}]{Maresca2022}
Maresca, J., Dye, S., Amvrosiadis, A., {et~al.} 2022, \mnras,
  \dodoi{10.1093/mnras/stac585}

\bibitem[{{Mart{\'\i}n} {et~al.}(2015){Mart{\'\i}n}, {Kohno}, {Izumi}, {Krips},
  {Meier}, {Aladro}, {Matsushita}, {Takano}, {Turner}, {Espada}, {Nakajima},
  {Terashima}, {Fathi}, {Hsieh}, {Imanishi}, {Lundgren}, {Nakai}, {Schinnerer},
  {Sheth}, \& {Wiklind}}]{Martin}
{Mart{\'\i}n}, S., {Kohno}, K., {Izumi}, T., {et~al.} 2015, \aap, 573, A116,
  \dodoi{10.1051/0004-6361/201425105}

\bibitem[{Massardi {et~al.}(2017)Massardi, Enia, Negrello, Mancuso, Lapi,
  Vignali, Gilli, Burkutean, Danese, \& Zotti}]{Massardi2017}
Massardi, M., Enia, A. F.~M., Negrello, M., {et~al.} 2017, \aap,
  \dodoi{10.1051/0004-6361/201731751}

\bibitem[{{Massardi} {et~al.}(2021){Massardi}, {Stoehr}, {Bendo}, {Bonato},
  {Brand}, {Galluzzi}, {Guglielmetti}, {Liuzzo}, {Marchili}, {Richards},
  {Rygl}, {Bedosti}, {Giannetti}, {Stagni}, {Knapic}, {Sponza}, {Fuller}, \&
  {Muxlow}}]{Massardi2021}
{Massardi}, M., {Stoehr}, F., {Bendo}, G.~J., {et~al.} 2021, \pasp, 133,
  085001, \dodoi{10.1088/1538-3873/ac159c}

\bibitem[{{McMullin} {et~al.}(2007){McMullin}, {Waters}, {Schiebel}, {Young},
  \& {Golap}}]{McMullin2007}
{McMullin}, J.~P., {Waters}, B., {Schiebel}, D., {Young}, W., \& {Golap}, K.
  2007, in Astronomical Society of the Pacific Conference Series, Vol. 376,
  Astronomical Data Analysis Software and Systems XVI, ed. R.~A. {Shaw},
  F.~{Hill}, \& D.~J. {Bell}, 127

\bibitem[{{Meijerink} \& {Spaans}(2005)}]{Meijerink}
{Meijerink}, R., \& {Spaans}, M. 2005, \aap, 436, 397,
  \dodoi{10.1051/0004-6361:20042398}

\bibitem[{{Mingozzi} {et~al.}(2018){Mingozzi}, {Vallini}, {Pozzi}, {Vignali},
  {Mignano}, {Gruppioni}, {Talia}, {Cimatti}, {Cresci}, \&
  {Massardi}}]{Mingozzi18}
{Mingozzi}, M., {Vallini}, L., {Pozzi}, F., {et~al.} 2018, \mnras, 474, 3640,
  \dodoi{10.1093/mnras/stx3011}

\bibitem[{{Murphy} {et~al.}(2011){Murphy}, {Condon}, {Schinnerer}, {Kennicutt},
  {Calzetti}, {Armus}, {Helou}, {Turner}, {Aniano}, {Beir{\~a}o}, {Bolatto},
  {Brandl}, {Croxall}, {Dale}, {Donovan Meyer}, {Draine}, {Engelbracht},
  {Hunt}, {Hao}, {Koda}, {Roussel}, {Skibba}, \& {Smith}}]{Murphy2011}
{Murphy}, E.~J., {Condon}, J.~J., {Schinnerer}, E., {et~al.} 2011, \apj, 737,
  67, \dodoi{10.1088/0004-637X/737/2/67}

\bibitem[{Nightingale {et~al.}(2021)Nightingale, Hayes, Kelly, Amvrosiadis,
  Etherington, He, Li, Cao, Frawley, Cole, Enia, Frenk, Harvey, Li, Massey,
  Negrello, \& Robertson}]{Nightingale2021}
Nightingale, J., Hayes, R., Kelly, A., {et~al.} 2021, The Journal of Open
  Source Software, 6, 2825, \dodoi{10.21105/joss.02825}

\bibitem[{Nightingale \& Dye(2015)}]{Nightingale2015}
Nightingale, J.~W., \& Dye, S. 2015, \mnras, 452, 2940,
  \dodoi{10.1093/mnras/stv1455}

\bibitem[{Nightingale {et~al.}(2018)Nightingale, Dye, \&
  Massey}]{Nightingale2018}
Nightingale, J.~W., Dye, S., \& Massey, R.~J. 2018, \mnras, 478, 4738,
  \dodoi{10.1093/mnras/sty1264}

\bibitem[{{Omont} {et~al.}(2013){Omont}, {Yang}, {Cox}, {Neri}, {Beelen},
  {Bussmann}, {Gavazzi}, {van der Werf}, {Riechers}, {Downes}, {Krips}, {Dye},
  {Ivison}, {Vieira}, {Wei{\ss}}, {Aguirre}, {Baes}, {Baker}, {Bertoldi},
  {Cooray}, {Dannerbauer}, {De Zotti}, {Eales}, {Fu}, {Gao}, {Gu{\'e}lin},
  {Harris}, {Jarvis}, {Lehnert}, {Leeuw}, {Lupu}, {Menten}, {Micha{\l}owski},
  {Negrello}, {Serjeant}, {Temi}, {Auld}, {Dariush}, {Dunne}, {Fritz},
  {Hopwood}, {Hoyos}, {Ibar}, {Maddox}, {Smith}, {Valiante}, {Bock},
  {Bradford}, {Glenn}, \& {Scott}}]{Omont2013}
{Omont}, A., {Yang}, C., {Cox}, P., {et~al.} 2013, \aap, 551, A115,
  \dodoi{10.1051/0004-6361/201220811}

\bibitem[{{Oteo} {et~al.}(2017){Oteo}, {Zhang}, {Yang}, {Ivison}, {Omont},
  {Bremer}, {Bussmann}, {Cooray}, {Cox}, {Dannerbauer}, {Dunne}, {Eales},
  {Furlanetto}, {Gavazzi}, {Gao}, {Greve}, {Nayyeri}, {Negrello}, {Neri},
  {Riechers}, {Tunnard}, {Wagg}, \& {Van der Werf}}]{Oteo2017}
{Oteo}, I., {Zhang}, Z.~Y., {Yang}, C., {et~al.} 2017, \apj, 850, 170,
  \dodoi{10.3847/1538-4357/aa8ee3}

\bibitem[{{Pantoni} {et~al.}(2019){Pantoni}, {Lapi}, {Massardi}, {Goswami}, \&
  {Danese}}]{Pantoni2019}
{Pantoni}, L., {Lapi}, A., {Massardi}, M., {Goswami}, S., \& {Danese}, L. 2019,
  \apj, 880, 129, \dodoi{10.3847/1538-4357/ab2adc}

\bibitem[{{Pensabene} {et~al.}(2022){Pensabene}, {van der Werf}, {Decarli},
  {Ba{\~n}ados}, {Meyer}, {Riechers}, {Venemans}, {Walter}, {Wei{\ss}},
  {Brusa}, {Fan}, {Wang}, \& {Yang}}]{Pensabene2022}
{Pensabene}, A., {van der Werf}, P., {Decarli}, R., {et~al.} 2022, \aap, 667,
  A9, \dodoi{10.1051/0004-6361/202243406}

\bibitem[{{Pereira-Santaella} {et~al.}(2017){Pereira-Santaella},
  {Gonz{\'a}lez-Alfonso}, {Usero}, {Garc{\'\i}a-Burillo},
  {Mart{\'\i}n-Pintado}, {Colina}, {Alonso-Herrero}, {Arribas}, {Cazzoli},
  {Rico}, {Rigopoulou}, \& {Storchi Bergmann}}]{Pereira-Santaella2017}
{Pereira-Santaella}, M., {Gonz{\'a}lez-Alfonso}, E., {Usero}, A., {et~al.}
  2017, \aap, 601, L3, \dodoi{10.1051/0004-6361/201730851}

\bibitem[{{Planck Collaboration} {et~al.}(2020){Planck Collaboration},
  {Aghanim}, {Akrami}, {Ashdown}, {Aumont}, {Baccigalupi}, {Ballardini},
  {Banday}, {Barreiro}, {Bartolo}, {Basak}, {Battye}, {Benabed}, {Bernard},
  {Bersanelli}, {Bielewicz}, {Bock}, {Bond}, {Borrill}, {Bouchet}, {Boulanger},
  {Bucher}, {Burigana}, {Butler}, {Calabrese}, {Cardoso}, {Carron},
  {Challinor}, {Chiang}, {Chluba}, {Colombo}, {Combet}, {Contreras}, {Crill},
  {Cuttaia}, {de Bernardis}, {de Zotti}, {Delabrouille}, {Delouis}, {Di
  Valentino}, {Diego}, {Dor{\'e}}, {Douspis}, {Ducout}, {Dupac}, {Dusini},
  {Efstathiou}, {Elsner}, {En{\ss}lin}, {Eriksen}, {Fantaye}, {Farhang},
  {Fergusson}, {Fernandez-Cobos}, {Finelli}, {Forastieri}, {Frailis},
  {Fraisse}, {Franceschi}, {Frolov}, {Galeotta}, {Galli}, {Ganga},
  {G{\'e}nova-Santos}, {Gerbino}, {Ghosh}, {Gonz{\'a}lez-Nuevo}, {G{\'o}rski},
  {Gratton}, {Gruppuso}, {Gudmundsson}, {Hamann}, {Handley}, {Hansen},
  {Herranz}, {Hildebrandt}, {Hivon}, {Huang}, {Jaffe}, {Jones}, {Karakci},
  {Keih{\"a}nen}, {Keskitalo}, {Kiiveri}, {Kim}, {Kisner}, {Knox},
  {Krachmalnicoff}, {Kunz}, {Kurki-Suonio}, {Lagache}, {Lamarre}, {Lasenby},
  {Lattanzi}, {Lawrence}, {Le Jeune}, {Lemos}, {Lesgourgues}, {Levrier},
  {Lewis}, {Liguori}, {Lilje}, {Lilley}, {Lindholm}, {L{\'o}pez-Caniego},
  {Lubin}, {Ma}, {Mac{\'\i}as-P{\'e}rez}, {Maggio}, {Maino}, {Mandolesi},
  {Mangilli}, {Marcos-Caballero}, {Maris}, {Martin}, {Martinelli},
  {Mart{\'\i}nez-Gonz{\'a}lez}, {Matarrese}, {Mauri}, {McEwen}, {Meinhold},
  {Melchiorri}, {Mennella}, {Migliaccio}, {Millea}, {Mitra},
  {Miville-Desch{\^e}nes}, {Molinari}, {Montier}, {Morgante}, {Moss}, {Natoli},
  {N{\o}rgaard-Nielsen}, {Pagano}, {Paoletti}, {Partridge}, {Patanchon},
  {Peiris}, {Perrotta}, {Pettorino}, {Piacentini}, {Polastri}, {Polenta},
  {Puget}, {Rachen}, {Reinecke}, {Remazeilles}, {Renzi}, {Rocha}, {Rosset},
  {Roudier}, {Rubi{\~n}o-Mart{\'\i}n}, {Ruiz-Granados}, {Salvati}, {Sandri},
  {Savelainen}, {Scott}, {Shellard}, {Sirignano}, {Sirri}, {Spencer},
  {Sunyaev}, {Suur-Uski}, {Tauber}, {Tavagnacco}, {Tenti}, {Toffolatti},
  {Tomasi}, {Trombetti}, {Valenziano}, {Valiviita}, {Van Tent}, {Vibert},
  {Vielva}, {Villa}, {Vittorio}, {Wandelt}, {Wehus}, {White}, {White},
  {Zacchei}, \& {Zonca}}]{PlanckCollaboration2020}
{Planck Collaboration}, {Aghanim}, N., {Akrami}, Y., {et~al.} 2020, \aap, 641,
  A6, \dodoi{10.1051/0004-6361/201833910}

\bibitem[{{Pozzi} {et~al.}(2017){Pozzi}, {Vallini}, {Vignali}, {Talia},
  {Gruppioni}, {Mingozzi}, {Massardi}, \& {Andreani}}]{Pozzi17}
{Pozzi}, F., {Vallini}, L., {Vignali}, C., {et~al.} 2017, \mnras, 470, L64,
  \dodoi{10.1093/mnrasl/slx077}

\bibitem[{{Pratap} {et~al.}(1997){Pratap}, {Dickens}, {Snell}, {Miralles},
  {Bergin}, {Irvine}, \& {Schloerb}}]{Pratap}
{Pratap}, P., {Dickens}, J.~E., {Snell}, R.~L., {et~al.} 1997, \apj, 486, 862,
  \dodoi{10.1086/304553}

\bibitem[{{Privon} {et~al.}(2020){Privon}, {Ricci}, {Aalto}, {Viti}, {Armus},
  {D{\'\i}az-Santos}, {Gonz{\'a}lez-Alfonso}, {Iwasawa}, {Jeff}, {Treister},
  {Bauer}, {Evans}, {Garg}, {Herrero-Illana}, {Mazzarella}, {Larson}, {Blecha},
  {Barcos-Mu{\~n}oz}, {Charmandaris}, {Stierwalt}, \&
  {P{\'e}rez-Torres}}]{Privon2020}
{Privon}, G.~C., {Ricci}, C., {Aalto}, S., {et~al.} 2020, \apj, 893, 149,
  \dodoi{10.3847/1538-4357/ab8015}

\bibitem[{{Rangwala} {et~al.}(2011){Rangwala}, {Maloney}, {Glenn}, {Wilson},
  {Rykala}, {Isaak}, {Baes}, {Bendo}, {Boselli}, {Bradford}, {Clements},
  {Cooray}, {Fulton}, {Imhof}, {Kamenetzky}, {Madden}, {Mentuch}, {Sacchi},
  {Sauvage}, {Schirm}, {Smith}, {Spinoglio}, \& {Wolfire}}]{Rangwala2011}
{Rangwala}, N., {Maloney}, P.~R., {Glenn}, J., {et~al.} 2011, \apj, 743, 94,
  \dodoi{10.1088/0004-637X/743/1/94}

\bibitem[{Ravasi \& Vasconcelos(2020)}]{Ravasi2020}
Ravasi, M., \& Vasconcelos, I. 2020, SoftwareX, 11, 100361,
  \dodoi{https://doi.org/10.1016/j.softx.2019.100361}

\bibitem[{{Riechers} {et~al.}(2006){Riechers}, {Walter}, {Carilli}, {Weiss},
  {Bertoldi}, {Menten}, {Knudsen}, \& {Cox}}]{Riechers}
{Riechers}, D.~A., {Walter}, F., {Carilli}, C.~L., {et~al.} 2006, \apjl, 645,
  L13, \dodoi{10.1086/505908}

\bibitem[{{Sakamoto} {et~al.}(2010){Sakamoto}, {Aalto}, {Evans}, {Wiedner}, \&
  {Wilner}}]{Sakamoto}
{Sakamoto}, K., {Aalto}, S., {Evans}, A.~S., {Wiedner}, M.~C., \& {Wilner},
  D.~J. 2010, \apjl, 725, L228, \dodoi{10.1088/2041-8205/725/2/L228}

\bibitem[{{Sani} {et~al.}(2012){Sani}, {Davies}, {Sternberg},
  {Graci{\'a}-Carpio}, {Hicks}, {Krips}, {Tacconi}, {Genzel}, {Vollmer},
  {Schinnerer}, {Garc{\'\i}a-Burillo}, {Usero}, \& {Orban de Xivry}}]{Sani2012}
{Sani}, E., {Davies}, R.~I., {Sternberg}, A., {et~al.} 2012, \mnras, 424, 1963,
  \dodoi{10.1111/j.1365-2966.2012.21333.x}

\bibitem[{{Schleicher} {et~al.}(2010){Schleicher}, {Spaans}, \&
  {Klessen}}]{Schleicher2010}
{Schleicher}, D.~R.~G., {Spaans}, M., \& {Klessen}, R.~S. 2010, \aap, 513, A7,
  \dodoi{10.1051/0004-6361/200913467}

\bibitem[{{Sch{\"o}ier} {et~al.}(2005){Sch{\"o}ier}, {van der Tak}, {van
  Dishoeck}, \& {Black}}]{LAMDA}
{Sch{\"o}ier}, F.~L., {van der Tak}, F.~F.~S., {van Dishoeck}, E.~F., \&
  {Black}, J.~H. 2005, \aap, 432, 369, \dodoi{10.1051/0004-6361:20041729}

\bibitem[{{Simpson} {et~al.}(2014){Simpson}, {Swinbank}, {Smail}, {Alexander},
  {Brandt}, {Bertoldi}, {de Breuck}, {Chapman}, {Coppin}, {da Cunha},
  {Danielson}, {Dannerbauer}, {Greve}, {Hodge}, {Ivison}, {Karim}, {Knudsen},
  {Poggianti}, {Schinnerer}, {Thomson}, {Walter}, {Wardlow}, {Wei{\ss}}, \&
  {van der Werf}}]{Simpson2014}
{Simpson}, J.~M., {Swinbank}, A.~M., {Smail}, I., {et~al.} 2014, \apj, 788,
  125, \dodoi{10.1088/0004-637X/788/2/125}

\bibitem[{{Snell} {et~al.}(2011){Snell}, {Narayanan}, {Yun}, {Heyer}, {Chung},
  {Irvine}, {Erickson}, \& {Liu}}]{Snell2011}
{Snell}, R.~L., {Narayanan}, G., {Yun}, M.~S., {et~al.} 2011, \aj, 141, 38,
  \dodoi{10.1088/0004-6256/141/2/38}

\bibitem[{{Soifer} {et~al.}(1999){Soifer}, {Neugebauer}, {Matthews}, {Becklin},
  {Ressler}, {Werner}, {Weinberger}, \& {Egami}}]{Soifer1999}
{Soifer}, B.~T., {Neugebauer}, G., {Matthews}, K., {et~al.} 1999, \apj, 513,
  207, \dodoi{10.1086/306855}

\bibitem[{{Solomon} {et~al.}(1997){Solomon}, {Downes}, {Radford}, \&
  {Barrett}}]{Solomon1997}
{Solomon}, P.~M., {Downes}, D., {Radford}, S.~J.~E., \& {Barrett}, J.~W. 1997,
  \apj, 478, 144, \dodoi{10.1086/303765}

\bibitem[{{Solomon} \& {Vanden Bout}(2005)}]{Solomon2005}
{Solomon}, P.~M., \& {Vanden Bout}, P.~A. 2005, \araa, 43, 677,
  \dodoi{10.1146/annurev.astro.43.051804.102221}

\bibitem[{Speagle(2020)}]{Speagle2020}
Speagle, J.~S. 2020, \mnras, 493, 3132, \dodoi{10.1093/mnras/staa278}

\bibitem[{{Vallini} {et~al.}(2018){Vallini}, {Pallottini}, {Ferrara},
  {Gallerani}, {Sobacchi}, \& {Behrens}}]{Vallini2018}
{Vallini}, L., {Pallottini}, A., {Ferrara}, A., {et~al.} 2018, \mnras, 473,
  271, \dodoi{10.1093/mnras/stx2376}

\bibitem[{{van der Werf} {et~al.}(2011){van der Werf}, {Berciano Alba},
  {Spaans}, {Loenen}, {Meijerink}, {Riechers}, {Cox}, {Wei{\ss}}, \&
  {Walter}}]{vanderWerf2011}
{van der Werf}, P.~P., {Berciano Alba}, A., {Spaans}, M., {et~al.} 2011, \apjl,
  741, L38, \dodoi{10.1088/2041-8205/741/2/L38}

\bibitem[{{van Dishoeck} {et~al.}(2021){van Dishoeck}, {Kristensen}, {Mottram},
  {Benz}, {Bergin}, {Caselli}, {Herpin}, {Hogerheijde}, {Johnstone}, {Liseau},
  {Nisini}, {Tafalla}, {van der Tak}, {Wyrowski}, {Baudry}, {Benedettini},
  {Bjerkeli}, {Blake}, {Braine}, {Bruderer}, {Cabrit}, {Cernicharo}, {Choi},
  {Coutens}, {de Graauw}, {Dominik}, {Fedele}, {Fich}, {Fuente}, {Furuya},
  {Goicoechea}, {Harsono}, {Helmich}, {Herczeg}, {Jacq}, {Karska}, {Kaufman},
  {Keto}, {Lamberts}, {Larsson}, {Leurini}, {Lis}, {Melnick}, {Neufeld},
  {Pagani}, {Persson}, {Shipman}, {Taquet}, {van Kempen}, {Walsh}, {Wampfler},
  {Y{\i}ld{\i}z}, \& {WISH Team}}]{vanDishoeck2021}
{van Dishoeck}, E.~F., {Kristensen}, L.~E., {Mottram}, J.~C., {et~al.} 2021,
  \aap, 648, A24, \dodoi{10.1051/0004-6361/202039084}

\bibitem[{{Vishwas} {et~al.}(2018){Vishwas}, {Ferkinhoff}, {Nikola},
  {Parshley}, {Schoenwald}, {Stacey}, {Higdon}, {Higdon}, {Weiss},
  {G{\"u}sten}, \& {Menten}}]{Vishwas2018}
{Vishwas}, A., {Ferkinhoff}, C., {Nikola}, T., {et~al.} 2018, \apj, 856, 174,
  \dodoi{10.3847/1538-4357/aab354}

\bibitem[{Warren \& Dye(2003)}]{Warren2003}
Warren, S.~J., \& Dye, S. 2003, \apj, 590, 673, \dodoi{10.1086/375132}

\bibitem[{{Wei{\ss}} {et~al.}(2007){Wei{\ss}}, {Downes}, {Neri}, {Walter},
  {Henkel}, {Wilner}, {Wagg}, \& {Wiklind}}]{Weiss2007}
{Wei{\ss}}, A., {Downes}, D., {Neri}, R., {et~al.} 2007, \aap, 467, 955,
  \dodoi{10.1051/0004-6361:20066117}

\bibitem[{{Wei{\ss}} {et~al.}(2010){Wei{\ss}}, {Requena-Torres}, {G{\"u}sten},
  {Garc{\'\i}a-Burillo}, {Harris}, {Israel}, {Klein}, {Kramer}, {Lord},
  {Martin-Pintado}, {R{\"o}llig}, {Stutzki}, {Szczerba}, {van der Werf},
  {Philipp-May}, {Yorke}, {Akyilmaz}, {Gal}, {Higgins}, {Marston}, {Roberts},
  {Schl{\"o}der}, {Schultz}, {Teyssier}, {Whyborn}, \& {Wunsch}}]{Weiss2010}
{Wei{\ss}}, A., {Requena-Torres}, M.~A., {G{\"u}sten}, R., {et~al.} 2010, \aap,
  521, L1, \dodoi{10.1051/0004-6361/201015078}

\bibitem[{{Yang} {et~al.}(2013){Yang}, {Gao}, {Omont}, {Liu}, {Isaak},
  {Downes}, {van der Werf}, \& {Lu}}]{Yang2013}
{Yang}, C., {Gao}, Y., {Omont}, A., {et~al.} 2013, \apjl, 771, L24,
  \dodoi{10.1088/2041-8205/771/2/L24}

\bibitem[{{Yang} {et~al.}(2020){Yang}, {Gonz{\'a}lez-Alfonso}, {Omont},
  {Pereira-Santaella}, {Fischer}, {Beelen}, \& {Gavazzi}}]{Yang2020}
{Yang}, C., {Gonz{\'a}lez-Alfonso}, E., {Omont}, A., {et~al.} 2020, \aap, 634,
  L3, \dodoi{10.1051/0004-6361/201937319}

\bibitem[{{Yang} {et~al.}(2016){Yang}, {Omont}, {Beelen},
  {Gonz{\'a}lez-Alfonso}, {Neri}, {Gao}, {van der Werf}, {Wei{\ss}}, {Gavazzi},
  {Falstad}, {Baker}, {Bussmann}, {Cooray}, {Cox}, {Dannerbauer}, {Dye},
  {Gu{\'e}lin}, {Ivison}, {Krips}, {Lehnert}, {Micha{\l}owski}, {Riechers},
  {Spaans}, \& {Valiante}}]{Yang2016}
{Yang}, C., {Omont}, A., {Beelen}, A., {et~al.} 2016, \aap, 595, A80,
  \dodoi{10.1051/0004-6361/201628160}

\bibitem[{{Yang} {et~al.}(2017){Yang}, {Omont}, {Beelen}, {Gao}, {van der
  Werf}, {Gavazzi}, {Zhang}, {Ivison}, {Lehnert}, {Liu}, {Oteo},
  {Gonz{\'a}lez-Alfonso}, {Dannerbauer}, {Cox}, {Krips}, {Neri}, {Riechers},
  {Baker}, {Micha{\l}owski}, {Cooray}, \& {Smail}}]{Yang2017}
---. 2017, \aap, 608, A144, \dodoi{10.1051/0004-6361/201731391}

\end{thebibliography}

\bibliographystyle{aasjournal}

\end{document}